\documentclass{article}

\usepackage{amssymb}
\usepackage{amsmath}
\usepackage{amsthm}
\usepackage{epubtk}
\usepackage{ifpdf}

\showlistoftablesfalse

\newtheorem{theorem}{Theorem}[section]

\newtheorem{cor}{Corollary}[section]
\newtheorem{prop}{Proposition}[section]

\newcommand{\be}{\begin{equation}}
\newcommand{\ee}{\end{equation}}
\newcommand{\bea}{\begin{eqnarray}}
\newcommand{\eea}{\end{eqnarray}}
\newcommand{\td}{\,\mathrm{d}}

\newcommand{\AdS}[1]{AdS\sub{#1}}
\newcommand{\CFT}[1]{CFT\sub{#1}}
\newcommand{\dS}[1]{dS\sub{#1}}

\begin{document}

\title{ {\bf Classification of Near-Horizon Geometries of Extremal Black Holes}}

\author{%
\epubtkAuthorData{{\bf Hari K.\ Kunduri}}{%
Department of Mathematics and Statistics, \\ 
Memorial University of Newfoundland, \\
St John's NL A1C 4P5, Canada}{%
hkkunduri@mun.ca} {%
}%
\and
\epubtkAuthorData{{\bf James Lucietti}}{%
School of Mathematics and Maxwell Institute of Mathematical Sciences \\ 
University of Edinburgh, \\
King's Buildings, Edinburgh, EH9 3JZ, UK}{%
j.lucietti@ed.ac.uk}{%
}%
}

\date{}
\maketitle

\begin{abstract}
Any spacetime containing a degenerate Killing horizon, such as an extremal black hole, possesses a well-defined notion of a near-horizon geometry. We review such near-horizon geometry solutions in a variety of dimensions and theories in a unified manner. We discuss various general results including horizon topology and near-horizon symmetry enhancement. We also discuss the status of the classification of near-horizon geometries in theories ranging from vacuum gravity to Einstein--Maxwell theory and supergravity theories. Finally, we discuss applications to the classification of extremal black holes and various related topics. Several new results are presented and open problems are highlighted throughout.
\end{abstract}

\epubtkKeywords{Extremal black holes, Near-horizon geometry}

 \newpage

 \tableofcontents



\newpage
\section{Introduction}
\label{section:introduction}

Equilibrium black-hole solutions to Einstein's equations have been known since the advent of general relativity. The most obvious reason such solutions are of physical interest is the expectation that they arise as the end state of catastrophic gravitational collapse of some suitably localised matter distribution. A less obvious reason such solutions are important is that they have played a key role in guiding studies of quantum gravity. 

Classically equilibrium black holes are inert objects. However, the laws of black hole mechanics have a formal similarity with the laws of thermodynamics~\cite{Bardeen:1973gs}. By studying quantum fields in a black-hole background, Hawking demonstrated that this is not a mere analogy and in fact quantum mechanically black holes are a thermodynamic system~\cite{Hawking:1974sw}. The black hole radiates at small temperature
\begin{equation}
T_H = \frac{\hbar \kappa}{2\pi} \label{temp}
\end{equation}
proportional to the surface gravity $\kappa$ of the horizon, and possesses a large entropy
\begin{equation}
S_{\mathrm{BH}} = \frac{A_H}{4\hbar} \label{entropy}
\end{equation}
proportional to the area $A_H$ of spatial cross sections of the horizon.\epubtkFootnote{We work in geometrised units throughout.}

Deriving these semi-classical thermodynamic formulae from statistical mechanics requires a microscopic understanding of the ``degrees of freedom'' of the black hole. This has been a major motivation and driving force for quantum gravity research over the last four decades, although it is fair to say this is still poorly understood. It is in this context that \emph{extremal black holes} are central. By definition, an equilibrium black hole is extremal (or extreme or degenerate), if the surface gravity 
\begin{equation}
\kappa = 0.
\end{equation}
It immediately follows that the Hawking temperature vanishes -- extremal black holes do not radiate after all! Hence, even semi-classically, extremal black holes are inert objects\epubtkFootnote{Of course, a charged extremal black hole can always discharge in the presence of charged matter.} and as such are expected to have a simpler quantum description. 

The main purpose of this review is to discuss the classification of the \emph{near-horizon geometries} of extremal black holes. There are a number of different motivations for considering this, which we will briefly review. As alluded to above, the principle reasons stem from studies in quantum gravity. In Section~\ref{sec1:string} and \ref{sec1:AdS} we discuss the various ways extremal black holes and their near-horizon geometries have appeared in modern studies of quantum gravity. In Section~\ref{sec1:classification} we discuss the more general black hole classification problem (which is also partly motivated by quantum gravity), and how near-horizon geometries provide a systematic tool for investigating certain aspects of this problem for extremal black holes.

\subsection{Black holes in string theory}
\label{sec1:string}

To date, the most promising candidate for a theory of quantum gravity is string theory. Famously, this predicts the existence of extra spatial dimensions. As discussed above, an important test for any candidate theory of quantum gravity is that it is able to explain the semi-classical formulae (\ref{temp}) and (\ref{entropy}). A major breakthrough of Strominger and Vafa~\cite{Strominger:1996sh} was to use string theory to supply a microscopic derivation of Eq.~(\ref{entropy}) for certain \emph{five dimensional} extremal black holes. 

The black holes in question are higher-dimensional counterparts of the extremal Reissner--Nordstr\"om (RN) black holes. These are \emph{supersymmetric} solutions to a supergravity theory, which can be obtained as a consistent Kaluza--Klein (KK) reduction of the ten/eleven dimensional supergravity that describes string theory at low energies. Supersymmetry was crucial for their calculation, since non-renormalisation results allowed them to perform a weak-coupling string calculation (involving certain D-brane configurations) to deduce the entropy of the semi-classical black holes that exist in the strong coupling regime (see~\cite{David:2002wn} for a review). This was quickly generalised to supersymmetric black holes with angular momentum~\cite{Breckenridge:1996is} and supersymmetric four dimensional black holes~\cite{Maldacena:1996gb}.

An important assumption in these string theory calculations is that a given black hole is uniquely specified by its conserved charges: its mass/energy, electric charge and angular momentum. For four dimensional Einstein--Maxwell theory this follows from the black-hole uniqueness theorem (see~\cite{Chrusciel:2012jk} for a review). However, Emparan and Reall demonstrated that black hole uniqueness is violated for five-dimensional asymptotically-flat vacuum spacetimes~\cite{Emparan:2001wn}. This was via the construction of an explicit counterexample, the \emph{black ring}, which is a black hole whose spatial horizon topology is $S^1\times S^2$. Together with the higher-dimensional analogues of the Kerr black hole (which have spherical topology) found by Myers and Perry~\cite{Myers:1986un}, this established that the conserved charges are not sufficient to specify a black hole uniquely and also that other horizon topologies are possible. Indeed, this remarkable result motivated the study of stationary black holes in higher dimensional spacetimes. Subsequently, a supersymmetric black ring was constructed~\cite{Elvang:2004rt, Elvang:2004ds} that coexists with the spherical topology black holes used in the original entropy calculations. Although microscopic descriptions for the black rings have been proposed~\cite{Bena:2004tk, Cyrier:2004hj}, it is fair to say that the description of black hole non-uniqueness within string theory is not properly understood (see~\cite{Emparan:2006mm} for a brief review).

Any supersymmetric black hole is necessarily extremal. Since the Strominger--Vafa calculation, a substantial amount of work has been directed at removing the assumption of supersymmetry and extremality, with the ultimate goal being a string theory derivation of the thermodynamics of realistic black holes such as the four dimensional Schwarzschild or Kerr black holes. Although little progress has been made in the description of such non-extremal black holes, significant progress has been made for extremal non-supersymmetric black holes. In particular, this has been via the black-hole \emph{attractor mechanism.}

The attractor mechanism is the phenomenon that the entropy of certain extremal black holes in string theory does not depend on the moduli of the theory (typically scalar fields in the supergravity theory). This was first observed for supersymmetric static black holes~\cite{Ferrara:1995ih,Strominger:1996kf,Ferrara:1996dd} although later it was realised it is valid for generic extremal black holes~\cite{Goldstein:2005hq, Sen:2005wa, Astefanesei:2006dd}. The key idea is that extremal black holes have a well-defined \emph{near-horizon geometry} that typically possesses an \AdS{2} symmetry. Assuming this symmetry, it was argued that the entropy must be independent of the moduli of the theory. Motivated by this, it was then proved that the near-horizon geometry of any extremal black hole in this context must in fact possess an \AdS{2} symmetry~\cite{Kunduri:2007vf}. This general attractor mechanism thus ensures the black hole entropy is independent of the string coupling, so it can be safely computed at weak coupling. This shows that in fact it is extremality, rather than supersymmetry, that is behind the success of the string theory microscopic calculations~\cite{Dabholkar:2006tb, Astefanesei:2006sy}. This also explains the success of the entropy calculations for extremal, non-supersymmetric black holes in four and five dimensions, e.g., \cite{Reall:2007jv, Emparan:2006it, Horowitz:2007xq}.

\subsection{Gauge/gravity duality}
\label{sec1:AdS}

A significant breakthrough in the study of quantum gravity is the Anti de Sitter/Conformal Field Theory (AdS/CFT) duality \cite{Maldacena:1997re, Witten:1998qj,Witten:1998zw, Gubser:1998bc}. In principle, AdS/CFT asserts a fully non-perturbative equivalence of quantum gravity in asymptotically AdS spacetimes with a conformally invariant quantum field theory in one lower spatial dimension. This is an explicit realisation of a `holographic principle' underlying quantum gravity \cite{tHooft:1993gx,Susskind:1998dq}. 

A crucial feature of the duality is that classical gravity in AdS spacetimes is dual to the strongly-coupled regime in the CFT. This provides a precise framework to analyse the microscopic description of black holes in terms of well-defined quantum field theories. The duality was originally proposed~\cite{Maldacena:1997re} in the context of string theory on $\mathrm{AdS}_5 \times S^5$, in which case the CFT is the maximally supersymmetric four dimensional $SU(N)$ Yang--Mills gauge theory. However, the original idea has subsequently been generalised to a number of dimensions and theories, and such \emph{gauge/gravity dualities} are believed to hold more generally.

Classical non-extremal AdS black holes represent high-energy thermal states in the dual theory at large $N$ and strong coupling~\cite{Witten:1998zw}. Strong coupling poses the main obstacle to providing a precise entropy counting for such black holes, although excellent qualitative agreement can be found via extrapolating weak coupling calculations~\cite{Gubser:1996de, Berman:1999mh, Hawking:1999dp}. Precise agreement has been achieved \cite{Strominger:1997eq} for the asymptotically \AdS{3} Ba\~nados--Teitelboim--Zanelli (BTZ) black hole (even for the non-extremal case) \cite{Banados:1992wn}. This is because generically any theory of quantum gravity in \AdS{3} must be described by a two dimensional \CFT{2} with a specific central charge~\cite{Brown:1986nw}. This allows one to compute the entropy from Cardy's formula, without requiring an understanding of the microscopic degrees of freedom. In fact the string theory calculations described in Section \ref{sec1:string} can be thought of as applications of this method. This is because the black holes in question can be viewed (from a higher-dimensional viewpoint) as black strings with an \AdS{3} factor in the near-horizon geometry, allowing \AdS{3}/\CFT{2} to be applied. 

A major open problem is to successfully account for black-hole entropy using a higher dimensional CFT. The best understood case is when the CFT is four dimensional, in which case the black holes are asymptotically \AdS{5}. As in the original string calculations, a strategy to overcome the strong-coupling problem is to focus on supersymmetric AdS black holes. The dual CFT states then belong to certain Bogomol'nyi--Prasad--Sommerfield (BPS) representations, and so weak-coupling calculations may not receive quantum corrections. It turns out that such black-hole solutions must rotate and hence are difficult to construct. In fact the first examples of supersymmetric \AdS{5} black holes~\cite{Gutowski:2004ez,Gutowski:2004yv} were found via a classification of near-horizon geometries. Subsequently, a more general four-parameter family of black-hole solutions were found~\cite{Chong:2005hr, Kunduri:2006ek}. The problem of classifying all supersymmetric AdS black holes motivated further classifications of near-horizon geometries, which have ruled out the possibility of other types of black hole such as supersymmetric AdS black rings~\cite{Kunduri:2008tk, Kunduri:2006uh, GGS}. Despite significant effort, a microscopic derivation of the entropy from the CFT has not yet been achieved in this context. Due to the low amount of supersymmetry preserved by the black holes, it appears that non-zero coupling effects must be taken into account~\cite{Kinney:2005ej,Berkooz:2006wc,Kim:2006he,Berkooz:2008gc,Chang:2013fba}.

The original AdS/CFT duality was established by arguing that there exist two complementary descriptions of the low energy physics of the string theory of a stack of $N$ extremal D3 black branes. Near the horizon of the D-brane only low energy excitations survive, which are thus described by string theory in the $\mathrm{AdS}_5 \times S^5$ near-horizon geometry. On the other hand, the massless degrees of freedom on a D-brane arrange themselves into (super) $SU(N)$ Yang--Mills theory. It is natural to extend this idea to extremal black holes. Since extremal black holes typically possess an \AdS{2} factor in their near-horizon geometry, one may then hope that an \AdS{2}/\CFT{1} duality~\cite{Strominger:1998yg} could provide a microscopic description of such black holes. Unfortunately this duality is not as well understood as the higher-dimensional cases. However, it appears that the black-hole entropy can be reproduced from the degeneracy of the ground states of the dual conformal mechanics~\cite{Maldacena:1998uz, Sen:2008vm}.

Another recently-developed approach is to generalise the \AdS{3}/\CFT{2} derivation of the BTZ entropy to describe more general black holes. This involves finding an asymptotic symmetry group of a given near-horizon geometry that contains a Virasoro algebra and then applying Cardy's formula. This was applied to the extremal Kerr black hole and led to the Kerr/CFT correspondence~\cite{Guica:2008mu}, which is a proposal that quantum gravity in the near-horizon geometry of extremal Kerr is described by a chiral \CFT{2} (see the reviews \cite{Bredberg:2011hp, Compere:2012jk}). This technique has provided a successful counting of the entropy of many black holes. However, as in the \AdS{2}/\CFT{1} case, the duality is poorly understood and it appears that non-trivial excitations of the near-horizon geometry do not exist \cite{Dias:2009ex, Amsel:2009ev}. The relation between these various approaches has been investigated in the special case of BTZ~\cite{Balasubramanian:2009bg}. Furthermore, a \CFT{2} description has been proposed for a certain class of near-extremal black holes, which possess a local near-horizon \AdS{3} factor but a vanishing horizon area in the extremal limit~\cite{SheikhJabbaria:2011gc, Johnstone:2013ioa}.\epubtkFootnote{The emergence of an \AdS{3} factor for solutions with certain null singularities has been previously observed~\cite{Emparan:2001rp, Bardeen:1999px}. We will only consider regular horizons, so the area of spatial cross sections of the horizon is necessarily non-zero.}

Recently, ideas from the gauge/gravity duality have been used to model certain phase transitions that occur in condensed-matter systems, such as superfluids or superconductors~\cite{Gubser:2008px, Hartnoll:2008vx}. The key motivation to this line of research, in contrast to the above, is to use knowledge of the gravitational system to learn about strongly-coupled field theories. Charged black holes in AdS describe the finite temperature phases. The non-superconducting phase is dual to the standard planar RN-AdS black hole of Einstein--Maxwell theory, which is stable at high enough temperature. However, at low enough temperatures this solution is unstable to the formation of a charged scalar condensate. The dominant phase at low temperatures is a charged black hole with scalar hair, which describes the superconducting phase. This instability of (near)-extremal RN-AdS can be understood as occurring due to the violation of the Breitenlohner--Freedmann bound in the \AdS{2} factor of the near-horizon geometry. A similar result has also been shown for neutral rotating AdS black holes~\cite{Dias:2010ma}. The near-horizon \AdS{2} has also been used to provide holographic descriptions of quantum critical points and Fermi surfaces~\cite{Faulkner:2009wj}.

\subsection{Black hole classification}
\label{sec1:classification}

The classification of higher-dimensional stationary black-hole solutions to Einstein's equations is a major open problem in higher dimensional general relativity (see \cite{Emparan:2008eg, Hollands:2012xy} for reviews). As explained above, the main physical motivation stems from studies of quantum gravity and high energy physics. However, its study is also of intrinsic value both physically and mathematically. On the physical side we gain insight into the behaviour of gravity in higher-dimensional spacetimes, which in turn often provides renewed perspective for the classic four-dimensional results. On the mathematical side, solutions to Einstein's equation have also been of interest in differential geometry~\cite{Besse}.\epubtkFootnote{Although spacetimes correspond to Lorentzian metrics, one can often analytically continue these to complete Riemannian metrics. Indeed, the first example of an inhomogeneous Einstein metric on a compact manifold was found by Page, by taking a certain limit of the Kerr--de~Sitter metrics~\cite{Page}, giving a metric on $\mathbb{CP}^2 \# \overline{\mathbb{CP}}^2$. }

In four dimensions the black-hole uniqueness theorem provides an answer to the classification problem for asymptotically-flat black-hole solutions of Einstein--Maxwell theory (see~\cite{Chrusciel:2012jk} for a review).\epubtkFootnote{Albeit, under some technical assumptions such as analyticity of the metric.} However, in higher dimensions, uniqueness is violated even for asymptotically-flat vacuum black holes. To date, the explicit black-hole solutions known are the spherical horizon topology Myers--Perry black holes~\cite{Myers:1986un} and the black rings~\cite{Emparan:2001wn, Pomeransky:2006bd} that have $S^1\times S^2$ horizon topology (see~\cite{Emparan:2008eg} for a review). If one allows for more complicated boundary conditions, such as KK asymptotics, then uniqueness is violated even for static black holes (see, e.g., \cite{Horowitz:2011cq}). Although of less obvious physical relevance, the investigation of \emph{asymptotically-flat vacuum black holes} is the fundamental starting case to consider in higher dimensions, since such solutions can be viewed as limits of black holes with more general asymptotics such as KK, AdS and matter fields.

General results have been derived that constrain the \emph{topology} of black holes. By generalising Hawking's horizon topology theorem~\cite{Hawking:1971vc} to higher dimensions, Galloway and Schoen~\cite{Galloway:2005mf} have shown that the spatial topology of the horizon must be such that it admits a positive scalar curvature metric (i.e., positive Yamabe type). Horizon topologies are further constrained by topological censorship~\cite{Friedman:1993ty, Chrusciel:1994tr}. For asymptotically-flat (and globally AdS) black holes, this implies that there must be a simply connected (oriented) cobordism between cross sections of the horizon and the $(D-2)$-dimensional sphere at spatial infinity.\epubtkFootnote{Two oriented manifolds are said to be oriented-cobordant if there exists some other oriented manifold whose boundary (with the induced orientation) is their disjoint union.} In $D=4$, this rules out toroidal black holes, although for $D=5$ it imposes no constraint. For $D>5$ this does provide a logically independent constraint in addition to the positive Yamabe condition~\cite{Reall:2002bh, Kunduri:2010vg}. 

General results have also been derived that constrain the \emph{symmetries} of black hole spacetimes. Firstly, asymptotically-flat, static vacuum black holes must be spherically symmetric and hence are uniquely given by the higher dimensional Schwarzschild black hole~\cite{Gibbons:2002bh}.\epubtkFootnote{Similarly, any such black hole in Einstein--Maxwell-dilaton theory with a purely electric field strength must be given by the RN solution~\cite{Gibbons:2002av, Gibbons:2002ju}.} By generalising Hawking's rigidity theorem~\cite{Hawking:1971vc}, it was shown that asymptotically-flat and AdS stationary non-extremal rotating black holes must admit at least $\mathbb{R} \times U(1)$ isometry \cite{Hollands:2006rj,Moncrief:2008mr} (for partial results pertaining to extremal rotating black holes see \cite{Hollands:2008wn}). This additional isometry can be used to further refine the allowed set of $D=5$ black hole horizon topologies \cite{Hollands:2010qy}.

An important class of spacetimes, for which substantial progress towards classification has been made, are the generalised Weyl solutions~\cite{Emparan:2001wk, Harmark:2004rm}. By definition these possess an $\mathbb{R}\times U(1)^{D-3}$ symmetry group and generalise $D=4$ stationary and axisymmetric spacetimes. As in the $D=4$ case, it turns out that the vacuum Einstein equations for spacetimes with these symmetries are integrable. For $D=5$ this structure has allowed one to prove certain uniqueness theorems for asymptotically-flat black holes with $\mathbb{R} \times U(1)^2$ symmetry, using the same methods as for $D=4$~\cite{Hollands:2007aj}. Furthermore, this has led to the explicit construction of several novel asymptotically-flat, stationary, multi--black-hole vacuum solutions, the first example being a (non-linear) superposition of a black ring and a spherical black hole~\cite{Elvang:2007rd}. For $D>5$, the symmetry of these spacetimes is not compatible with asymptotic flatness, that would require the number of commuting rotational symmetries to not exceed $\lfloor \frac{D-1}{2} \rfloor $, the rank of the rotation group $SO(D-1)$. In this case, Weyl solutions are compatible with KK asymptotics and this has been used to prove uniqueness theorems for (uniform) KK black holes/strings~\cite{Hollands:2008fm}.

The general topology and symmetry constraints discussed above become increasingly weak as one increases the number of dimensions. Furthermore, there is evidence that black hole uniqueness will be violated much more severely as one increases the dimensions. For example, by an analysis of gravitational perturbations of the Myers--Perry black hole, evidence for a large new family of black holes was found~\cite{Reall:2012it}. Furthermore, the investigation of ``blackfolds'', where the long-range effective dynamics of certain types of black holes can be analysed, suggests that many new types of black holes should exist, see \cite{Emparan:2011zz} for a review. In the absence of new ideas, it appears that the general classification problem for asymptotically-flat black holes is hopelessly out of reach. 

As discussed in Section~\ref{sec1:AdS}, the black-hole--classification problem for asymptotically AdS black holes is of interest in the context of gauge/gravity dualities. The presence of a (negative) cosmological constant renders the problem even more complicated. Even in four dimensions, there is no analogue of the uniqueness theorems. One reason for this comes from the fact that Einstein's equations with a cosmological constant for stationary and axisymmetric metrics are not integrable. Hence the standard method used to prove uniqueness of Kerr cannot be generalised. This also means that constructing charged generalisations, in four and higher dimensions, from a neutral seed can not be accomplished using standard solution generating methods. In fact, perturbation analyses of known solutions (e.g., Kerr-AdS and higher dimensional generalisations), reveal that if the black holes rotate sufficiently fast, super-radiant instabilities exist~\cite{Hawking:1999dp, Cardoso:2004hs, Kunduri:2006qa, Cardoso:2006wa}. It has been suggested that the endpoint of these instabilities are new types of non-axisymmetric black-hole solutions that are not stationary in the usual sense, but instead invariant under a single Killing field co-rotating with the horizon~\cite{Kunduri:2006qa}. (Examples of such solutions have been constructed in a scalar gravity theory~\cite{Dias:2011at}). A further complication in AdS comes from the choice of asymptotic boundary conditions. In AdS there is the option of replacing the sphere on the conformal boundary with more general manifolds, in which case topological censorship permits more black-hole topologies~\cite{Galloway:1999bp}. 

It is clear that supersymmetry provides a technically-simplifying assumption to classifying spacetimes, since it reduces the problem to solving first-order Killing spinor equations rather than the full Einstein equations. A great deal of work has been devoted to developing systematic techniques for constructing supersymmetric solutions, most notably in five-dimensional ungauged supergravity \cite{Gauntlett:2002nw} and gauged supergravity \cite{Gauntlett:2003fk}. These have been used to construct new five-dimensional supersymmetric black-hole solutions, which are asymptotically flat~\cite{Elvang:2004ds, Elvang:2004rt} and AdS~\cite{Gutowski:2004ez, Gutowski:2004yv, Kunduri:2006ek}, respectively. Furthermore, the first uniqueness theorem for asymptotically-flat supersymmetric black holes was proved using these methods~\cite{Reall:2002bh}.

Less obviously, it turns out that the weaker assumption of extremality can also be used as a simplifying assumption, as follows. The event horizon of all known extremal black holes is a degenerate Killing horizon with compact spatial cross sections $H$. It turns out that restricting Einstein's equations for a $D$-dimensional spacetime to a degenerate horizon gives a set of geometric equations for the induced metric on such $(D-2)$-dimensional cross sections $H$, that depend only on quantities intrinsic to $H$. By studying solutions to this problem of Riemannian geometry on a compact manifold $H$, one can thus consider the possible horizon geometries (and topologies) independently of the full parent spacetime. This strategy often also works in cases where the standard black hole uniqueness/classification techniques do not apply (e.g., AdS, higher dimensions etc.).

One can understand this feature of degenerate horizons in terms of the near-horizon limit, which, as we explain in Section~\ref{section:nearhorizongeometry}, exists for any spacetime containing a degenerate horizon. This allows one to define an associated near-horizon geometry, which must also satisfy the full Einstein equations~\cite{Reall:2002bh, Kunduri:2007vf}, so classifying near-horizon geometries is then equivalent to classifying possible horizon geometries (and topologies). Indeed, the topic of this review is the \emph{classification of near-horizon geometries} in diverse dimensions and theories. 

The classification of near-horizon geometries allows one to explore in a simplified setup the main issues that appear in the general black-hole classification problem, such as the horizon topology, spacetime symmetry and the ``number'' of solutions. The main drawback of this approach is that the existence of a near-horizon geometry solution does not guarantee the existence of a corresponding black-hole solution (let alone its uniqueness).\epubtkFootnote{Indeed counterexamples are known in both senses.} Hence, one must keep this in mind when interpreting near-horizon classifications in the context of black holes, although definite statements can be learned. In particular, one can use this method to rule out possible black-hole horizon topologies, for if one can classify near-horizon geometries completely and a certain horizon topology does not appear, this implies there can be no extremal black hole with that horizon topology either. A notable example of this method has been a proof of the non-existence of supersymmetric AdS black rings in $D=5$ minimal gauged supergravity~\cite{Kunduri:2006uh, GGS}.

\subsection{This review}

\subsubsection{Scope}
In this review we will consider near-horizon geometries of solutions to Einstein's equations, in all dimensions $D \geq 3$, that contain smooth degenerate horizons. Our aim is to provide a unified treatment of such near-horizon solutions in diverse theories with matter content ranging from vacuum gravity, to Einstein--Maxwell theories and various (minimal) supergravity theories. 

We do not assume the near-horizon geometry arises as a near-horizon limit of a black-hole solution. However, due to the application to extremal black holes we will mostly consider horizons that admit a spatial cross section that is \emph{compact}. As we will see in various setups, compactness often allows one to avoid explicitly solving the full Einstein equations and instead use global arguments to constrain the space of solutions. As a result, the classification of near-horizon geometries with non-compact horizon cross sections is a much more difficult problem about which less is known. This is relevant to the classification of extremal black \emph{branes} and therefore lies outside the scope of this article. Nevertheless, along the way, we will point out cases in which classification has been achieved without the assumption of compactness, and in Section~\ref{app:branes} we briefly discuss extremal branes in this context.

Although this is a review article, we streamline some of the known proofs and we also present several new results that fill in various gaps in the literature. Most notably, we fully classify \emph{three dimensional} near-horizon geometries in vacuum gravity and Einstein--Maxwell--Chern--Simons theories, in Section~\ref{vac:3d} and \ref{gauge:3d} respectively, and classify \emph{homogeneous} near-horizon geometries in five dimensional Einstein--Maxwell--Chern--Simons theories in Section~\ref{gauge:homo}.

\subsubsection{Organisation}

In Section~\ref{section:nearhorizongeometry} we provide key definitions, introduce a suitably general notion of a near-horizon geometry and set up the Einstein equations for such near-horizon geometries.

In Section~\ref{section:generalresults} we review various general results that constrain the topology and symmetry of near-horizon geometries. This includes the horizon topology theorem and various near-horizon symmetry enhancement theorems. We also discuss the physical charges one can calculate from a near-horizon geometry.

In Section~\ref{sec:vac} we discuss the classification of near-horizon geometries in vacuum gravity, including a cosmological constant, organised by dimension. In cases where classification results are not known, we describe the known solutions.

In Section~\ref{sec:susy} we discuss the classification of supersymmetric near-horizon geometries in various supergravity theories, organised by dimension.

In Section~\ref{sec:gauge} we discuss the classification of general near-horizon geometries coupled to gauge fields. This includes $D=3,4,5$ Einstein--Maxwell theories, allowing for Chern--Simons terms where appropriate, and $D=4$ Einstein--Yang--Mills theory.

In Section~\ref{section:applications} we discuss various applications of near-horizon geometries and related topics. This includes uniqueness/classification theorems of the corresponding extremal black-hole solutions, stability of near-horizon geometries and extremal black holes, geometric inequalities, analytic continuation of near-horizon geometries, and extremal branes and their near-horizon geometries.



\newpage
\section{Degenerate Horizons and Near-Horizon Geometry}
\label{section:nearhorizongeometry}

\subsection{Coordinate systems and near-horizon limit}

In this section we will introduce a general notion of a near-horizon geometry. This requires us to first introduce some preliminary constructions. Let $\mathcal{N}$ be a smooth\epubtkFootnote{In fact our constructions only assume the metric is $C^2$ in a neighbourhood of the horizon. This encompasses certain examples of multi--black-hole spacetimes with non-smooth horizons~\cite{Candlish:2007fh}.} codimension-1 null hypersurface in a $D$ dimensional spacetime $(M,g)$. In a neighbourhood of any such hypersurface there exists an adapted coordinate chart called \emph{Gaussian null coordinates} that we now recall~\cite{IM, Friedrich:1998wq}. 

Let $N$ be a vector field normal to $\mathcal{N}$ whose integral curves are future-directed null geodesic generators of $\mathcal{N}$. In general these will be non-affinely parameterised so on $\mathcal{N}$ we have $\nabla_N N=\kappa N$ for some function $\kappa$. Now let $H$ denote a smooth $(D-2)$-dimensional spacelike submanifold of $\mathcal{N}$, such that each integral curve of $N$ crosses $H$ exactly once: we term $H$ a \emph{cross section} of $\mathcal{N}$ and assume such submanifolds exist. On $H$ choose arbitrary local coordinates $(x^a)$, for $a=1, \dots, D-2$, containing some point $p \in H$. Starting from $p \in H$, consider the point in $\mathcal{N}$ a parameter value $v$ along the integral curve of $N$. Now assign coordinates $(v,x^a)$ to such a point, i.e., we extend the functions $x^a$ into $\mathcal{N}$ by keeping them constant along such a curve. This defines a set of coordinates $(v,x^a)$ in a tubular neighbourhood of the integral curves of $N$ through $p \in H$, such that $N=\partial /\partial v$. Since $N$ is normal to $\mathcal{N}$ we have $N \cdot N=g_{vv}=0$ and $N \cdot \partial/ \partial x^a= g_{va}=0$ on $\mathcal{N}$. 

We now extend these coordinates into a neighbourhood of $\mathcal{N}$ in $M$ as follows. For any point $q \in \mathcal{N}$ contained in the above coordinates $(v,x^a)$, let $L$ be the unique past-directed null vector satisfying $L\cdot N =1$ and $L \cdot \partial/ \partial x^a=0$. Now starting at $q$, consider the point in $M$ an affine parameter value $r$ along the null geodesic with tangent vector $L$. Define the coordinates of such a point in $M$ by $(v,r,x^a)$, i.e., the functions $v,x^a$ are extended into $M$ by requiring them to be constant along such null geodesics. This provides coordinates $(v,r,x^a)$ defined in a neighbourhood of $\mathcal{N}$ in $M$, as required.

We extend the definitions of $N$ and $L$ into $M$ by $N=\partial/ \partial v$ and $L=\partial /\partial r$. By construction the integral curves of $L=\partial / \partial r$ are null geodesics and hence $g_{rr}=0$ everywhere in the neighbourhood of $\mathcal{N}$ in $M$ in question. Furthermore, using the fact that $N$ and $L$ commute (they are coordinate vector fields), we have
 \begin{equation}
 \nabla_L (L \cdot N)= L \cdot (\nabla_L N)= L \cdot (\nabla_N L)= \frac{1}{2}\nabla_N (L \cdot L)=0
 \end{equation}
and therefore $L \cdot N=g_{vr}=1$ for all $r$. A similar argument shows $L \cdot \partial /\partial x^a =g_{ra}=0$ for all $r$. 

These considerations show that, in a neighbourhood of $\mathcal{N}$ in $M$, the spacetime metric $g$ written in Gaussian null coordinates $(v,r,x^a)$ is of the form
\be
\label{GNC}
g= 2\, \td v \left( \td r + r h_a \td x^a +\tfrac{1}{2} rf \td v \right) +\gamma_{ab}\td x^a \td x^b \,,
\ee
where $\mathcal{N}$ is the hypersurface $\{ r=0 \}$, the metric components $f, h_a, \gamma_{ab}$ are smooth functions of all the coordinates, and $\gamma_{ab}$ is an invertible $(D-2)\times (D-2)$ matrix. This coordinate chart is unique up to choice of cross section $H$ and choice of coordinates $(x^a)$ on $H$. Upon a change of coordinates on $H$ the quantities $f, h_a, \gamma_{ab}$ transform as a function, 1-form and non-degenerate metric, respectively. Hence they may be thought of as components of a globally-defined function, 1-form and Riemannian metric on $H$.

The coordinates developed above are valid in the neighbourhood of any smooth null hypersurface $\mathcal{N}$. In this work we will in fact be concerned with smooth \emph{Killing horizons}. These are null hypersurfaces that possess a normal that is a Killing field $K$ in $M$. Hence we may set $N=K$ in the above construction. Since $K= \partial / \partial v$ we deduce that in the neighbourhood of a Killing horizon $\mathcal{N}$, the metric can be written as Eq.~(\ref{GNC}) where the functions $f,h_a, \gamma_{ab}$ are all independent of the coordinate $v$. Using the Killing property one can rewrite $\nabla_K K= \kappa K$ as $\td (K \cdot K)= -2\kappa K$ on $\mathcal{N}$, where $\kappa$ is now the usual surface gravity of a Killing horizon. 

We may now introduce the main objects we will study in this work: \emph{degenerate} Killing horizons. These are defined as Killing horizons $\mathcal{N}$ such that the normal Killing field $K$ is tangent to \emph{affinely} parameterised null geodesics on $\mathcal{N}$, i.e., $\kappa \equiv 0$. Therefore, $\td (K \cdot K)|_{\mathcal{N}} =0$, which implies that in Gaussian null coordinates $(\partial_r g_{vv})|_{r=0}=0$. It follows that $g_{vv}= r^2 F$ for some smooth function $F$. Therefore, in the neighbourhood of any smooth degenerate Killing horizon the metric in Gaussian null coordinates reads
\be
\label{degenerate}
g= 2\,\td v \left( \td r +r h_a(r,x) \td x^a +\tfrac{1}{2} r^2 F(r,x)\td v \right)+\gamma_{ab}(r,x)\td x^a \td x^b \,.
\ee

We are now ready to define the \emph{near-horizon geometry} of a $D$-dimensional spacetime $(M,g)$ containing such a degenerate horizon. Given any $\epsilon>0$, consider the diffeomorphism defined by $v \to v/ \epsilon$ and $r \to \epsilon r$. The metric in Gaussian null coordinates transforms $g \to g_\epsilon$ where $g_\epsilon$ is given by Eq.~(\ref{degenerate}) with the replacements $F(r,x)\to F(\epsilon r, x)$, $h_a(r, x) \to h_a (\epsilon r, x)$ and $\gamma_{ab}(r,x)\to \gamma_{ab}(\epsilon r, x)$. The near-horizon limit is then defined as the $\epsilon \to 0$ limit of $g_\epsilon$. It is clear this limit always exists since all metric functions are smooth at $r=0$. The resulting metric is called the \emph{near-horizon geometry} and is given by
\be
\label{nhg}
g_{\mathrm{NH}} = 2\, \td v \left( \td r + r h_a(x) \td x^a+ \tfrac{1}{2} r^2 F(x) \td v \right) +\gamma_{ab}(x) \td x^a \td x^b \,,
\ee
where $F(x)=F(0,x), h_a(x)=h_a(0,x)$ and $\gamma_{ab}(x)=\gamma_{ab}(0,x)$. Notice that the $r$ dependence of the metric is completely fixed. In fact the near-horizon geometry is completely specified by the following geometric data on the $(D-2)$-dimensional cross section $H$: a smooth function $F$, 1-form $h_a$ and Riemannian metric $\gamma_{ab}$. We will often refer to the triple of data $(F,h_a, \gamma_{ab})$ on $H$ as the near-horizon data. 

Intuitively, the near-horizon limit is a scaling limit that focuses on the spacetime near the horizon $\mathcal{N}$. We emphasise that the degenerate assumption $g_{vv}=O(r^2)$ is crucial for defining this limit and such a general notion of a near-horizon limit does not exist for a non-degenerate Killing horizon.

\subsection{Curvature of near-horizon geometry}

As we will see, geometric equations (such as Einstein's equations) for a near-horizon geometry can be equivalently written as geometric equations defined purely on a $(D-2)$-dimensional cross section manifold $H$ of the horizon. In this section we will write down general formulae relating the curvature of a near-horizon geometry to the curvature of the horizon $H$. For convenience we will denote the dimension of $H$ by $n=D-2$.

It is convenient to introduce a null-orthonormal frame for the near-horizon metric (\ref{nhg}), denoted by $(e^A)$, where $A=(+,-, a)$, $a=1, \dots, n$ and
\be
e^+=\td v, \qquad 
e^-= \td r+rh_a \hat{e}^a + \tfrac{1}{2}r^2F\td v, \qquad 
e^a=\hat{e}^a \,, \label{frame}
\ee
so that $g= \eta_{AB} e^A e^B = 2e^+e^- +e^ae^a$, where $\hat{e}^a$ are vielbeins for the horizon metric $\gamma=\hat{e}^a \hat{e}^a$.\epubtkFootnote{To avoid proliferation of indices we will denote both coordinate and vielbein indices on $H$ by lower case latin letters $a,b,\dots$.} The dual basis vectors are 
\be
e_+= \partial_v- \tfrac{1}{2}Fr^2\partial_r, \qquad 
e_-= \partial_r, \qquad 
e_a = \hat{\partial}_a- rh_a \partial_r,
\ee
where $\hat{\partial}_a$ denote the dual vectors to $\hat{e}^a$. 
The connection 1-forms satisfy $\td e^A= -\omega^A_{\phantom{A}B} \wedge e^B$ and are given by
\begin{eqnarray}
\nonumber \omega_{+-} &=& rFe^+ + \tfrac{1}{2}h_ae^a \,, \qquad
\omega_{+a}= \tfrac{1}{2}r^2(\hat{\partial}_aF-Fh_a)e^+-\tfrac{1}{2}h_ae^- +r \hat{\nabla}_{[a} h_{b]} e^b \,, \\ 
\omega_{-a} &=& -\tfrac{1}{2}h_a e^+ \,, \qquad \qquad \quad \omega_{ab} = \hat{\omega}_{ab}- r \hat{\nabla}_{[a} h_{b]} e^+ \,,
\end{eqnarray}
where $\hat{\omega}_{ab}$ and $\hat{\nabla}_a$ are the connection 1-forms and  Levi-Civita connection of the metric $\gamma_{ab}$ on $H$ respectively.
The curvature two-forms defined by $\Omega_{AB}= \td \omega_{AB}+ \omega_{AC} \wedge \omega^{C}_{\phantom{C}B}$ give the Riemann tensor in this basis using $\Omega_{AB}= \tfrac{1}{2} R_{ABCD} e^C \wedge e^D$.
The curvature two forms are:
\begin{eqnarray}
\Omega_{ab} &=& \hat{\Omega}_{ab} + e^+ \wedge e^- \hat{\nabla}_{[a} h_{b]} + r e^+ \wedge e^d \left( - h_d \hat{\nabla}_{[a} h_{b]}+ \hat{\nabla}_d \hat{\nabla}_{[a} h_{b]} + \tfrac{1}{2}h_a \hat{\nabla}_{[b} h_{d]} - \tfrac{1}{2}h_b \hat{\nabla}_{[a} h_{d]} \right), \nonumber \\ \nonumber \Omega_{+-} &=& \left( \tfrac{1}{4}h_ah_a-F \right)e^+\wedge e^- + r e^b \wedge e^+ \left( \hat{\partial}_bF-Fh_b+\tfrac{1}{2}h_a \hat{\nabla}_{[a} h_{b]} \right) + \tfrac{1}{2} \hat{\nabla}_{[a} h_{b]} e^a \wedge e^b, \\ \Omega_{+a} &=& r^2 e^+ \wedge e^d \left[ \left(-\tfrac{1}{2} \hat{\nabla}_d+ h_d \right)( \hat{\partial}_a F -Fh_a) + \tfrac{1}{2}F \hat{\nabla}_{[a} h_{d]} + \hat{\nabla}_{[c} h_{a]} \hat{\nabla}_{[c} h_{d]} + \tfrac{1}{2} h_{[a} \hat{\nabla}_{d]} F \right] \nonumber \\ &+& r e^+ \wedge e^- \left( h_aF- \hat{\partial}_aF- \tfrac{1}{2} h_b \hat{\nabla}_{[b} h_{a]} \right)+ \tfrac{1}{2}e^- \wedge e^b \left( \hat{\nabla}_a h_b- \tfrac{1}{2}h_ah_b \right) \nonumber \\ &+& r e^b \wedge e^d \left( - \hat{\nabla}_d \hat{\nabla}_{[a} h_{b]} + \tfrac{1}{2} h_a \hat{\nabla}_{[d} h_{b]} - \tfrac{1}{2} h_b \hat{\nabla}_{[a} h_{d]} \right), \nonumber \\ 
\Omega_{-a} &=& \tfrac{1}{2}e^+ \wedge e^b \left( \hat{\nabla}_b h_a - \tfrac{1}{2}h_ah_b \right) \, ,
\end{eqnarray}
where $\hat{\Omega}_{ab}$ is the curvature of $\hat{\omega}_{ab}$ on $H$.
The non-vanishing components of the Ricci tensor are thus given by:
\begin{eqnarray}
\label{RicciNH}
R_{+-}&=& F-\tfrac{1}{2}h_ah_a+ \tfrac{1}{2}\hat{\nabla}_ah_a \,, \nonumber \\
R_{ab} &=& \hat{R}_{ab} + \hat{\nabla}_{(a} h_{b)} - \tfrac{1}{2}h_ah_b \,, \nonumber \\
R_{++} &=& r^2\left[-\tfrac{1}{2}\hat{\nabla}^2 F + \tfrac{3}{2}h^{a}\hat{\nabla}_{a}F + \tfrac{1}{2}F\hat{\nabla}^{a}h_{a} - Fh_ah_a + \hat{\nabla}_{[c}h_{a]}\hat{\nabla}_{[c}h_{a]} \right] \equiv r^2 S_{++} \,, \nonumber \\
R_{+a} &=& r\left[\hat{\nabla}_{a}F - Fh_a -2h_b\hat{\nabla}_{[a}h_{b]} + \hat{\nabla}_{b}\hat{\nabla}_{[a}h_{b]} \right] \equiv r S_{+a} \,, \label{NHRicci}
\end{eqnarray}
where $\hat{R}_{ab}$ is the Ricci tensor of the metric $\gamma_{ab}$ on $H$. The spacetime contracted Bianchi identity implies the following identities on $H$:
\begin{eqnarray}
S_{++} &=& - \tfrac{1}{2}(\hat{\nabla}^a-2h^a) S_{+a} \,, \label{Sid1} \\
S_{+a} &=& -\hat{\nabla}^b [R_{ba} - \tfrac{1}{2}\gamma_{ba}( R^c_{~c}+2R_{+-})] + h^bR_{ba} - h_a R_{+-} \,, \label{Sid2}
\end{eqnarray}
which may also be verified directly from the above expressions.

It is worth noting that the following components of the Weyl tensor automatically vanish: $C_{-a-b}=0$ and $C_{-abc}=0$. This means that $e_-=\partial_r$ is a multiple Weyl aligned null direction and hence \emph{any} near-horizon geometry is at least algebraically special of type II within the classification of~\cite{Coley:2004jv}. In fact, it can be checked that the null geodesic vector field $\partial_r$ has vanishing expansion, shear and twist and therefore any near-horizon geometry is a Kundt spacetime.\epubtkFootnote{A Kundt spacetime is one that admits a null geodesic vector field with vanishing expansion, shear and twist.} Indeed, by inspection of Eq.~(\ref{nhg}) it is clear that near-horizon geometries are a subclass of the degenerate Kundt spacetimes\epubtkFootnote{A Kundt spacetime is said to be degenerate if the Riemann tensor and all its covariant derivatives are type II with respect to the defining null vector field~\cite{Ortaggio:2012jd}.} which are all algebraically special of at least type II~\cite{Ortaggio:2012jd}.

Henceforth, we will drop the ``hats'' on all horizon quantities, so $R_{ab}$ and $\nabla_a$ refer to the Ricci tensor and Levi-Civita connection of $\gamma_{ab}$ on $H$.

\subsection{Einstein equations and energy conditions}

We will consider spacetimes that are solutions to Einstein's equations:
\be
R_{\mu\nu} =\Lambda g_{\mu\nu}+ T_{\mu\nu} -\frac{1}{n}g^{\rho\sigma}T_{\rho\sigma}\, g_{\mu\nu}, \label{einstein}
\ee
where $T_{\mu\nu}$ is the energy-momentum and $\Lambda$ is the cosmological constant of our spacetime. We will be interested in a variety of possible energy momentum tensors and thus in this section we will keep the discussion general. 

An important fact is that if a spacetime containing a degenerate horizon satisfies Einstein's equations then so does its near-horizon geometry. This is easy to see as follows. If the metric $g$ in Eq.~(\ref{degenerate}) satisfies Einstein's equations, then so will the 1-parameter family of diffeomorphic metrics $g_\epsilon$ for any $\epsilon>0$. Hence the limiting metric $\epsilon \to 0$, which by definition is the near-horizon geometry, must also satisfy the Einstein equations. 

The near-horizon limit of the energy momentum tensor thus must also exist and takes the form
\begin{equation}
T = 2\, \td v \left( T_{+-} \td r + r (\beta_{a}+T_{+-} h_a) \td x^a + \tfrac{1}{2}r^2 (\alpha+T_{+-}F) \td v \right) + T_{ab} \td x^a \td x^b,
\end{equation} 
where $T_{+-}, \alpha$ are functions on $H$ and $\beta_a$ is a 1-form on $H$. Working in the vielbein frame (\ref{frame}), it is then straightforward to verify that the $ab$ and $+-$ components of the Einstein equations for the near-horizon geometry give the following equations on the cross section $H$:
\bea 
R_{ab} &=& \tfrac{1}{2} h_a h_b -\nabla_{(a} h_{b)} +\Lambda \gamma_{ab} +P_{ab}, \label{horizoneq} \, \\
F &=& \tfrac{1}{2}h^2 -\tfrac{1}{2} \nabla_ah^a + \Lambda- E, \label{Feq}
\eea 
where we have defined
\bea
P_{ab} &\equiv& T_{ab} - \frac{1}{n}(\gamma^{cd}T_{cd} +2T_{+-})\gamma_{ab}, \\ 
E &\equiv &- \left( \frac{n-2}{n} \right) T_{+-} + \frac{1}{n} \gamma^{ab}T_{ab} \,.
\eea
It may be shown that the rest of the Einstein equations are automatically satisfied as a consequence of Eqs.~(\ref{horizoneq}), (\ref{Feq}) and the matter field equations, as follows. 

The matter field equations must imply the spacetime conservation equation $\nabla^\mu T_{\mu\nu} =0$. This is equivalent to the following equations on $H$:
\begin{equation}
\alpha =-\tfrac{1}{2} (\nabla^a-2h^a )\beta_{a} \,, \qquad
\beta_a = - (\nabla^b-h^b)T_{ab} - h_a T_{+-} \,, \label{Hcons}
\end{equation} 
which thus determine the components of the energy momentum tensor $\alpha, \beta_a$ in terms of $T_{+-}, T_{ab}$.
The $++$ and $+a$ components of the Einstein equations are $S_{++} = \alpha$ and $S_{+a}= \beta_a$ respectively, where $S_{++}$ and $S_{+a}$ are defined in Eq.~(\ref{NHRicci}). The first equation in (\ref{Hcons}) and the identity (\ref{Sid1}) imply that the $++$ equation is satisfied as a consequence of the $+a$ equation. Finally, substituting Eqs.~(\ref{horizoneq}) and (\ref{Feq}) into the identity (\ref{Sid2}), and using the second equation in (\ref{Hcons}), implies the $+a$ equation. Alternatively, a tedious calculation shows that the $+a$ equation follows from Eqs.~(\ref{horizoneq}) and (\ref{Feq}) using the contracted Bianchi identity for Eq.~(\ref{horizoneq}), together with the second equation in (\ref{Hcons}).

Although the energy momentum tensor must have a near-horizon limit, it is not obvious that the matter fields themselves must. Thus, consider the full spacetime before taking the near-horizon limit. Recall that for any Killing horizon $R_{\mu\nu}K^\mu K^\nu|_{\mathcal{N}}=0$ and therefore $T_{\mu\nu}K^\mu K^\nu|_{\mathcal{N}}= 0$. This imposes a constraint on the matter fields. We will illustrate this for Einstein--Maxwell theory whose energy-momentum tensor is
\be
T_{\mu\nu} = 2 \left( {\cal F}_{\mu \rho} {\cal F}_{\nu}^{\phantom{\nu} \rho} - \tfrac{1}{4} {\cal F}^2 g_{\mu\nu} \right) \,, \label{TMax}
\ee
where ${\cal F}$ is the Maxwell 2-form, which must satisfy the Bianchi identity $\td {\cal F}=0$.
It can be checked that in Gaussian null coordinates $T_{\mu \nu}K^\mu K^\nu|_{\mathcal{N}} = 2 {\cal F}_{v a} {\cal F}_{v b} \gamma^{ab} |_{r=0}$ and hence we deduce that ${\cal F}_{va}|_{r=0}=0$. Thus, smoothness requires ${\cal F}_{va}={\cal O}(r)$, which implies the near-horizon limit of ${\cal F}$ in fact exists. Furthermore, imposing the Bianchi identity to the near-horizon limit of the Maxwell field relates ${\cal F}_{vr}$ and ${\cal F}_{va}$, allowing one to write 
\be
\label{NHmaxwell}
{\cal F}_{\mathrm{NH}} = \td \left( r \Delta(x) \td v \right) + \tfrac{1}{2} B_{ab}(x) \td x^a \wedge \td x^b \,,
\ee
where $\Delta$ is a function on $H$ and $B$ is a closed 2-form on $H$. The 2-form $B$ is the Maxwell field induced on $H$ and locally can be written as $B=\td A$ for some 1-form potential $A$ on $H$. It can be checked that for the near-horizon limit
\bea
E &=&\frac{2(n-1)}{n} \Delta^2 + \frac{1}{n} B_{ab}B^{ab}, \\
P_{ab} &=& 2B_{ac}B_{b}^{\phantom{b} c} +\left(\frac{2}{n} \Delta^2 - \frac{B_{cd}B^{cd}}{n}\right)\gamma_{ab} \,.
\eea
We will present the Maxwell equations in a variety of dimensions in Section~\ref{sec:gauge}.

It is worth remarking that the above naturally generalises to $p$-form electrodynamics, with $ p\geq 2$, for which the energy momentum tensor is
\be
T_{\mu\nu} = \frac{2}{(p-1)!} \left( {\cal F}_{\mu \rho_1 \dots \rho_{p-1}} {\cal F}_{\nu}^{\phantom{\nu} \rho_1 \dots \rho_{p-1}} - \frac{1}{2p} {\cal F}^2 g_{\mu\nu} \right) \,, \label{Tpform}
\ee
where ${\cal F}$ is a $p$-form field strength satisfying the Bianchi identity $\td {\cal F}=0$. It is then easily checked that $T_{\mu \nu}K^\mu K^\nu|_{\mathcal{N}} =0$ implies ${\cal F}_{v a_1\dots a_{p-1}} {\cal F}_{v}^{~a_1\dots a_{p-1}} |_{r=0}=0$ and hence ${\cal F}_{va_1 \dots a_{p-1}}|_{r=0}=0$. Thus, smoothness requires ${\cal F}_{va_1 \dots a_{p-1}} ={\cal O}(r)$, which implies that the near-horizon limit of the $p$-form exists. The Bianchi identity then implies that the most general form for the near-horizon limit is
\begin{equation}
\label{NHpform}
{\cal F}_{\mathrm{NH}} = \td \left( Y \wedge r \td v \right) + X,
\end{equation}
where $Y$ is a $(p-2)$-form on $H$ and $X$ is a closed $p$-form on $H$.

The Einstein equations for a near-horizon geometry can also be interpreted as geometrical equations arising from the restriction of the Einstein equations for the full spacetime to a degenerate horizon, without taking the near-horizon limit, as follows.
The near-horizon limit can be thought of as the $\epsilon \to 0$ limit of the ``boost'' transformation $(K,L) \to (\epsilon K, \epsilon^{-1} L)$. This implies that restricting the boost-invariant components of the Einstein equations for the full spacetime to a degenerate horizon is equivalent to the boost invariant components of the Einstein equations for the near-horizon geometry. The boost-invariant components are $+-$ and $ab$ and hence we see that Eqs.~(\ref{horizoneq}) and (\ref{Feq}) are also valid for the full spacetime quantities restricted to the horizon. We deduce that the restriction of these components of the Einstein equations depends only on data \emph{intrinsic} to $H$: this special feature only arises for degenerate horizons.\epubtkFootnote{The remaining components of the Einstein equations for the full spacetime restricted to $\mathcal{N}$ give equations for extrinsic data (i.e., $r$-derivatives of $F,h_a,\gamma_{ab}$ that do not appear in the near-horizon geometry). On the other hand, the rest of the Einstein equations for the near-horizon geometry restricted to the horizon vanish.} It is worth noting that the horizon equations~(\ref{horizoneq}) and (\ref{Feq}) remain valid in the more general context of extremal isolated horizons~\cite{Lewandowski:2002ua, Wu:2009di, Booth:2012xm} and Kundt metrics~\cite{Jezierski:2008ur}. 

The positivity of $E$ and $P_{ab}$ can be related to standard energy conditions. For a near-horizon geometry $R_{\mu\nu} (K-L)^\mu(K-L)^\nu|_{\mathcal{N}} = -2 R_{\mu\nu} K^\mu L^\nu|_{\mathcal{N}}$. Since $K-L$ is timelike on the horizon, the strong energy condition implies $R_{\mu \nu} K^\mu L^\nu|_{\mathcal{N}} \leq 0$. Hence, noting that $R_{\mu \nu} K^\mu L^\nu|_{\mathcal{N}}= R_{+-}= -E +\Lambda$ we deduce that the strong energy condition implies
\be
E \geq \Lambda \,. \label{SEC}
\ee
On the other hand the dominant energy condition implies $T_{\mu\nu} K^{\mu} L^{\nu}|_{\mathcal{N}} \leq 0$. 
One can show $T_{\mu\nu} K^{\mu} L^\nu |_{\mathcal{N}} =-\frac{1}{2} P_{ab}\gamma^{ab}$. Therefore, the dominant energy condition implies
\be
P_\gamma \equiv P_{ab}\gamma^{ab} \geq 0 \,. \label{DEC}
\ee
Since $P_\gamma= -2 T_{+-}$, if $n\geq 2$ the dominant energy condition implies $E\geq 0$: hence, if $\Lambda \leq 0$ the dominant energy condition implies both Eqs.~(\ref{SEC}) and (\ref{DEC}).
Observe that Einstein--Maxwell theory with $\Lambda \leq 0$ satisfies both of these conditions.

In this review, we describe the current understanding of the space of solutions to the basic horizon equation (\ref{horizoneq}), together with the appropriate horizon matter field equations, in a variety of dimensions and theories. 

\subsection{Physical charges} 

So far we have considered near-horizon geometries independently of any extremal--black-hole solutions. In this section we will assume that the near-horizon geometry arises from a near-horizon limit of an extremal black hole. This limit discards the asymptotic data of the parent--black-hole solution. As a result, only a subset of the physical properties of a black hole can be calculated from the near-horizon geometry alone. In particular, information about the asymptotic stationary Killing vector field is lost and hence one cannot compute the mass from a Komar integral, nor can one compute the angular velocity of the horizon with respect to infinity. Below we discuss physical properties that can be computed purely from the near-horizon geometry~\cite{Hanaki:2007mb, Figueras:2008qh, Kunduri:2008tk}.

\vspace{6pt}
\noindent \textit{Area}. The area of cross sections of the horizon $H$ is defined by
\begin{equation}
A_H = \int_H \epsilon_\gamma \,,
\end{equation}
where $\epsilon_\gamma$ is the volume form associated to the induced Riemannian metric $\gamma_{ab}$ on $H$. 

For definiteness we now assume the parent black hole is \emph{asymptotically flat}.

\vspace{6pt}
\noindent \textit{Angular momentum}. The conserved angular momentum associated with a rotational symmetry, generated by a Killing vector $m$, is given by a Komar integral on a sphere at spacelike infinity $S_\infty$:\epubtkFootnote{The Komar integral associated with the null generator of the horizon $\partial/\partial v$ vanishes identically. In fact, one can show that for a general non-extremal Killing horizon, this integral is merely proportional to $\kappa$.} 
\begin{equation}
J = \frac{1}{16\pi} \int_{S_\infty} \star \td m \,.
\end{equation} 
This expression can be rewritten as an integral of the near-horizon data over $H$, by applying Stokes' theorem to a spacelike hypersurface $\Sigma$ with boundary $S_\infty \cup H$. The field equations can be used to evaluate the volume integral that is of the form $\int_\Sigma \star R(m)$, where $R(m)_\mu = R_{\mu\nu} m^{\nu}$. In particular, for vacuum gravity one simply has:
\begin{equation}
J = \frac{1}{16\pi}\int_H (h \cdot m)\, \epsilon_\gamma \,.
\end{equation} 
For Einstein--Maxwell theories the integral $\int_\Sigma \star R(m)$ can also be written as an integral over $H$, giving extra terms that correspond to the contribution of the matter fields to the angular momentum. For example, consider pure Einstein--Maxwell theory in any dimension so the Maxwell equation is $\td \star {\cal F}=0$. Parameterising the near-horizon Maxwell field by \eqref{NHmaxwell} one can show that, in the gauge ${\cal L}_m A = 0$,
\begin{equation}
J = \frac{1}{16\pi} \int_H \left( h \cdot m + 4(m \cdot {A}) \Delta\right)\epsilon_\gamma \,,
\end{equation} 
so the angular momentum is indeed determined by the near-horizon data. 

In five spacetime dimensions it is natural to couple Einstein--Maxwell theory to a Chern--Simons (CS) term. While the Einstein equations are unchanged, the Maxwell equation now becomes 
\be
\td \star {\cal F} + \frac{2\xi}{\sqrt{3}} {\cal F} \wedge {\cal F}=0 \,,
\ee
where $\xi$ is the CS coupling constant.
The angular momentum in this case can also be written purely as an integral over $H$:
\begin{equation}\label{J5dCS}
J = \frac{1}{16 \pi} \int_H\left(h \cdot m + 4(m\cdot {A})\Delta\right) \epsilon_\gamma + \tfrac{16}{3\sqrt{3}} \xi (m\cdot {A}) {A} \wedge B \,.
\end{equation} 
Of particular interest is the theory defined by CS coupling $\xi=1$, since this corresponds to the bosonic sector of minimal supergravity.

\vspace{6pt}
\noindent \textit{Gauge charges}. For Einstein--Maxwell theories there are also electric, and possibly magnetic, charges. For example, in pure Einstein--Maxwell theory in any dimension, the electric charge is written as an integral over spatial infinity:
\begin{equation}\label{echarge}
Q_e = \frac{1}{4\pi} \int_{S_\infty} \star {\cal F} \,.
\end{equation} 
By applying Stokes' Theorem to a spacelike hypersurface $\Sigma$ as above, and using the Maxwell equation, one easily finds
\begin{equation}
Q_e = \frac{1}{4\pi}\int_H \Delta \epsilon_\gamma \,.
\end{equation}
For $D=5$ Einstein--Maxwell--CS theory one instead gets
\begin{equation}
\label{echargeCS}
Q_e = \frac{1}{4\pi}\int_H \Delta \epsilon_\gamma + \tfrac{2}{\sqrt{3}} \xi {A} \wedge B \,.
\end{equation}
For $D=4$ one also has a conserved magnetic charge $Q_m = \tfrac{1}{4\pi} \int_{S_\infty} {\cal F}$. Using the Bianchi identity this can be written as
\begin{equation}
Q_m = \frac{1}{4\pi}\int_H B \,.
\end{equation}
For $D>4$ asymptotically-flat black holes there is no conserved magnetic charge. However, for $D=5$ black rings $H \cong S^1 \times S^2$, one can define a quasi-local dipole charge over the $S^2$
\begin{equation}
\mathcal{D} = \frac{1}{2\pi} \int_{S^2} {\cal F} = \frac{1}{2\pi} \int_{S^2} B \,,
\end{equation} 
where in the second equality we have expressed it in terms of the horizon Maxwell field. 

Note that in general the gauge field $A$ will not be globally defined on $H$, so care must be taken to evaluate expressions such as \eqref{J5dCS} and \eqref{echargeCS}, see~\cite{Hanaki:2007mb, Kunduri:2011zr}.



\newpage
\section{General Results}
\label{section:generalresults}

In this section we describe a number of general results concerning the topology and symmetry of near-horizon geometries under various assumptions. 

\subsection{Horizon topology theorem}
\label{section:topology}

Hawking's horizon topology theorem is one of the fundamental ingredients of the classic four-dimensional black-hole uniqueness theorems~\cite{Hawking:1971vc}. It states that cross sections of the event horizon of an asymptotically-flat, stationary, black-hole solution to Einstein's equations, satisfying the dominant energy condition, must be homeomorphic to $S^2$. The proof is an elegant variational argument that shows that any cross section with negative Euler characteristic can be deformed outside the event horizon such that its outward null geodesics converge. This means one has an outer trapped surface outside the event horizon, which is not allowed by general results on black holes.\epubtkFootnote{The borderline case of $T^2$ topology was only excluded later using topological censorship~\cite{Chrusciel:1994tr}. Furthermore, these results were generalised to the non-stationary case and asymptotically AdS case~\cite{Galloway:1999bp}.} 

Galloway and Schoen have shown how to generalise Hawking's horizon topology theorem to higher dimensional spacetimes~\cite{Galloway:2005mf}. Their theorem states if the dominant energy condition holds, a cross section $H$ of the horizon of a black hole, or more generally a marginally outer trapped surface, must have positive Yamabe invariant.\epubtkFootnote{A borderline case also arises in this proof, corresponding to the induced metric on $H$ being Ricci flat. This was in fact later excluded~\cite{Galloway:2006ws}.} The positivity of the Yamabe invariant, which we define below, is equivalent to the existence of a positive scalar curvature metric and is well known to impose restrictions on the topology, see, e.g., \cite{Galloway:2011np}. For example, when $H$ is three dimensional, the only possibilities are connected sums of $S^3$ (and their quotients) and $S^1\times S^2$, consistent with the known examples of black-hole solutions.

In the special case of degenerate horizons a simple proof of this topology theorem can be given directly from the near-horizon geometry~\cite{Lucietti:2012sa}. This is essentially a specialisation of the simplified proof of the Galloway--Schoen theorem given in~\cite{Racz:2008tf}. However, we note that since we only use properties of the near-horizon geometry, in particular only the horizon equation~(\ref{horizoneq}), we do not require the existence of a black hole.

For four dimensional spacetimes, so dim $H=2$, the proof is immediate, see, e.g., \cite{Jezierski:2008ur,Kunduri:2008rs}.

\begin{theorem}
Consider a spacetime containing a degenerate horizon with a compact cross section $H$ and assume the dominant energy condition holds. If $\Lambda \geq 0$ then $H \cong S^2$, except for the special case where the near-horizon geometry is flat (so $\Lambda=0$) and $H \cong T^2$. If $\Lambda <0$ and $\chi(H)<0$ the area of $H$ satisfies $A_H \geq 2\pi \Lambda^{-1} \chi(H)$ with equality if and only if the near-horizon geometry is $\mathrm{AdS}_2 \times \Sigma_g$, where $\Sigma_g$ is a compact quotient of hyperbolic space of genus $g$. 
\end{theorem} 

The proof is elementary. The Euler characteristic of $H$ can be calculated by integrating the trace of Eq.~(\ref{horizoneq}) over $H$ to get
\begin{equation}
\chi(H) = \frac{1}{4\pi}\int_H R_\gamma \epsilon_\gamma = \frac{1}{8\pi}\int_H (h \cdot h +4\Lambda +2P_\gamma) \epsilon_\gamma \,,
\end{equation}
where $\epsilon_\gamma$ is the volume form of the horizon metric $\gamma$. Therefore, for $\Lambda \geq 0$ and matter satisfying the dominant energy condition $P_\gamma \geq 0$ (see Eq.~(\ref{DEC})), it follows that $\chi(H)\geq 0$. Equality can only occur if $\Lambda=0$, $P_\gamma \equiv 0$, $h_{a}\equiv 0$: using Eqs.~(\ref{horizoneq}) and (\ref{Feq}) this implies $R_{ab}=0$ and $F=0$, so the near-horizon geometry is the trivial flat solution $\mathbb{R}^{1,1}\times T^2$. For the $\Lambda<0$ case the above argument fails and one finds no restriction on the topology of $H$. Instead, for $\chi(H)<0$, one can derive a lower bound for the area of $H$:
\begin{equation}
\label{area4}
A_H = \frac{2\pi | \chi(H) |}{|\Lambda|} +\frac{1}{4 |\Lambda|}\int_H (h \cdot h +2P_\gamma)\epsilon_\gamma \geq \frac{2\pi | \chi(H) |}{|\Lambda|} \,.
\end{equation}
This agrees with the lower bounds found in~\cite{Gibbons:1998zr, Woolgar:1999yi} in the more general context of apparent horizons. The lower bound in Eq.~(\ref{area4}) is saturated if and only if $h_a\equiv 0$, $P_\gamma\equiv 0$, which implies $R_{ab}=-|\Lambda |\gamma_{ab}$ and $F=\Lambda$, so the near-horizon geometry is $\mathrm{AdS}_2 \times \Sigma_g$. 

It is of interest to generalise these results to higher dimensions along the lines of Galloway and Schoen. As is well known, the total integral of the scalar curvature in itself does not constrain the topology of $H$ in this case. An analogue of this invariant for dim $H = n \geq 3$ is given by the Yamabe invariant $\sigma(H)$. This is defined via the Yamabe constant associated to a given conformal class of metrics $[\gamma]$ on $H$. First consider the volume-normalised Einstein--Hilbert functional
\be
E[\gamma'] \equiv \frac{\int_H R_{\gamma'} \epsilon_{\gamma'}}{(\int_H \epsilon_{\gamma'})^{\frac{n-2}{n}}} \,,
\ee 
where $\gamma'$ is a Riemannian metric on $H$ and $\epsilon_{\gamma'}$ is the associated volume form. As is well known, this functional is neither bounded from above or below. However, the restriction of $E$ to any conformal class of metrics is always bounded from below: the Yamabe constant $Y(H,[\gamma])$ for a given conformal class is then defined as the infimum of this functional. Parameterising the conformal class by $\gamma'=\phi^{\frac{4}{n-2}} \gamma$, for smooth positive functions $\phi$, we have $Y(H,[\gamma])\equiv \inf_{\phi>0} E_\gamma[\phi] $, where
\bea
 \label{yamabe}
 E_\gamma[\phi] \equiv \frac{\int_{H} \left( \frac{4(n-1)}{n-2} |\nabla \phi|^2 + R_{\gamma} \phi^2\right) \epsilon_\gamma}{ \left(\int_{H} \phi^{\frac{2n}{n-2}} \epsilon_\gamma \right)^{\frac{n-2}{n}}} \,.
\eea
The Yamabe invariant $\sigma(H)$ is defined by $\sigma(H) =\sup_{[\gamma]} Y(H,[\gamma])$, where the supremum is taken over all possible conformal classes. 
The solution to the \emph{Yamabe problem} states the following remarkable fact: for every conformal class $[\gamma]$ on compact $H$, the functional $E_\gamma[\phi]$ achieves its infimum and this occurs for a constant scalar curvature metric.

We are now ready to present the degenerate horizon topology theorem.

\begin{theorem}
Consider a spacetime containing a degenerate horizon with a compact cross section $H$ and assume the dominant energy condition holds. If $\Lambda \geq 0$, then either $\sigma(H)>0$ or the induced metric on the horizon is Ricci flat. If $\Lambda <0$ and $\sigma(H)<0$ the area of $H$ satisfies 
\begin{equation}
\label{Abound}
A_H \geq \left( \frac{\sigma(H)}{n\Lambda} \right)^{n/2} \,.
\end{equation}
\end{theorem}

A simple proof exploits the solution to the Yamabe problem mentioned above~\cite{Racz:2008tf, Lucietti:2012sa}. First observe that if there exists a conformal class of metrics $[\gamma]$ for which the Yamabe constant $Y(H,[\gamma])>0$ then it follows that $\sigma(H)>0$. Therefore, to establish that $H$ has positive Yamabe invariant, it is sufficient to show that for some $\gamma_{ab}$ the functional $E_\gamma[\phi]>0$ for all $\phi>0$, since the solution to the Yamabe problem then tells us that $Y(H,[\gamma])=E_\gamma[\phi_0]>0$ for some $\phi_0>0$. 

For our Riemannian manifolds $(H,\gamma)$ it is easy to show, except for one exceptional circumstance, that $E_\gamma[\phi]>0$ for all $\phi>0$ and thus $Y(H,[\gamma])>0$. The proof is as follows. The horizon equation~(\ref{horizoneq}) can be used to establish the identity
\be
\label{id}
2 | \nabla \phi |^2 +R_\gamma \phi^2 = 2| \mathrm{D}\phi |^2 -\nabla \cdot (\phi^2 h) +(\Lambda n+P_\gamma) \phi^2
\ee
for all $\phi$, where we have defined the differential operator $\mathrm{D}_a \equiv \nabla_a +\frac{1}{2} h_a$. It is worth noting that this identity relies crucially on the precise constants appearing in Eq.~(\ref{horizoneq}). This implies the following integral identity over $H$:
\be
\label{intid}
\int_{H} \left[ \frac{4(n-1)}{n-2} |\nabla \phi|^2 + R_{\gamma} \phi^2\right] \epsilon_\gamma=\int_H \left[ 2 | \mathrm{D} \phi |^2 + \frac{2n}{n-2} | \nabla \phi |^2 +(n\Lambda +P_\gamma) \phi^2 \right] \epsilon_\gamma \,.
\ee
If $\Lambda \geq 0$, the dominant energy condition $P_\gamma \geq 0$ implies $E_\gamma[\phi]\geq 0$ for all $\phi>0$ with equality only if $h_a \equiv 0$, $P_\gamma \equiv 0$ and $\Lambda=0$. The exceptional case $h_a \equiv 0$, $P_\gamma \equiv 0$ and $\Lambda=0$ implies $R_\gamma=0$, which allows one to infer~\cite{Galloway:2005mf} that either $R_{ab}=0$ or $H$ admits a metric of positive scalar curvature (and is thus positive Yamabe after all). 

As in four spacetime dimensions the above argument fails for $\Lambda<0$, and thus provides no restriction of the topology of $H$. Instead, assuming the dominant energy condition, Eq.~(\ref{intid}) implies
\be
E_\gamma[\phi] \geq -n |\Lambda| \frac{\int_H \phi^2 \epsilon_\gamma}{ (\int_H \phi^{\frac{2n}{n-2}} \epsilon_\gamma)^{\frac{n-2}{n}}} \geq - n |\Lambda| A_H^{2/n}
\ee
for any $\phi>0$, where the second inequality follows from H\"older's inequality. Therefore, by the definition of $Y(H,[ \gamma ])$, we deduce that $Y(H,[\gamma]) \geq - n |\Lambda| A_H^{2/n}$. It follows that if $\sigma(H)<0$, we get the stated non-trivial lower bound on the area of $H$. We note that the lower bound can only be achieved if $h_a \equiv 0$ and $P_\gamma \equiv 0$, which implies $R_{\gamma}=-n |\Lambda |$, in which case $\gamma$ necessarily minimises the functional $E$ in the conformal class $[\gamma]$ so that $Y(H,[\gamma]) = - n |\Lambda| A_H^{2/n}$. However, since $-Y(H,[\gamma]) \geq |\sigma(H)|$, it need not be the case that the lower bound in Eq.~(\ref{Abound}) is saturated by such horizon metrics (in contrast to the $n=2$ case above).

We note that the above topology theorems in fact only employ the scalar curvature of the horizon metric and not the full horizon equation~(\ref{horizoneq}). It would be interesting if one could use the non-trace part of the horizon equation to derive further topological restrictions.

\subsection{\AdS{2}-structure theorems}
\label{sec:ads2theorems}

It is clear that a general near-horizon geometry, Eq.~(\ref{nhg}), possesses enhanced symmetry: in addition to the translation symmetry $v \to v +c$ one also has a dilation symmetry $(v,r) \to (\lambda v, \lambda^{-1} r)$ where $\lambda \neq 0$ and together these form a two-dimensional non-Abelian isometry group. In this section we will discuss various near-horizon symmetry theorems that guarantee further enhanced symmetry.

\subsubsection{Static near-horizon geometries}
\label{ads2:static}

A static near-horizon geometry is one for which the normal Killing field $K$ is hypersurface orthogonal, i.e., $K \wedge \td K \equiv 0$ everywhere. 

\begin{theorem}[\cite{Kunduri:2007vf}]
\label{thrm:static}
Any static near-horizon geometry is locally a warped product of \AdS{2}, \dS{2} or $\mathbb{R}^{1,1}$ and $H$. If $H$ is simply connected this statement is global. In this case if $H$ is compact and the strong energy conditions holds it must be the \AdS{2} case or the direct product $\mathbb{R}^{1,1}\times H$.
\end{theorem}

\noindent \textit{Proof}: As a 1-form $K= K_\mu \td x^\mu= \td r+ r h_a \td x^a + r^2 F \td v$. A short calculation then reveals that $K \wedge \td K=0$ if and only if
\be
\td h = 0 \qquad \td F = h F \label{eq:static} \,,
\ee
which are the staticity conditions for a near-horizon geometry. Locally they can be solved by
\be
h= \td \lambda \qquad F= A_0 e^\lambda \,, \label{eq:staticsol}
\ee
where $\lambda(x)$ is a function on $H$ and $A_0$ is a constant. Substituting these into the near-horizon geometry and changing the affine parameter $r \to e^{-\lambda(x)} r$ gives:
\be
g= e^{-\lambda(x)} [ A_0 r^2 \td v^2 +2\td v \td r] + \gamma_{ab}(x) \td x^a \td x^b \,.
\ee
The metric in the square bracket is a maximally symmetric space: \AdS{2} for $A_0<0$, \dS{2} for $A_0>0$ and $\mathbb{R}^{1,1}$ for $A_0=0$. If $H$ is simply connected then $\lambda$ is globally defined on $H$. Now consider Eq.~(\ref{Feq}), which in this case reduces to
\be
A_0 = \tfrac{1}{2} \nabla^2 e^{-\lambda} - e^{-\lambda}( E-\Lambda) \,.
\ee
Assume $\lambda$ is a globally-defined function. Integrating over $H$ shows that if the strong energy condition~(\ref{SEC}) holds then $A_0 \leq 0$. The equality $A_0=0$ occurs if and only if $E=\Lambda$, in which case $\lambda$ is harmonic and hence a constant.

\subsubsection{Near-horizon geometries with rotational symmetries} 
\label{ads2:rot}

We begin by considering near-horizon geometries with a $U(1)^{D-3}$ rotational symmetry, whose orbits are generically cohomogeneity-1 on cross sections of the horizon $H$. The orbit spaces $H/U(1)^{D-3}$ have been classified and are homeomorphic to either the closed interval or a circle, see, e.g., \cite{Hollands:2008fm}. The former corresponds to $H$ of topology $S^2, S^3, L(p,q)$ times an appropriate dimensional torus, whereas the latter corresponds to $H \cong {T}^{D-3}$. Unless otherwise stated we will assume non-toroidal topology.

It turns out to be convenient to work with a geometrically-defined set of coordinates as introduced in~\cite{Kunduri:2008rs}. Let $m_i$ for $i=1 \dots D-3$ be the Killing vector fields generating the isometry. Define the 1-form $\Sigma=-i_{m_1} \cdots i_{m_{D-3}} \epsilon_\gamma$, where $\epsilon_\gamma$ is the volume form associated with the metric $\gamma$ on $H$. Note that $\Sigma$ is closed and invariant under the Killing fields $m_i$ and so defines a closed one-form on the orbit space. Hence there exists a globally-defined invariant function $x$ on $H$ such that 
\begin{equation}
\td x = -i_{m_1} \cdots i_{m_{D-3}} \epsilon_\gamma \,.
\end{equation} 
It follows that $|\td x |^2 = \det B$, where $B_{ij} \equiv \gamma(m_i,m_j)$, so $\td x$ vanishes precisely at the endpoints of the closed interval where the matrix $B_{ij}$ has rank $D-4$. As a function on $H$, $x$ has precisely one minimum $x_1$ and one maximum $x_2$, which must occur at the endpoints of the orbit space. Hence $H/U(1)^{D-3} \cong [x_1,x_2]$. 

Introducing coordinates adapted to the Killing fields $m_i = \partial / \partial \phi^i$, we can use $(x,\phi^i)$ as a chart on $H$ everywhere except the endpoints of the orbit space. The metric for $x_1 < x < x_2$ then reads
\begin{equation}
\gamma_{ab} \td x^a \td x^b = \frac{\td x^2}{\det B} + B_{ij}(x) \td \phi^i \td \phi^j \,. 
\end{equation} 
By standard results, the one-form $h$ may be decomposed globally on $H$ as
\begin{equation}
h = \beta + \td \lambda \,,
\end{equation}
where $\beta$ is co-closed. Since $h$ is invariant under the $m_i$, it follows that $\beta$ and $\td \lambda $ are as well. Further, periodicity of the orbits of the $m_i$ implies $m_i \cdot \td \lambda=0$, i.e., $\lambda = \lambda(x)$. It is convenient to define the globally-defined positive function 
\begin{equation}
\Gamma(x) \equiv e^{-\lambda} \,.
\end{equation} Next, writing $\beta = \beta_x \td x + \beta_i \td \phi^i$ and imposing that $\beta$ is co-closed, implies $\beta^x \left(\det B\right)^{1/2} = c$, where $c$ is a constant. It follows $i_\beta \Sigma= c$ and since $\Sigma$ vanishes at the fixed points of the $m_i$, this implies $c$ must vanish. Hence we may write
\begin{equation}
h = \frac{k_i \td \phi^i}{\Gamma} - \frac{\Gamma' \td x}{\Gamma} \,,
\end{equation}
where we define $k_i(x) \equiv \Gamma h_i$. It is worth noting that in the toroidal case one can introduce coordinates $(x,\phi)$ so that the horizon metric takes the same form, with $x$ now periodic and $\det B>0$ everywhere, although now the one-form $h$ may have an extra term since the constant $c$ need not vanish~\cite{Holland:2010bd}. 

We are now ready to state the simplest of the \AdS{2} near-horizon symmetry enhancement theorems:

\begin{theorem}[\cite{Kunduri:2007vf}]\label{Theorem:SO(2,1)}
Consider a $D$-dimensional spacetime containing a degenerate horizon, invariant under an $\mathbb{R}\times U(1)^{D-3}$ isometry group, and satisfying the Einstein equations $R_{\mu\nu}=\Lambda g_{\mu\nu}$. Then the near-horizon geometry has a global $G \times U(1)^{D-3}$ symmetry, where $G$ is either $O(2,1)$ or the 2D Poincar\'e group. Furthermore, if $\Lambda \leq 0$ and the near-horizon geometry is non-static the Poincar\'e case is excluded.
\end{theorem}

\noindent \textit{Proof:}
For the non-toroidal case we use the above coordinates. By examining the $(x v)$ and $(x i)$ components of the spacetime Einstein equations and changing the affine parameter $r \to \Gamma(x) r$, one can show
\begin{eqnarray}
\label{enhancedNH}
 g &=& \Gamma(x) \left[ A_0 r^2 \td v^2 + 2 \td v \td r \right] + \frac{\td x^2}{\det B} + B_{ij}(x) \left( \td\phi^i + k^i r \td v \right)\left( \td\phi^j + k^j r \td v \right) ,
\end{eqnarray}
where $A_0$ and $k^i$ are constants. The metric in the square bracket is a maximally-symmetric space: \AdS{2} for $A_0<0$, \dS{2} for $A_0>0$ and $\mathbb{R}^{1,1}$ for $A_0=0$. Any isometry of these 2D base spaces transforms $r \td v \to r \td v + \td \psi$, for some function $\psi(v,r)$. Therefore, by simultaneously transforming $\phi^i \to \phi^i- k^i \psi$, the full near-horizon geometry inherits the full isometry group $G$ of the 2D base, which for $A_0 \neq 0$ is $O(2,1)$ and for $A_0=0$ is the 2D Poincar\'e group.

The toroidal case can in fact be excluded~\cite{Holland:2010bd}, although as remarked above the coordinate system needs to be developed differently. 

In fact as we will see in Section~\ref{sec:vac} one can completely solve for the near-horizon geometries of the above form in the $\Lambda=0$ case.

The above result has a natural generalisation for $D=4,5$ Einstein--Maxwell theories. For the sake of generality, consider a general 2-derivative theory describing Einstein gravity coupled to Abelian vectors $A^I$ ($I=1\ldots N$) and uncharged scalars $\Phi^A$ ($A=1 \ldots M$) in $D=4,5$ dimensions, with action
\be
\label{gentheory}
 S = \int \td^D x \sqrt{-g} \left(R - \frac{1}{2} f_{AB}(\Phi)
 \partial_\mu \Phi^A \partial^\mu \Phi^B - V(\Phi) - \frac{1}{4}
 g_{IJ}(\Phi) F^I_{\mu\nu} F^{J\mu\nu} \right) + S_{\mathrm{top}} \,,
\ee
where $F^I \equiv \td A^I$, $V(\Phi)$ is an arbitrary scalar potential (which allows for a cosmological constant), and
\be
\label{top4d}
 S_{\mathrm{top}} = \frac{1}{2} \int h_{IJ}(\Phi) F^I \wedge F^J
 \qquad \text{if}\ D=4 \,,
\ee
or
\be
\label{top5d}
 S_{\mathrm{top}} = \frac{1}{6} \int C_{IJK}F^I \wedge F^J \wedge A^K
 \qquad \text{if}\ D=5 \,,
\ee
where $C_{IJK}$ are constants. This encompasses many theories of interest, e.g., vacuum gravity
with a cosmological constant, Einstein--Maxwell theory, and various
(possibly gauged) supergravity theories arising from
compactification from ten or eleven dimensions. 

\begin{theorem}[\cite{Kunduri:2007vf}] \label{Theorem:genth}
Consider an extremal--black-hole solution of the above $D=4,5$ theory with $\mathbb{R} \times U(1)^{D-3}$ symmetry. The near-horizon limit of this solution has a global $G \times U(1)^{D-3}$ symmetry, where $G$ is either $SO(2,1)$ or (the orientation-preserving subgroup of) the 2D Poincar\'e group. The Poincar\'e-symmetric case is excluded if $f_{AB}(\Phi)$ and $g_{IJ}(\Phi)$ are positive definite, the scalar potential is non-positive, and the horizon topology is not ${T}^{D-2}$. 
\end{theorem}

\noindent \textit{Proof:} The Maxwell fields $F^I$ and the scalar fields are invariant under the Killing fields $m_i$, hence $\Phi^I= \Phi^I(x)$. By examining the $(x v)$ and $(x i)$ components of Einstein's equations for the above general theory, and changing the affine parameter $r \to \Gamma(x) r$, one can show that the near-horizon metric is given by Eq.~(\ref{enhancedNH}) and the Maxwell fields are given by
\begin{eqnarray}
\label{enhancedF}
 F^{I} =\td\left[ e^I r \td v + b^I_i(x)\left( \td\phi^i + k^i r \td v\right) \right] ,
\end{eqnarray}
where $e^I$ are constants. Hence both the near-horizon metric and Maxwell fields are invariant under $G$. Generic orbits of the symmetry group have the structure of $T^{D-3}$ fibred over a 2D maximally-symmetric space, i.e., \AdS{2}, \dS{2} or $\mathbb{R}^{1,1}$. \AdS{2} and \dS{2} give $SO(2,1)$ symmetry, whereas $\mathbb{R}^{1,1}$ gives Poincar\'e symmetry. The \dS{2} and $\mathbb{R}^{1,1}$ cases are excluded subject to the additional assumptions mentioned, which ensure that the theory obeys the strong energy condition. 

\noindent \textit{Remarks}:

\begin{itemize}
\item In the original statement of this theorem asymptotically-flat or AdS boundary conditions were assumed~\cite{Kunduri:2007vf}. These were only used at one point in the proof, where the property that the generator of each rotational symmetry must vanish somewhere in the asymptotic region (on the ``axis'' of the symmetry) was used to constrain the Maxwell fields. In fact, using the general form for the near-horizon limit of a Maxwell field~(\ref{NHmaxwell}) and the fact that for non-toroidal topology at least one of the rotational Killing fields must vanish somewhere, allows one to remove any assumptions on the asymptotics of the black-hole spacetime. 
\item In the context of black holes, toroidal topology is excluded for $\Lambda =0$ when the dominant energy condition holds by the black-hole--topology theorems.
\end{itemize} 
An important corollary of the above theorems is:
\begin{cor}
\label{cor:ads3}
Consider a $D \geq 5$ spacetime with a degenerate horizon invariant under a $\mathbb{R}\times U(1)^{D-3}$ symmetry as in Theorem~\ref{Theorem:SO(2,1)} and \ref{Theorem:genth}. The near-horizon geometry is static if it is either a warped product of \AdS{2} and $H$, or it is a warped product of locally \AdS{3} and a $(D-3)$-manifold.
\end{cor}
The static conditions (\ref{eq:static}) for Eq.~(\ref{enhancedNH}) occur if and only if: (i) $k^i=0$ for all $i=1, \dots, D-3$, or (ii) $k_i= \text{const}\; \Gamma$ and $k^ik_i= \ell^{-2} \Gamma$ with $A_0=-\ell^{-2}<0$. Case (i) gives a warped product of a 2D maximally-symmetric space and $H$ as in Theorem~(\ref{thrm:static}). For case (ii) one can introduce $U(1)^{D-3}$ coordinates $(y^1, y^I)$ where $I=2, \dots, D-3$, not necessarily periodic, so that $k =\ell^{-1} \partial / \partial y^1$ and
\begin{equation}
g= \Gamma(x) \left[ -\frac{r^2}{\ell^2} \td v^2 + 2 \td v \td r + \left(\td y^1+ \frac{r}{\ell} \td v \right)^2 \right] + \frac{\td x^2}{B(x)} + B_{IJ}(x) \td y^I \td y^J \,.
\end{equation}
The metric in the square brackets is locally isometric to \AdS{3}. If $y^1$ is a periodic coordinate the horizon topology is $S^1\times B$, where $B$ is some $(D-3)$-dimensional manifold. 

Theorem~(\ref{Theorem:genth}) can be extended to higher-derivative theories of gravity as follows. Consider a general theory of gravity coupled to Abelian vectors $A^I$ and uncharged scalars $\Phi^A$ with action
\be
\label{eqn:hdaction}
 S = S_2 + \sum_{m \ge 1} \lambda^m \int \sqrt{-g} {\cal L}_m \,,
\ee
where $S_2$ is the 2-derivative action above, $\lambda$ is a coupling constant, and ${\cal L}_m$ is constructed by contracting (derivatives of) the Riemann tensor, volume form, scalar fields and Maxwell fields in such a way that the action is diffeomorphism and gauge-invariant.

\begin{prop}[\cite{Kunduri:2007vf}] 
\label{Theorem2}
Consider an extremal black-hole solution of the above higher-derivative theory, obeying the same assumptions as in Theorem~\ref{Theorem:genth}. Assume there is a regular horizon when $\lambda=0$ with $SO(2,1) \times U(1)^{D-3}$ near-horizon symmetry, and the near-horizon solution is analytic in $\lambda$. Then the near-horizon solution has $SO(2,1) \times U(1)^{D-3}$ symmetry to all orders in $\lambda$.
\end{prop}
Hence Theorem~\ref{Theorem:genth} is stable with respect to higher-derivative corrections. However, it does not apply to ``small'' black holes (i.e., if there is no regular black hole for $\lambda=0$). 

So far, the results described all assume $D-3$ commuting rotational Killing fields. For $D=4,5$ this is the same number as the rank of the rotation group $SO(D-1)$, so the above results are applicable to asymptotically-flat or globally-AdS black holes. For $D>5$ the rank of this rotation group is $\lfloor \frac{D-1}{2}\rfloor$, which is smaller than $D-3$, so the above theorems do not apply to asymptotically-flat or AdS black holes. An important open question is whether the above theorems generalise when fewer than $D-3$ commuting rotational isometries are assumed, in particular the case with $\lfloor \frac{D-1}{2}\rfloor$ commuting rotational symmetries. To this end, partial results have been obtained assuming a certain non-Abelian cohomogeneity-1 rotational isometry.

\begin{prop}[\cite{Figueras:2008qh}]
\label{Theorem3}
Consider a near-horizon geometry with a rotational isometry group $U(1)^m \times K$, whose generic orbit on $H$ is a cohomogeneity-1 $T^m$-bundle over a $K$-invariant homogeneous space $B$. Furthermore, assume $B$ does not admit any $K$-invariant one-forms. If Einstein's equations $R_{\mu \nu} = \Lambda g_{\mu \nu}$ hold, then the near-horizon geometry possesses a $G\times U(1)^m \times K$ isometry group, where $G$ is either $O(2,1)$ or the 2D Poincar\'e group. Furthermore, if $\Lambda \leq 0$ and the near-horizon geometry is non-static then the Poincar\'e group is excluded.
\end{prop} 
The assumptions in the above result reduce the Einstein equations for the near-horizon geometry to ODEs, which can be solved in the same way as in Theorem~\ref{Theorem:SO(2,1)}.
The special case $K=SU(q)$, $B=\mathbb{CP}^{q-1}$ with $m=1$ and $m=2$, gives a near-horizon geometry of the type that occurs for a Myers--Perry black hole with all the angular momenta of set equal in $2q + 2$ dimensions, or all but one set equal in $2q + 3$ dimensions, respectively. 

The preceding results apply only to cohomogeneity-1 near-horizon geometries. As discussed above, this is too restrictive to capture the generic case for $D>5$. The following result for higher-cohomogeneity near-horizon geometries has been shown.

\ifpdf
\begin{theorem}[\cite{Lucietti:2012sa}] Consider a spacetime containing a degenerate horizon invariant under orthogonally transitive\epubtkFootnote{An isometry group whose surfaces of transitivity are $p<D$ dimensional is said to be orthogonally transitive if there exists $D-p$ dimensional surfaces orthogonal to the surfaces of transitivity at every point.} isometry group $\mathbb{R} \times U(1)^N$, where $1 \leq N \leq D-3$, such that the surfaces orthogonal to the surfaces of transitivity are simply connected. Then the near-horizon geometry has an isometry group $G \times U(1)^N$, where $G$ is either $SO(2,1)$ or the 2D Poincar\'e group. Furthermore, if the strong energy condition holds and the near-horizon geometry is non-static, the Poincar\'e case is excluded. 
\end{theorem}
\else
\begin{theorem}[\cite{Lucietti:2012sa}] Consider a spacetime containing a degenerate horizon invariant under orthogonally transitive isometry group $\mathbb{R} \times U(1)^N$, where $1 \leq N \leq D-3$, such that the surfaces orthogonal to the surfaces of transitivity are simply connected. Then the near-horizon geometry has an isometry group $G \times U(1)^N$, where $G$ is either $SO(2,1)$ or the 2D Poincar\'e group. Furthermore, if the strong energy condition holds and the near-horizon geometry is non-static, the Poincar\'e case is excluded.
\end{theorem}%
\epubtkFootnote{An isometry group whose surfaces of transitivity are $p<D$ dimensional is said to be orthogonally transitive if there exists $D-p$ dimensional surfaces orthogonal to the surfaces of transitivity at every point.}
\fi

The near-horizon geometry in this case can be written as
\begin{equation}
\label{nhads2gen}
g= \Gamma(y)[ A_0 r^2 \td v^2 +2\td v \td r] + \gamma_{IJ}(y)( \td\phi^I+k^I r\td v)(\td\phi^J+k^Jr \td v) +\gamma_{mn}(y) \td y^m \td y^n \,,
\end{equation}
where as in the above cases we have rescaled the affine parameter $r \to \Gamma r$.
For the $N=D-3$ case, orthogonal transitivity follows from Einstein's equations~\cite{Emparan:2001wk}, which provides another proof of Theorem~\ref{Theorem:SO(2,1)} and \ref{Theorem:genth}. For $N < D-3$ this result guarantees an \AdS{2} symmetry for all known extremal--black-hole solutions, since all known explicit solutions possess orthogonally-transitive symmetry groups. In these higher cohomogeneity cases, the relation between Einstein's equations and orthogonal transitivity is not understood. It would be interesting to investigate this further.



\newpage
\section{Vacuum Solutions}
\label{sec:vac}

The Einstein equations for a near-horizon geometry \eqref{nhg} in the absence of matter fields are equivalent to the following equation on $H$
\be
\label{vacabeq}
R_{ab} = \tfrac{1}{2} h_a h_b -\nabla_{(a} h_{b)} + \Lambda \gamma_{ab} \,,
\ee
with the function $F$ determined by
\be
\label{vacFeq}
F= \tfrac{1}{2} h^a h_a - \tfrac{1}{2} \nabla_a h^a +\Lambda \,,
\ee
see Eqs.~\eqref{horizoneq}, \eqref{Feq}.
In this section we will explore solutions to Eq.~(\ref{vacabeq}) in various dimensions. It is useful to note that the contracted Bianchi identity for the horizon metric is equivalent to
\be
\label{vacbianchi}
\nabla_{a}F - Fh_a -2h^b\nabla_{[a}h_{b]} + \nabla^{b}{\nabla}_{[a}h_{b]} =0 \,.
\ee

\subsection{Static: all dimensions}
\label{vac:static}

A complete classification is possible for $\Lambda \leq 0$. Recall from Section~\ref{ads2:static}, the staticity conditions for a near-horizon geometry are $\td h=0$ and $\td F= h F$.

\begin{theorem}[\cite{Chrusciel:2005pa}]
The only vacuum static near-horizon geometry for $\Lambda \leq 0$ and compact $H$ is given by $h_a \equiv 0$, $F=\Lambda$ and $R_{ab}=\Lambda \gamma_{ab}$. For $D=4$ this result is also valid for $\Lambda>0$.
\end{theorem}

\noindent \textit{Proof:} A simple proof of the first statement is as follows. Substituting the staticity conditions (\ref{eq:staticsol}) into (\ref{vacFeq}) gives
\be
\nabla^2 \psi^2 + 2 \Lambda \psi^2 = 2A_0 \,,
\ee
where $\psi \equiv e^{-\lambda/2}$. Irrespective of the topology of $H$ one can argue that $\psi$ is a globally-defined function (for simply connected $H$ this is automatic, otherwise it can be shown by working in patches on $H$ and exploiting the fact that $\lambda$ in each patch is only defined up to an additive constant). For $\Lambda=0$ it is then clear that compactness implies $\psi$ must be a constant. For $\Lambda<0$, if one assumes $\psi$ is non-constant one can easily derive a contradiction by evaluating the above equation at the maximum and minimum of $\psi$ (which must exist by compactness). Hence in either case $h_a \equiv 0$, which gives the claimed near-horizon data.

In four dimensions one can solve Eq.~(\ref{vacabeq}) in general without assuming compactness of $H$. For non-constant $\psi$ and any $\Lambda$ one obtains the near-horizon geometry (sending $r \to \psi^2 r$)
\be
\label{4dgenstatic}
g = \psi^2 [ A_0 r^2 \td v^2+2\td v \td r] +\frac{\td \psi^2}{P(\psi)} + P(\psi) \td \chi^2,
\ee
where $P(\psi)= A_0+\beta \psi^{-1} - \frac{1}{3} \Lambda \psi^2$. (This is an analytically-continued Schwarzschild with a cosmological constant.) This local form of the metric can be used to show that for $\Lambda>0$ and $\psi$ non-constant, there are also no smooth horizon metrics on a compact $H$.

\subsection{Three dimensions}
\label{vac:3d}

The classification of near-horizon geometries in $D=3$ vacuum gravity with a cosmological constant can be completely solved. Although very simple, to the best of our knowledge this has not been presented before, so for completeness we include it here. 

The main simplification comes from the fact that cross sections of the horizon $H$ are one-dimensional, so the horizon equations are automatically ODEs. Furthermore, there is no intrinsic geometry on $H$ and so the only choice concerns its global topology, which must be either $H\cong S^1$ or $H\cong \mathbb{R}$.

\begin{theorem} Consider a near-horizon geometry with compact cross section $H\cong S^1$, which satisfies the vacuum Einstein equations including cosmological constant $\Lambda$. If $\Lambda < 0$ the near-horizon geometry is given by the quotient of \AdS{3} in Eq.~(\ref{ads3}). For $\Lambda=0$ the only solution is the trivial flat geometry $\mathbb{R}^{1,1} \times S^1$. There are no solutions for $\Lambda > 0$. 
\end{theorem}

\noindent \textit{Proof:} We may choose a periodic coordinate $x$ on $H$ so the horizon metric is simply $\gamma = dx^2$ and the 1-form $h=h(x)dx$. Observe that $h(x)$ must be a globally-defined function and hence must be a periodic function of $x$. Since the curvature and metric connection trivially vanish, the horizon equations~(\ref{vacabeq}) and (\ref{vacFeq}) simplify to
\begin{eqnarray}
&& h'=\tfrac{1}{2} h^2 +\Lambda \label{3dReq}, \\
&& F = \tfrac{1}{2} h^2 - \tfrac{1}{2} h' +\Lambda \label{3dFeq} \,.
\end{eqnarray} 
This system of ODEs can be explicitly integrated as we explain below. Instead, we will avoid this and employ a global argument on $H$. If $\Lambda \geq 0$, integrate Eq.~(\ref{3dReq}) over $H$ to deduce that $h\equiv 0$ and $\Lambda=0$, which gives the trivial flat near-horizon geometry $\mathbb{R}^{1,1}\times S^1$. For $\Lambda =- \frac{2}{\ell^2}<0$ we argue as follows. Multiply Eq.~(\ref{3dReq}) by $h'$ and integrate over $H$ to find $\int_{S^1} h'^2 = 0$. Hence $h$ must be a constant and substituting into the horizon equations gives $h = \frac{2}{\ell}$ and $F= 0$ (without loss of generality we have chosen a sign for $h$). The near-horizon geometry is then
\begin{equation}
g= 2\td v \td r + \frac{4r}{\ell} \td v \td x +\td x^2 = -\frac{4r^2}{\ell^2} \td v^2 + 2\td v \td r+ \left( \td x+ \frac{2 r}{\ell} \td v \right)^2 \,. \label{ads3}
\end{equation}
This metric is locally \AdS{3} and in the second equality we have written it as a fibration over \AdS{2}.

It is worth remarking that the ODE~(\ref{3dReq}) can be completely integrated without assuming compactness. For $\Lambda<0$ this reveals a second solution
$h = -\frac{2}{\ell} \tanh \left(\frac{x}{\ell} \right)$ and $F = -\frac{2}{\ell^2}\, \mathrm{sech}^2\left(\frac{x }{\ell} \right)$, where we have set the integration constant to zero by translating the coordinate $x$. Upon changing $r \to r \cosh^2 \left( \frac{x}{\ell} \right)$ the resulting near-horizon geometry is:
\begin{equation}
g = \cosh^2 \left( \frac{x}{\ell} \right) \left( -\frac{r^2}{\ell^2} \td v^2 + 2\td v \td r \right) +\td x^2 \,. \label{ads3R}
\end{equation}
Again, this metric is locally \AdS{3}. Unlike the previous case though, $H$ cannot be taken to be compact and hence we must have $H \cong \mathbb{R}$. For $\Lambda=0$ there is also a second solution given by $h= -\frac{2}{x}$ and $F= \frac{1}{x^2}$, although this is singular. For $\Lambda>0$ there is a unique solution given by $h = \frac{2}{\ell} \tan \left( \frac{ x}{\ell}\right)$ and $F=\frac{1}{\ell^2}\, \mathrm{sec}^2\left(\frac{ x}{\ell}\right)$, although this is also singular.\epubtkFootnote{This can be obtained by analytically continuing Eq.~(\ref{ads3R}) by $\ell \to i \ell$.}

\subsection{Four dimensions}
\label{vac:4d}

The general solution to Eq.~(\ref{vacabeq}) is not known in this case. In view of the rigidity theorem it is natural to assume axisymmetry. If one assumes such a symmetry, the problem becomes of ODE type and it is possible to completely solve it. The result is summarised by the following theorem, first proved in~\cite{Hajicek,Lewandowski:2002ua} for $\Lambda=0$ and in~\cite{Kunduri:2008tk} for $\Lambda<0$.

\begin{theorem}[\cite{Hajicek,Lewandowski:2002ua,Kunduri:2008tk}]
Consider a spacetime containing a degenerate horizon, invariant under an $\mathbb{R}\times U(1)$ isometry, satisfying the vacuum Einstein equations including a cosmological constant. Any non-static near-horizon geometry, with compact cross section, is given by the near-horizon limit of the extremal Kerr or Kerr-(A)dS black hole.
\end{theorem}

\noindent \textit{Proof:} We present a streamlined version of the proof in~\cite{Kunduri:2008tk}. As described in Section~\ref{ads2:rot}, axisymmetry implies one can introduce coordinates on $H$ so that
\be
\gamma_{ab} \td x^a \td x^b = \frac{ \td x^2}{B(x)} + B(x) \td \phi^2, \qquad \qquad h= \frac{B k(x)}{\Gamma} \td \phi - \frac{\Gamma'}{\Gamma} \td x \,.
\ee
The $x\phi$ component of Eq.~(\ref{vacabeq}) implies $k(x) \equiv k$ is a constant. The $x$ component of Eq.~(\ref{vacbianchi}) then implies
\be
F= \frac{A_0}{\Gamma} +\frac{B k^2}{\Gamma^2},
\ee
where $A_0$ is a constant. Substituting this into Eq.~(\ref{vacFeq}) gives
\be
\label{4dA0}
A_0 = \frac{1}{2} \nabla^2 \Gamma - \frac{B k^2}{2\Gamma} +\Lambda \Gamma \,.
\ee
Now subtracting the $xx$ component from the $\phi\phi$ component of Eq.~(\ref{vacabeq}) gives
\be
2\Gamma'' - \frac{\Gamma'^2}{\Gamma}-\frac{k^2}{\Gamma} = 0 \,.
\ee
A non-static near-horizon geometry must have $k \neq 0$ and therefore from the above equation $\Gamma$ is non-constant. Using this, one can write Eq.~(\ref{4dA0}) as
\be
\left( \frac{B\Gamma}{\Gamma'} \right)' = \frac{2(A_0-\Lambda \Gamma)\Gamma}{\Gamma'^2} \,.
\ee
The solution to the ODE for $\Gamma$ is given by
\be
\Gamma = \frac{k^2}{\beta}+ \frac{\beta x^2}{4} \,,
\ee
where $\beta$ is a positive constant, which can then be used to solve the ODE for $B$:
\be
 B= \frac{P(x)}{\Gamma} \,,
\ee
where 
\be
P(x)= -\frac{\beta \Lambda x^4}{12}+(A_0-2\Lambda k^2 \beta^{-1})x^2+c_1x - \frac{4k^2}{\beta^2} \left( A_0-\Lambda k^2\beta^{-1} \right)
\ee
and $c_1$ is a constant. Changing affine parameter $r \to \Gamma(x) r$ in the full near-horizon geometry finally gives
\be
g = \Gamma(x) [A_0 r^2 \td v^2+2\td v\td r] + \frac{\Gamma(x)}{P(x)} \td x^2 +\frac{P(x)}{\Gamma(x)} (\td \phi+ k r \td v)^2 \,, \label{NHKerrAdS}
\ee
with $\Gamma, P$ determined above. Observe that this derivation is purely local and does not assume anything about the topology of $H$ (unlike the derivation in \cite{Kunduri:2008tk}, which assumed compactness). If $k=0$ we recover the general static solution (\ref{4dgenstatic}), hence let us now assume $k \neq 0$.

Now assume $H$ is compact, so by axisymmetry one must have either $S^2$ or $T^2$. Integrating Eq.~(\ref{4dA0}) over $H$ then shows that if $\Lambda \leq 0$ then $A_0<0$ and so the metric in square brackets is \AdS{2}. The horizon metric extends to a smooth metric on $H \cong S^2$ if and only if $c_1=0$. It can then be checked the near-horizon geometry is isometric to that of extremal Kerr for $\Lambda=0$ or Kerr-AdS for $\Lambda<0$~\cite{Kunduri:2008tk}. It is also easy to check that for $\Lambda>0$ it corresponds to extremal Kerr-dS. In the non-static case, the horizon topology theorem excludes the possibility of $H \cong T^2$ for $\Lambda \geq 0$. If $\Lambda<0$ the non-static possibility with $H\cong T^2$ can also be excluded~\cite{Li:2013gca}.

It would be interesting to remove the assumption of axisymmetry in the above theorem. In~\cite{Jezierski:2012ue} it is shown that regular non-axisymmetric \emph{linearised} solutions of Eq.~(\ref{vacabeq}) about the extremal Kerr near-horizon geometry do not exist. This supports the conjecture that any smooth solution of Eq.~(\ref{vacabeq}) on $H \cong S^2$ must be axisymmetric and hence given by the above theorem.

\subsection{Five dimensions}
\label{vac:5d}

In this case there are several different symmetry assumptions one could make. Classifications are known for homogeneous horizons and horizons invariant under a $U(1)^2$-rotational symmetry. 

We may define a homogeneous near-horizon geometry to be one for which the Riemannian manifold $(H, \gamma_{ab})$ is a homogeneous space whose transitive isometry group $K$ also leaves the rest of the near-horizon data $(F,h_a)$ invariant. Since any near-horizon geometry (\ref{nhg}) possesses the 2D symmetry generated by $v \to v +c$ and $(v,r) \to (\lambda v, \lambda^{-1} r)$ where $\lambda \neq 0$, it is clear that this definition guarantees the near-horizon geometry itself is a homogeneous spacetime. Conversely, if the near-horizon geometry is a homogeneous spacetime, then any cross section $(H, \gamma_{ab})$ must be a homogeneous space under a subgroup $K$ of the spacetime isometry group, which commutes with the 2D symmetry in the $(v,r)$ plane (since $H$ is a constant $(v,r)$ submanifold). It follows that $(F,h_a)$ must also be invariant under the isometry $K$, showing that our original definition is indeed equivalent to the near-horizon geometry being a homogeneous spacetime.

Homogeneous geometries can be straightforwardly classified without assuming compactness of $H$ as follows.
\begin{theorem}[\cite{Kunduri:2012uq}] \label{vachomo}
Any vacuum, homogeneous, non-static near-horizon geometry is locally isometric to
\begin{equation}
g = \left( -\tfrac{1}{2} k^2 +\Lambda \right) r^2 \td v^2 + 2 \td v
\td r + (\hat{\omega}+ k r\td v)^2 + \hat{g} \,,
\end{equation}
where $\hat{\omega}$ is a $U(1)$-connection over a 2D base space satisfying $\mathrm{Ric}(\hat{g}) = \hat{\lambda} \hat{g}$ with $\hat{\lambda} = \tfrac{1}{2}k^2+2\Lambda$. The curvature of the connection is $\td \hat{\omega} = \sqrt{k^2 + 2\Lambda} \,\hat{\epsilon}$, where $\hat{\epsilon}$ is the volume form of the 2D base, and $k^2+2\Lambda \geq 0$.
\end{theorem}

The proof uses the fact that homogeneity implies $h$ must be a Killing field and then one reduces the problem onto the 2D orbit space. Observe that for $k\to 0$ one recovers the static near-horizon geometries. For $k \neq 0 $ and $\Lambda \geq 0$ we see that $\hat{\lambda}>0$ so that the 2D metric $\hat{g}$ is a round $S^2$ and the horizon metric is locally isometric to a homogeneously squashed $S^3$. Hence we have:
\begin{cor}
Any vacuum, homogeneous, non-static near-horizon geometry is locally isometric to the near-horizon limit of the extremal Myers--Perry black hole with $SU(2)\times U(1)$ rotational symmetry (i.e., equal angular momenta). For $\Lambda>0$ one gets the same result with the Myers--Perry black hole replaced by its generalisation with a cosmological constant~\cite{Hawking:1998kw}.
\end{cor}
For $\Lambda<0$ we see that there are more possibilities depending on the sign of $\hat{\lambda}$. If $\hat{\lambda}>0$ we again have a horizon geometry locally isometric to a homogeneous $S^3$. If $\hat{\lambda}=0$, we can write $\hat{g}= \td x^2+\td y^2$ and the $U(1)$-connection $\hat{\omega} = \sqrt{2|\Lambda |} ( x\td y-y\td x)$ is non-trivial, so the cross sections $H$ are the Nil group manifold with its standard homogeneous metric. For $\hat{\lambda}<0$, we can write $|\hat{\lambda}|\hat{g} =(\td x^2+ \td y^2)/ y^{2}$ and the connection $|\hat{\lambda}|\hat{\omega} = \sqrt{k^2+2\Lambda} \,\td x / y$, so the cross sections $H$ are the $SL(2,\mathbb{R})$ group manifold with its standard homogeneous metric, unless $k^2=-2\Lambda$, which gives $H=\mathbb{R} \times \mathbb{H}^2$. Hence we have:
\begin{cor} \label{cor:vachomo}
For $\Lambda<0$ any vacuum, homogeneous, non-static near-horizon geometry is locally isometric to either the near-horizon limit of the extremal rotating black hole~\cite{Hawking:1998kw} with $SU(2)\times U(1)$ rotational symmetry, or a near-horizon geometry with: (i) $H=\text{Nil}$ and its standard homogenous metric, (ii) $H=SL(2,\mathbb{R})$ and its standard homogeneous metric or (iii) $H=\mathbb{R} \times \mathbb{H}^2$. 
\end{cor}
This is analogous to a classification first obtained for supersymmetric near-horizon geometries in gauged supergravity, see Proposition~(\ref{Thm:homogauged}). 

We now consider a weaker symmetry assumption, which allows for inhomogeneous horizons. A $U(1)^2$-rotational isometry is natural in five dimensions and all known explicit black-hole solutions have this symmetry. The following classification theorem has been derived:
\begin{theorem}[\cite{Kunduri:2008rs}]
Consider a vacuum non-static near-horizon geometry with a $U(1)^2$-rotational isometry and a compact cross section $H$. It must be globally isometric to the near-horizon geometry of one of the following families of black-hole solutions: 
\begin{enumerate} 
\item $H \cong S^1 \times S^2$: the 3-parameter boosted extremal Kerr string.
\item $H\cong S^3$: the 2-parameter extremal Myers--Perry black hole or the 3-parameter `fast' rotating extremal KK black hole~\cite{Rasheed:1995zv}.
\item $H \cong L(p,q)$: the Lens space quotients of the above $H \cong S^3$ solutions.
\end{enumerate}
\end{theorem}

\noindent \textit{Remarks}:
\begin{itemize} 
\item The near-horizon geometry of the vacuum extremal black ring~\cite{Pomeransky:2006bd} is a 2-parameter subfamily of case 1, corresponding to a Kerr string with vanishing tension \cite{Kunduri:2007vf}. 
\item The near-horizon geometry of the `slowly' rotating extremal KK black hole~\cite{Rasheed:1995zv} is identical to that of the 2-parameter extremal Myers--Perry in case 2.
\item The $H \cong S^3$ cases can be written as a single 3-parameter family of near-horizon geometries~\cite{Hollands:2009ng}.
\item The $H \cong {T}^3$ case has been ruled out \cite{Holland:2010bd}.
\end{itemize}
For $\Lambda \neq 0$, the analogous problem has not been solved. The only known solution in this case is the $H \cong S^3$ near-horizon geometry of the rotating black hole with a cosmological constant~\cite{Hawking:1998kw}, which generalises the Myers--Perry black hole. It would be interesting to classify near-horizon geometries with $H\cong S^1 \times S^2$ in this case since this would capture the near-horizon geometry of the yet-to-be-found asymptotically-\AdS{5} black ring. A perturbative attempt at constructing such a solution is discussed in \cite{Kunduri:2008rs}.

\subsection{Higher dimensions}
\label{vac:higherdims}

For spacetime dimension $D \geq 6$, so the horizon cross section dim$\,H \geq 4$, the horizon equation~(\ref{vacabeq}) is far less constraining than in lower dimensions. Few general classification results are known, although several large families of vacuum near-horizon geometries have been constructed.

\subsubsection{Weyl solutions}
\label{vac:weyl}

The only known classification result for vacuum $D>5$ near-horizon geometries is for $\Lambda=0$ solutions with $U(1)^{D-3}$-rotational symmetry. These generalise the $D=4$ axisymmetric solutions and $D=5$ solutions with $U(1)^2$-symmetry discussed in Sections \ref{vac:4d} and \ref{vac:5d} respectively. By performing a detailed study of the orbit spaces $H/U(1)^{D-3}$ it has been shown that the only possible topologies for $H$ are: $S^2 \times {T}^{D-4}$, $S^3 \times T^{D-5}$, $L(p,q) \times {T}^{D-5}$, and ${T}^{D-2}$~\cite{Hollands:2008fm}. 

An explicit classification of the possible near-horizon geometries (for the non-toroidal case) was derived in \cite{Hollands:2009ng}, see their Theorem 1. Using their theorem it is easy to show that the most general solution with $H \cong S^2\times T^{D-4}$ is in fact isometric to the near-horizon geometry of a boosted extremal Kerr-membrane (i.e., perform a general boost of Kerr$\times \mathbb{R}^{D-4}$ along the $\{ t \} \times \mathbb{R}^{D-4}$ coordinates and then compactify $\mathbb{R}^{D-4} \to T^{D-4}$). Non-static near-horizon geometries with $H \cong T^{D-2}$ have been ruled out \cite{Holland:2010bd} (including a cosmological constant).

\subsubsection{Myers--Perry metrics}

The Myers--Perry (MP) black-hole solutions~\cite{Myers:1986un} generically have isometry groups $\mathbb{R}\times U(1)^s$ where $s= \lfloor \frac{D-1}{2} \rfloor$. Observe that if $D>5$ then $s<D-3$ and hence these solutions fall outside the classification discussed in Section~\ref{vac:weyl}. They are parameterised by their mass parameter $\mu$ and angular momentum parameters $a_i$ for $i=1,\dots s$. The topology of the horizon cross section $H \cong S^{D-2}$. A generalisation of these metrics with non-zero cosmological constant has been found~\cite{Gibbons:2004js}. We will focus on the $\Lambda=0$ case, although analogous results hold for the $\Lambda \neq 0$ solutions.

The location of the horizon is determined by the largest positive number $r_+$ such that in odd and even dimensions $\Pi(r_+)-\mu \, r_+^2=0$ and $\Pi(r_+)-\mu\, r_+=0$, respectively, where
\begin{equation}
\Pi(r) = \prod_{i=1}^{s}\, (r^2+a_i^2) \,.
\label{Pidef}
\end{equation}
The extremal limit of these black holes in odd and even dimensions is given by $\Pi'(r_+)=2\, \mu \, r_+$ and $\Pi'(r_+)=\mu$, respectively. These conditions hold only when the black hole is spinning in all the two planes available, i.e., we need $a_i \neq 0$ for all $ i = 1, \cdots , s$.
Without loss of generality we will henceforth assume $a_i>0$ and use the extremality condition to eliminate the mass parameter $\mu$. The near-horizon geometry of the
extremal MP black holes can be written in a unified form~\cite{Figueras:2008qh}:
\begin{eqnarray}
g_{\mathrm{MP}} &=& F_+ \, \left( -\frac{\Pi''(r_+)}{2 \, \Pi(r_+)} \, r^2\,\td v^2 + 2\, \td v \td r \right) + \gamma_{\mu_i \mu_j} \, \td \mu^i \td \mu^j \nonumber \\
& & + \, \gamma_{ij} \, \left( \td \phi^i+\frac{2\, r_+\, a_i}{(r_+^2+a_i^2)^2}\, r\, \td v \right)\, \left(\td \phi^j+\frac{2\, r_+\, a_j}{(r_+^2+a_j^2)^2} \, r\, \td v \right) ,
\label{mpbhmet}
\end{eqnarray}
where
\begin{equation}
F_+= 1- \sum_{i=1}^s\frac{a_i^2\, \mu_i^2}{r_+^2+a_i^2}, \qquad 
\gamma_{ij}= (r^2_++a_i^2) \, \mu_i^2 \, \delta_{ij}+ \frac{1}{F_+} \, a_i\, \mu_i^2\, a_j\, \mu_j^2 \,,
\end{equation}
and in odd and even dimensions 
\begin{equation}
\gamma_{\mu_i\mu_j}\, \td \mu^i\,\td \mu^j= \sum_{i=1}^s \; (r_+^2+a_i^2)\, d\mu_i^2, \qquad \gamma_{\mu_i\mu_j}\, \td \mu^i\, \td \mu^j=r_+^2\, \td \alpha^2 + \sum_{i=1}^s \;(r_+^2+a_i^2)\, \td\mu_i^2 \,,
\end{equation}
respectively. The direction cosines $\mu_i$ and $\alpha$ take values in the range $0 \leq \mu_i \leq 1$ with $-1\leq \alpha \leq 1$ and in odd and even dimensions satisfy
\begin{equation}
\sum_{i=1}^{s} \mu_i^2=1 \,, \qquad 
\sum_{i=1}^{s}\mu_i^2+\alpha^2=1 \,,
\label{dircosconst}
\end{equation}
respectively. The generalisation of these near-horizon geometries for $\Lambda \neq 0$ was given in~\cite{Lu:2008jk}. It is worth noting that if subsets of the angular momentum parameters $a_i$ are set equal, the rotational symmetry enhances to a non-Abelian unitary group. 

Since these are vacuum solutions one can trivially add flat directions to generate new solutions. For example, by adding one flat direction one can generate a boosted MP string, whose near-horizon geometries have $H\cong S^1\times S^{D-3}$ topology. Interestingly, for odd dimensions $D$ the resulting geometry has $\lfloor \frac{D-1}{2} \rfloor$ commuting rotational isometries. For this reason, it was conjectured that a special case of this is also the near-horizon geometry of yet-to-be-found asymptotically-flat black rings (as is known to be the case in five dimensions)~\cite{Figueras:2008qh}.

\subsubsection{Exotic topology horizons}
\label{exotic}

Despite the absence of explicit $D>5$ black-hole solutions, a number of solutions to Eq.~(\ref{vacabeq}) are known. It is an open problem as to whether there are corresponding black-hole solutions to these near-horizon geometries.

All the constructions given below employ the following data.
Let $K$ be a compact Fano K\"ahler--Einstein manifold\epubtkFootnote{A complex manifold is Fano if its first Chern class is positive, i.e., $c_1(K)>0$. It follows that any K\"ahler--Einstein metric on such a manifold must have positive Einstein constant.} of complex dimension $q-1$ and $a \in H^2(K, \mathbb{Z})$ is the indivisible class given by $c_1(K)=I a$ with $I \in \mathbb{N}$ (the Fano index $I$ and satisfies $I \leq q$ with equality iff $K=\mathbb{CP}^{q-1}$). The K\"ahler--Einstein metric $\bar{g}$ on $K$ is normalised as $\mathrm{Ric}(\bar{g})=2q \bar{g}$ and we denote its isometry group by $G$. The simplest example occurs for $q=2$, in which case $K =\mathbb{CP}^1 \cong S^2$ with $\bar{g} = \tfrac{1}{4}(\td \theta^2+\sin^2\theta \td \chi^2)$.

In even dimensions greater than four, an infinite class of near-horizon geometries is revealed by the following result.
\begin{prop}[\cite{Kunduri:2010vg}]\label{infclass1}
Let $m \in \mathbb{Z}$ and $P_{m}$ be the principal $S^1$-bundle over any Fano K\"ahler--Einstein manifold $K$, specified by the characteristic class $m a$. For each $m>I$ there exists a 1-parameter family of smooth solutions to Eq.~(\ref{vacabeq}) on the associated $S^2$-bundles $H \cong P_m \times_{S^1} S^2$.
\end{prop}

The dim $H=2q$ ansatz used for the near-horizon data is the $U(1)\times G$ invariant form
\bea
\gamma_{ab} \td x^a \td x^b &=& \frac{\td x^2}{B(x)} + B(x) \omega \otimes \omega + A(x)^2 \bar{g}, \\
h &=& \frac{k B(x) \omega}{\Gamma} -\frac{\Gamma'(x)}{\Gamma} \td x \,,
\eea
where $\omega$ is a $U(1)$-connection over $K$ with curvature $2\pi m a$. The solutions depend on one continuous parameter $L>0$ and the integer $m>I$. The continuous parameter corresponds to the angular momentum $J[\partial_\phi]$ where $\partial_\phi$ generates the $U(1)$-isometry in the $S^2$-fibre. The various functions are given by $A(x)^2= L^2(1-x^2)$, $\Gamma(x)= \xi +x^2$, $k= \pm 2\sqrt{\xi} $ and $B(x) = P(x)/(A(x)^{2q-2} \Gamma(x))$ where $P$ is a polynomial in $x^2$ and smoothness fixes $\xi$ to be a function of $m$. 

The simplest example is the $q=2$, $\Lambda=0$ solution, for which the near-horizon data takes the explicit form
\bea
\gamma_{ab} \td x^a \td x^b &=& \frac{L^2(\xi_m+x^2) (1-x^2) dx^2}{(4-m^2x^2)\left(\xi_m - \frac{4 x^2}{3m^2}\right)}
+ \frac{L^2(4-m^2x^2)\left(\xi_m - \frac{4 x^2}{3m^2}\right)}{(\xi_m+x^2)(1-x^2)} \left(\td \phi+ \tfrac{1}{2}\cos\theta \td \chi \right)^2 \nonumber \\ && +\tfrac{1}{4}L^2(1-x^2)( \td \theta^2+\sin^2\theta \td \chi^2), \\
h_a \td x^a &=& \pm \frac{2\sqrt{\xi_m} L^2(4-m^2x^2)\left(\xi_m -
 \frac{4 x^2}{3m^2}\right)}{(\xi_m+x^2)^2(1-x^2)} \left(\td \phi+
\tfrac{1}{2}\cos\theta \td \chi \right) -\frac{2x}{\xi_m+x^2} \td x \,,
\eea
where
\be
\xi_m =\frac{4}{3} \left( \frac{3-\frac{4}{m^2}}{4+m^2}\right),
\ee
and $m>2$. The coordinate ranges are $-2/m \leq x \leq 2/m$, $0\leq \theta \leq \pi$, $\phi \sim \phi +2\pi/m$, $\chi\sim\chi+2\pi$. Cross sections of the horizon $H$, are homeomorphic to $S^2 \times S^2$ if $m$ is even, or the non-trivial bundle $S^2 \tilde{\times} S^2 \cong \mathbb{CP}^2 \# \overline{\mathbb{CP}}^2$ if $m$ is odd. For $\Lambda \neq 0$ the solutions are analogous. 

For $q>2$ the Fano base $K$ is higher dimensional and there are more choices available. The topology of the total space is always a non-trivial $S^2$-bundle over $K$ and in fact different $m$ give different topologies, so there are an infinite number of horizon topologies allowed. Furthermore, one can choose $K$ to have no continuous isometries giving examples of near-horizon geometries with a single $U(1)$-rotational isometry. Hence, if there are black holes corresponding to these horizon geometries they would saturate the lower bound in the rigidity theorem.

It is worth noting that the local form of the above class of near-horizon metrics includes as a special case that of the extremal MP metrics $H\cong S^{2q}$ with equal angular momenta (for $m=I$).
The above class of horizon geometries are of the same form as the Einstein metrics on complex line bundles~\cite{Page:1985bq}, which in four dimensions corresponds to the Page metric~\cite{Page}, although we may of course set $\Lambda\leq 0$.

Similar constructions of increasing complexity can be made in odd dimensions, again revealing an infinite class of near-horizon geometries.

\begin{prop}
Let $m \in \mathbb{Z}$ and $P_{m}$ be the principal $S^1$-bundle over $K$ specified by the characteristic class $m a$. There exists a 1-parameter family of Sasakian solutions to Eq.~(\ref{vacabeq}) on $H \cong P_m$.
\end{prop}

As a simple example consider $K= \mathbb{CP}^1\times \mathbb{CP}^1$. This leads to an explicit homogeneous near-horizon geometry with
\bea
\gamma_{ab}\td x^a \td x^b &=& \left( \frac{ k^2+2\Lambda }{2 \lambda^2} \right) \left( \td\psi+ \cos \theta_1 \td\phi_1 +\cos \theta_2 \td \phi_2 \right)^2 \nonumber \\ && \qquad \qquad + \frac{1}{\lambda} \left( \td \theta_1^2 +\sin^2 \theta_1 \td \phi_1^2+ \td \theta_2^2+\sin^2 \theta_2 \td \phi_2^2 \right) \nonumber \\
h^a \partial_a &=& k \sqrt{ \frac{2\lambda^2}{ k^2+2\Lambda }} \;
\frac{\partial}{\partial\psi} \,,
\eea
where for convenience we have written $h$ is a vector field, $k$ is a constant and $\lambda=(k^2+6\Lambda)/4$. Regularity requires that the Chern number $m$ of the $U(1)$-fibration over each $S^2$ to be the same and the period $\Delta \psi = 2\pi/m$. The total space is a Lens space $S^3/\mathbb{Z}_m$-bundle over $S^2$ and is topologically $H \cong S^3 \times S^2$. For $k=0$ and $\Lambda>0$ this corresponds to a Sasaki--Einstein metric on $S^3\times S^2$ sometimes known as $T^{1,1}$.

The above proposition can be generalised as follows.
\begin{prop}[\cite{Kunduri:2012uq}] \label{sasaki}
Given any Fano K\"ahler--Einstein manifold $K$ of complex dimension $q-1$ and coprime $p_1,p_2 \in \mathbb{N}$ satisfying $1< Ip_1/p_2<2$, there exists a 1-parameter family of solutions to Eq.~(\ref{vacabeq}) where $H$ is a compact Sasakian $(2q+1)$-manifold.
\end{prop}

These examples have $U(1)^2\times G$ symmetry, although possess only one independent angular momentum along the $T^2$-fibres. These are deformations of the Sasaki--Einstein $Y^{p,q}$ manifolds~\cite{Gauntlett:2004hh}.

There also exist a more general class of non-Sasakian horizons in odd dimensions.
\begin{prop}[\cite{Kunduri:2012hi}]
Let $P_{m_1,m_2}$ be the principal $T^2$-bundle over any Fano K\"ahler--Einstein manifold $K$, specified by the characteristic classes $(m_1 a, m_2 a)$ where $m_1,m_2 \in \mathbb{Z}$. For a countably infinite set of non-zero integers $(m_1,m_2, j,k)$, there exists a two-parameter family of smooth solutions to Eq.~(\ref{vacabeq}) on the associated Lens space bundles $H \cong P_{m_1,m_2} \times_{T^2} L(j,k)$.
\end{prop}
The dim $H=2q+1$ form of the near-horizon data in the previous two propositions is the $U(1)^2 \times G$ invariant form
\bea
\gamma_{ab} \td x^a \td x^b &=& \frac{\td x^2}{\det B} + B_{ij}(x) \omega^i \omega^j + A(x)^2 \bar{g}, \\
h_a \td x^a &=& \frac{B_{ij}k^j \omega^i}{\Gamma} -\frac{\Gamma'(x)}{\Gamma} \td x \,, \nonumber 
\eea
where $\omega^i$ is a principal $T^2$-connection over $K$ whose curvature is $2\pi m_i a$. The explicit functions $A(x)^2, \Gamma(x)$ are linear in $x$ and $B_{ij}(x)$ are ratios of various polynomials in $x$. Generically these solutions possess two independent angular momenta along the $T^2$-fibres. The Sasakian horizon geometries of Proposition~\ref{sasaki} arise as a special case with $m_1+m_2 = Ij$ and possess only one independent angular momentum. The base $K=\mathbb{CP}^1$ gives horizon topologies $H \cong S^3 \times S^2$ or $H \cong S^3 \tilde{\times} S^2$ depending on whether $m_1+m_2$ is even or odd respectively. 

It is worth noting that the local form of this class of near-horizon metrics includes as a special case that of the extremal MP metrics $H\cong S^{2q+1}$ with all but one equal angular momenta. The above class of horizon geometries are of the same form as the Einstein metrics found in~\cite{Chen:2011dh,Kunduri:2012hi}.



\newpage
\section{Supersymmetric Solutions}
\label{sec:susy}

By definition, a supersymmetric solution of a supergravity theory is a solution that also admits a Killing spinor, i.e., a spinor field $\psi$ that satisfies $D_\mu \psi=0$, where $D_\mu$ is a spinorial covariant derivative that depends on the matter fields of the theory. Given such a Killing spinor $\psi$, the bilinear $K^\mu= \bar{\psi} \Gamma^{\mu} \psi$ is a non-spacelike Killing field. By definition, a supersymmetric horizon is invariant under the Killing field $K^\mu$ and thus $K^\mu$ must be tangent to the horizon. Hence $K^\mu$ must be null on the horizon; in other words the horizon is a Killing horizon of $K^\mu$. Furthermore, since $K^2 \leq 0$ both outside and inside the horizon, it follows that on the horizon $\td K^2 =0$, i.e., it must be a degenerate horizon. Hence any supersymmetric horizon is necessarily a degenerate Killing horizon.

\subsection{Four dimensions}

The simplest supergravity theory in four dimensions admitting supersymmetric black holes is minimal ${\cal N}=2$ supergravity, whose bosonic sector is simply standard Einstein--Maxwell theory. The general supersymmetric solution in this theory is given by the Israel--Wilson--Perj\'es metrics. Using this fact, the following near-horizon uniqueness theorem has been proved:

\begin{theorem}[\cite{Chrusciel:2005ve}]
Any supersymmetric near-horizon geometry in ${\cal N}=2$ minimal supergravity is one of the maximally supersymmetric solutions $\mathbb{R}^{1,1}\times T^2$ or $\mathrm{AdS}_2 \times S^2$. 
\end{theorem}

Notice that staticity here follows from supersymmetry. In Section~\ref{section:applications} we will discuss the implications of this result for uniqueness of supersymmetric black holes in four dimensions. 

For ${\cal N}=2$ gauged supergravity, whose bosonic sector is Einstein--Maxwell theory with a negative cosmological constant, an analogous classification of supersymmetric near-horizon geometries has not been performed. Nevertheless, one may deduce the following result, from a classification of \emph{all} near-horizon geometries of this theory under the additional assumption of axisymmetry:
\begin{prop}[\cite{Kunduri:2008tk}]
Any supersymmetric, axisymmetric, near-horizon geometry in ${\cal N}=2$ gauged supergravity, is given by the near-horizon limit of the 1-parameter family of supersymmetric Kerr--Newman-\AdS{4} black holes~\cite{Kostelecky:1995ei}.
\end{prop}
Note that the above near-horizon geometry is non-static. This is related to the fact that supersymmetric AdS black holes must carry angular momentum. It would be interesting to remove the assumption of axisymmetry. Some related work has been done in the context of supersymmetric isolated horizons~\cite{Booth:2008ru}. 

Supersymmetric black holes are not expected to exist in ${\cal N}=1$ supergravity. For the general ${\cal N}=1$ supergravity the following result, supporting this expectation, has been established.
\begin{prop}[\cite{Gutowski:2010gv}]
A supersymmetric near-horizon geometry of ${\cal N}=1$ supergravity is either the trivial solution $\mathbb{R}^{1,1}\times T^2$, or $\mathbb{R}^{1,1} \times S^2$ where $S^2$ may be a non-round sphere.
\end{prop}
An example of a supersymmetric near-horizon geometry of the form $\mathbb{R}^{1,1} \times S^2$, with a round $S^2$, was given in~\cite{Meessen:2010ph}.

\subsection{Five dimensions}

The simplest $D=5$ supergravity theory admitting supersymmetric black holes is ${\cal N}=1$ minimal supergravity. The bosonic sector is Einstein--Maxwell theory with a Chern--Simons term given by Eq.~(\ref{EMCS2}) with a specific coupling $\xi=1$. Supersymmetric solutions to this theory were classified in~\cite{Gauntlett:2002nw}. This was used to obtain a complete classification of supersymmetric near-horizon geometries in this theory. 
\begin{theorem}[\cite{Reall:2002bh}]
Any supersymmetric near-horizon geometry of minimal supergravity is locally isometric to one of the following maximally supersymmetric solutions: $\mathrm{AdS}_3 \times S^2$, $\mathbb{R}^{1,1}\times T^3$, or the near-horizon geometry of the Breckenridge--Myers--Peet--Vafa (BMPV) black hole (of which $\mathrm{AdS}_2 \times S^3$ is a special case). 
\end{theorem} 
Note that here supersymmetry implies homogeneity. The $\mathrm{AdS}_3 \times S^2$ near-horizon geometry has cross sections $H \cong S^1 \times S^2$ and arises as the near-horizon limit of supersymmetric black rings~\cite{Elvang:2004rt, Elvang:2005sa} and supersymmetric black strings~\cite{Gauntlett:2002nw, Bena:2004wv}. Analogous results have been obtained in $D=5$ minimal supergravity coupled to an arbitrary number of vector multiplets \cite{Gutowski:2004bj}. As discussed in Section~\ref{section:applications}, the above theorem can be used to prove a uniqueness theorem for topologically spherical supersymmetric black holes. 

The corresponding problem for minimal \emph{gauged} supergravity has proved to be more difficult. The bosonic sector of this theory is Einstein--Maxwell--Chern--Simons theory with a negative cosmological constant. The theory admits asymptotically \AdS{5} black-hole solutions that are relevant in the context of the AdS/CFT correspondence \cite{Gutowski:2004ez, Chong:2005hr}. The following partial results have been shown.
\begin{prop}[\cite{Gutowski:2004ez}]
Consider a supersymmetric, homogeneous near-horizon geometry of minimal gauged supergravity. Cross sections of the horizon must be one of the following: a homogeneously squashed $S^3$, $\text{Nil}$ or $SL(2,\mathbb{R})$ manifold. \label{Thm:homogauged}
\end{prop} The near-horizon geometry of the $S^3$ case was used to construct the first example of an asymptotically \AdS{5} supersymmetric black hole~\cite{Gutowski:2004ez}. Analogous results in gauged supergravity coupled to an arbitrary number of vector multiplets (this includes $U(1)^3$ gauged supergravity) were obtained in \cite{Kunduri:2007qy}. Unlike the ungauged theory, homogeneity is not implied by supersymmetry, and indeed there are more general solutions. 
\begin{prop}[\cite{Kunduri:2006uh}]The most general supersymmetric near-horizon geometry in minimal gauged supergravity, admitting a $U(1)^2$-rotational symmetry and a compact horizon section, is the near-horizon limit of the topologically-spherical supersymmetric black holes of~\cite{Chong:2005hr}.
\end{prop} 
 The motivation for assuming this isometry group is that all known black-hole solutions in five dimensions possess this. Interestingly, this result implies the non-existence of supersymmetric \AdS{5} black rings with $\mathbb{R} \times U(1)^2$ isometry. 

In fact, recent results allow one to remove all assumptions and obtain a complete classification. Generic supersymmetric solutions of minimal gauged supergravity preserve $\tfrac{1}{4}$-supersymmetry. 
\begin{prop}[\cite{Gutowski:2008ca, GGS}]
Any $\tfrac{1}{2}$-supersymmetric near-horizon geometry in minimal gauged supergravity must be invariant under a local $U(1)^2$-rotational isometry.
\end{prop}
Furthermore, the following has also been shown.
\begin{prop}[\cite{Grover:2013ima}]
Any supersymmetric near-horizon geometry in minimal gauged supergravity with a compact horizon section must preserve $\tfrac{1}{2}$-supersymmetry.
\end{prop}
This latter result is proved using a Lichnerowicz type identity to establish a correspondence between Killing spinors and solutions to a horizon Dirac equation, and then applying an index theorem. Therefore, combining the previous three propositions gives a complete classification theorem for near-horizon geometries in minimal gauged supergravity.

\begin{theorem}[\cite{Kunduri:2006uh, Grover:2013ima, GGS}]
A supersymmetric near-horizon geometry in minimal gauged supergravity, with a compact horizon section, must be locally isometric to the near-horizon limit of the topologically-spherical supersymmetric black holes~\cite{Chong:2005hr}, or the homogeneous near-horizon geometries with the $\text{Nil}$ or $SL(2,\mathbb{R})$ horizons.
\end{theorem}

This theorem establishes a striking corollary for the corresponding black hole classification theorem.
\begin{cor}
Supersymmetric black rings in minimal gauged supergravity do not exist.
\end{cor}
We emphasise that the absence of supersymmetric \AdS{5} black rings is rather suprising, given asymptotically-flat counterparts are known to exist~\cite{Elvang:2004rt}.

Parts of the above analysis have been generalised by $U(1)^n$ gauged supergravity, although the results are slightly different.
\begin{prop} [\cite{Kunduri:2007qy}] Consider a supersymmetric near-horizon geometry in $U(1)^n$ minimal gauged supergravity with $U(1)^2$-rotational symmetry and a compact horizon section. It must be either: (i) the near-horizon limit of the topologically spherical black holes of~\cite{Kunduri:2006ek}; or (ii) $\mathrm{AdS}_3 \times S^2$ with $H \cong S^1 \times S^2$ or (iii) $\mathrm{AdS}_3 \times T^2$ with $H \cong {T}^3$. These latter two cases have constant scalars and only exist in certain regions of the scalar moduli space (not including the minimal theory).
\end{prop} 
Therefore, in this theory one cannot rule out the existence of supersymmetric \AdS{5} black rings (although as argued in~\cite{Kunduri:2007qy} they would not be connected to the asymptotically-flat black rings~\cite{Elvang:2004ds}).
It would be interesting to complete the classification of near-horizon geometries in this more general theory, along the lines of the minimal theory.

\subsection{Six dimensions}

The simplest supergravity in six dimensions is minimal supergravity. The bosonic field context of this theory is a metric and a 2-form potential with self-dual field strength. The classification of supersymmetric solutions to this theory was given in~\cite{Gutowski:2003rg}. This was used to work out a complete classification of supersymmetric near-horizon geometries.

\begin{theorem}[\cite{Gutowski:2003rg}]
Any supersymmetric near-horizon geometry of $D=6$ minimal supergravity, with a compact horizon cross section, is either $\mathbb{R}^{1,1}\times T^4$, $\mathbb{R}^{1,1}\times K_3$ or locally $\mathrm{AdS}_3 \times S^3$.
\end{theorem}

The $\mathrm{AdS}_3\times S^3$ solution has $H \cong S^1\times S^3$ and arises as the near-horizon limit of a supersymmetric rotating black string.

Analogous results have been obtained for minimal supergravity coupled to an arbitrary number of scalar and tensor multiplets~\cite{Akyol:2011mh}.

\subsection{Ten dimensions}

Various results have been derived for heterotic supergravity and type IIB supergravity. 

The bosonic field content of $D=10$ heterotic supergravity consist of the metric, a 2-form gauge potential and a scalar field (dilaton). The full theory is invariant under 16 supersymmetries. There are two classes of supersymmetric near-horizon geometries~\cite{Gutowski:2009wm}. One is the direct product $\mathbb{R}^{1,1}\times H$, with vanishing flux and constant dilaton, where $H$ is Spin(7) holonomy manifold, which generically preserves one supersymmetry (there are solutions in this class which preserve more supersymmetry provided $H$ has certain special holonomy). In the second class, the near-horizon geometry is a fibration of \AdS{3} over a base $B_7$ (with a $U(1)$-connection) with a $G_2$ structure, which must preserve $2$ supersymmetries. This class may preserve $4,6,8$ supersymmetries if $B_7$ is further restricted. In particular, an explicit classification for $\tfrac{1}{2}$-supersymmetric near-horizon geometries is possible.
\begin{prop}[\cite{Gutowski:2009wm}] Any supersymmetric near-horizon geometry of heterotic supergravity invariant under 8 supersymmetries, with a compact horizon cross section, must be locally isometric to one of $\mathrm{AdS}_3 \times S^3 \times {T}^4$, $\mathrm{AdS}_3 \times S^3 \times K_3$ or $\mathbb{R}^{1,1}\times {T}^4 \times K_3$ (with constant dilaton).
\end{prop} 
\noindent A large number of heterotic horizons preserving $4$ supersymmetries have been constructed, including explicit examples where $H$ is $SU(3)$ and $S^3\times S^3 \times T^2$~\cite{Gutowski:2010jk}. 

The bosonic field content of type IIB supergravity consists of a
metric, a complex scalar, a complex 2-form potential and a self-dual
5-form field strength. The theory is invariant under 32
supersymmetries. A variety of results concerning the classification of
supersymmetric near-horizon geometries in this theory have been
derived~\cite{Gran:2011ix, Gran:2011qr, Gran:2013kfa}. Certain explicit classification results are known for near-horizon geometries with just a 5-form flux preserving more than two supersymmetries, albeit under certain restrictive assumptions~\cite{Gran:2011qr}. More generally, the existence of one supersymmetry places rather weak geometric constraints on the horizon cross sections $H$: generically $H$ may be any almost Hermitian spin$_c$ manifold~\cite{Gran:2013kfa}. There are also special cases for which $H$ has a $Spin(7)$ structure and those for which it has an $SU(4)$ structure (where the Killing spinor is pure). More recently, it has been shown that any supersymmetric near-horizon geometry in type IIB supergravity must preserve an even number of supersymmetries, and furthermore, if a certain horizon Dirac operator has non-trivial kernel the bosonic symmetry group must contain $SO(2,1)$~\cite{Gran:2013wca}.

\subsection{Eleven dimensions}

The bosonic field content of $D=11$ supergravity consists of a metric $g_{\mu\nu}$ and a 3-form potential $C_{\mu\nu\rho}$. The near-horizon limit of the 4-form field strength ${\cal F}= \td C$ can be written as Eq.~(\ref{NHpform}), where $Y$ and $X$ are a 2-form and closed 4-form respectively on the 9-dimensional horizon cross sections $H$. The near-horizon Einstein equations are Eqs.~(\ref{horizoneq}) and (\ref{Feq}) with $\Lambda=0$ and the matter field terms are given by~\cite{Gutowski:2012eq}
\begin{eqnarray}
P_{ab} &=& -\tfrac{1}{2} Y_{ac} Y_{b}^{~c} + \tfrac{1}{12} X_{ac_1 c_2 c_3}X_{b}^{~c_1c_2c_3} +\gamma_{ab} \left( \tfrac{1}{12} Y^2 - \tfrac{1}{144} X^2 \right), \\
E & =& \tfrac{1}{6}Y^2 + \tfrac{1}{144} X^2 \,.
\end{eqnarray}
We note that the dominant and strong energy conditions are satisfied: $P_{ab}\gamma^{ab} = \tfrac{1}{4} Y^2+ \tfrac{1}{48} X^2 \geq 0$ and $E\geq 0$. Therefore, the general results established under these assumptions, discussed in Section~\ref{section:generalresults}, are all valid, including most notably the horizon topology theorem.

Various classification results have been derived for \emph{supersymmetric} near-horizon geometry solutions under the assumption that cross sections of the horizon are compact. Static supersymmetric near-horizon geometries are warped products of either $\mathbb{R}^{1,1}$ or \AdS{2} with $M_9$, where $M_9$ admits a particular $G$-structure~\cite{Gutowski:2011xx}. We note that these warped product forms are guaranteed by the general analysis of static near-horizon geometries in Section~\ref{ads2:static}. Supersymmetric near-horizon geometries have been studied more generally in \cite{Gutowski:2012eq}. Most interestingly, a near-horizon (super)symmetry enhancement theorem has been established.

\begin{theorem}[\cite{Gutowski:2013kma}]
Any supersymmetric near-horizon geometry solution to eleven dimensional supergravity, with compact horizon cross sections, must preserve an even number of supersymmetries. Furthermore, the bosonic symmetry group must contain $SO(2,1)$.
\end{theorem}

The proof of this follows by first establishing a Lichnerowicz type identity for certain horizon Dirac operators and then application of an index theorem. As far as bosonic symmetry is concerned, the above result is a direct analogue of the various near-horizon symmetry theorems discussed in Section~\ref{sec:ads2theorems}, which are instead established under various assumptions of rotational symmetry.



\newpage
\section{Solutions with Gauge Fields}
\label{sec:gauge}

In this section we will consider general near-horizon geometries coupled to non-trivial gauge fields. We will mostly focus on theories that are in the bosonic sector of minimal supergravity theories (since these are the best understood cases). 
Extremal, non-supersymmetric, near-horizon geometries may be thought of as interpolating between vacuum and supersymmetric solutions. They consist of a much larger class of solutions, which, at least in higher dimensions, are much more difficult to classify. In particular, we consider $D=3,4,5$ Einstein--Maxwell theory, possibly coupled to a Chern--Simons term in odd dimensions, and $D=4$ Einstein--Yang--Mills theory.

\subsection{Three dimensional Einstein--Maxwell--Chern--Simons theory}
\label{gauge:3d}

The classification of near-horizon geometries in $D=3$ Einstein--Maxwell theory with a cosmological constant can be completely solved. To the best of our knowledge this has not been presented before, so for completeness we include it here. It should be noted though that partial results which capture the main result were previously shown in~\cite{Matyjasek:2004pg}.

The method parallels the vacuum case in Section~\ref{vac:3d} closely. As in that case the near-horizon metric data reads $h = h(x) \td x$ and $\gamma = \td x^2$. Since cross sections $H$ of the horizon are one-dimensional, the Maxwell 2-form induced on $H$ must vanish identically. Hence, the most general near-horizon Maxwell field (\ref{NHmaxwell}) in three dimensions is ${\cal F}_{\mathrm{NH}} = \td (r \Delta(x) \td v)$. It is straightforward to show that the 3D Maxwell equation $\td\star {\cal F}=0$, where $\star$ is the Hodge dual with respect to the spacetime metric, is equivalent to the following equation on $H$:
\begin{equation}
\Delta' = h \Delta \label{eq:3dmax} \,.
\end{equation}
The near-horizon Einstein equations~(\ref{horizoneq}) and (\ref{Feq}) are simply
\begin{eqnarray}
h'= \tfrac{1}{2} h^2 + 2\Delta^2 +\Lambda, \label{eq:3dEM} \\
F = \tfrac{1}{2} h^2 - \tfrac{1}{2}h' +\Lambda \,.
\end{eqnarray}
\begin{theorem}
Consider a near-horizon geometry with a compact horizon cross section $H\cong S^1$ in Einstein--Maxwell-$\Lambda$ theory. If $\Lambda<0$ the near-horizon geometry must be either $\mathrm{AdS}_2 \times S^1$ with a constant \AdS{2} Maxwell field, or the quotient of \AdS{3} Eq.~(\ref{ads3}) with a vanishing Maxwell field. If $\Lambda=0$ the only solution is the trivial flat near-horizon geometry $\mathbb{R}^{1,1}\times S^1$. If $\Lambda>0$ there are no solutions.
\end{theorem}

\noindent \textit{Proof:} Rather that solving the above ODEs we may use a global argument. Compactness requires $x$ to be a periodic coordinate on $H\cong S^1$ and since $h, \Delta$ are globally defined they must be periodic functions of $x$. For $\Lambda \geq 0$ simply integrate Eq.~(\ref{eq:3dEM}) over $H$ to find that the only solution is the trivial flat one $h\equiv 0, \Delta \equiv 0$ and $\Lambda=0$. For $\Lambda \equiv -\frac{2}{\ell^2}<0$ we may argue as follows. Multiply Eq.~(\ref{eq:3dEM}) by $h'$ and integrate over $H$ to obtain
\begin{equation}
0=\int_{S^1} ( h'^2 - 2\Delta^2 h') \td x = \int_{S^1} ( h'^2 +4\Delta^2 h^2) \td x \,,
\end{equation}
where in the second equality we have integrated by parts and used Eq.~(\ref{eq:3dmax}). Hence $h$ must be a constant and $h\Delta \equiv 0$. Equation (\ref{eq:3dEM}) then implies $\Delta$ is also a constant. We deduce the only possible solutions are $h=0, \Delta =\pm \frac{1}{\ell}$, or $h=\pm \frac{2}{\ell}, \Delta =0$. The former gives a near-horizon geometry $\mathrm{AdS}_2\times S^1$ and the latter is the vacuum solution locally isometric to \AdS{3}. 

This result implies that the near-horizon limit of \emph{any} charged rotating black-hole solution to 3D Einstein--Maxwell-$\Lambda$ theory either has vanishing charge or angular momentum. 
The $\mathrm{AdS}_2\times S^1$ solution is the near-horizon limit of the non-rotating extremal charged BTZ black hole, whereas the \AdS{3} solution is the near-horizon limit of the  vacuum rotating extremal BTZ~\cite{Banados:1992wn}. 
Charged rotating black holes were first obtained within a wide class of stationary and axisymmetric solutions to Einstein--Maxwell-$\Lambda$ theory~\cite{Clement:1993kc}, and later by applying a solution generating technique to the charged non-rotating black hole~\cite{Clement:1995zt, Martinez:1999qi}. We have checked that in the extremal limit their near-horizon geometry is the $\mathrm{AdS}_2\times S^1$ solution, so the angular momentum is lost in the near-horizon limit, in agreement with the above analysis. It would be interesting to investigate whether charged rotating black holes exist that instead possess a locally \AdS{3} near-horizon geometry. In this case, charge would not be captured by the near-horizon geometry, a phenomenon that is known to occur for five-dimensional supersymmetric black rings whose near-horizon geometry is locally $\mathrm{AdS}_3 \times S^2$. 

In 2\,+\,1 dimensions an Abelian gauge field monopole is not isolated. Electric charge can be ``screened'' by adding a mass term to the gauge field. A natural way to do this is to add a Chern--Simons term $\mu \int {\cal A} \wedge {\cal F}$ to the spacetime action, resulting in a topologically-massive gauge theory. This only modifies the Maxwell equation:
\begin{equation}
\td \star {\cal F} + \mu {\cal F}=0 \,,
\end{equation}
where $\mu$ is the mass parameter of the gauge field. For the near-horizon Maxwell field it can be shown that this is equivalent to 
\begin{equation}
\Delta' = (h+\mu)\Delta \,. \label{eq:3dmaxcs}
\end{equation}
As in the pure Einstein--Maxwell case, a complete classification of near-horizon geometries to this theory is possible. To the best of our knowledge this has not been presented before.
\begin{theorem}
Consider a near-horizon geometry with a cross section $H\cong S^1$ in Einstein--Maxwell-$\Lambda$ theory with a Chern--Simons term and mass $\mu$. If $\Lambda<0$ the functions $\Delta$ and $h$ are constant and the near-horizon geometry is the homogeneous $S^1$-bundle over \AdS{2} (\ref{warpedads3}). If $\Lambda=0$ the only solution is the trivial flat near-horizon geometry $\mathbb{R}^{1,1}\times S^1$. If $\Lambda>0$ there are no solutions.
\end{theorem}
For $\Lambda \geq 0$ the proof of this is identical to the Einstein--Maxwell case above. For $\Lambda = -\tfrac{2}{\ell^2}$ one can also use the same argument as the Einstein--Maxwell case. Using the horizon equation~(\ref{eq:3dEM}) and Maxwell equation~(\ref{eq:3dmaxcs}) one can show
\begin{equation}
0=\int_{S^1} ( h'^2 - 2\Delta^2 h') \td x = \int_{S^1} ( h'^2 +4\Delta^2 (h+\mu)^2) \td x \,,
\end{equation}
which implies $h$ must be a constant and $\Delta (h+\mu) \equiv 0$. The horizon equation then implies $\Delta$ is a constant. If $\Delta=0$ one gets the vacuum \AdS{3} solution. If $\Delta \neq 0$ then $h=-\mu$ and $\tfrac{1}{2}\mu^2 + 2\Delta^2 =\tfrac{2}{\ell^2}$. The rest of the near-horizon data is given by $F= \tfrac{1}{2}\mu^2 +\Lambda$ and hence the near-horizon geometry is
\begin{eqnarray}
g &=& -2\Delta^2 r^2 \td v^2 +2 \td v \td r - 2\mu r \td v \td x+ \td x^2 \nonumber \\ &=&- \left( \tfrac{1}{2}\mu^2+\tfrac{2}{\ell^2} \right) r^2 \td v^2 + 2 \td v \td r + ( \td x- \mu r \td v)^2 \label{warpedads3}
\end{eqnarray}
and ${\cal F} = \Delta \td r\wedge \td v$. Note that if $\mu=0$ we recover the $\mathrm{AdS}_2 \times S^1$ solution, whereas if $\Delta \to 0$ we recover the vacuum \AdS{3} solution. 

This implies that the near-horizon geometry of any charged rotating black-hole solution to this theory is either the vacuum \AdS{3} solution, or the non-trivial solution~(\ref{warpedads3}), which is sometimes referred to as ``warped \AdS{3}''. Examples of charged rotating black-hole solutions in this theory have been found~\cite{Moussa:2008sj}.

\subsection{Four dimensional Einstein--Maxwell theory}
\label{gauge:4d}

The spacetime Einstein--Maxwell equations are Eqs.~(\ref{einstein}), (\ref{TMax}) with $n=2$ and $\td \star {\cal F}=0$, where $\star$ is the Hodge dual with respect to the spacetime metric, and the Bianchi identity $\td {\cal F}=0$. The near-horizon Maxwell field is given by (\ref{NHmaxwell}). The near-horizon geometry Einstein--Maxwell equations are given by Eqs.~(\ref{horizoneq}) and (\ref{Feq}), where 
\begin{eqnarray}
 P_{ab} &=& 2{B}_{ac}{B}_{bd}\gamma^{cd} + \Delta^2\gamma_{ab} - \frac{\gamma_{ab}}{2}{B}^2, \label{EMabeq} \\
E &=& \Delta^2 +\frac{B^2}{2}, \label{EMFeq} \\ 
\label{4dmaxwell}
\td\star_2 B &=& \star_2 i_hB+ \star_2 (\td\Delta-\Delta h) \,,
\end{eqnarray}
where $\star_2$ is the Hodge dual with respect to the horizon metric $\gamma_{ab}$. Observe that $\star_2 B$ is a function on $H$.

Static near-horizon geometries have been completely classified. For $\Lambda=0$ this was first derived in~\cite{Chrusciel:2006pc}, and generalised to $\Lambda \neq 0$ in~\cite{Kunduri:2008tk}.

\begin{theorem}[\cite{Chrusciel:2006pc, Kunduri:2008tk}]
Consider a static near-horizon geometry in $D=4$ Einstein--Maxwell-$\Lambda$ theory, with compact horizon cross section $H$. For $\Lambda \geq 0$ it must be $\mathrm{AdS}_2\times S^2$. For $\Lambda<0$ it must be $\mathrm{AdS}_2\times H$ where $H$ is one of the constant curvature surfaces $S^2, T^2, \Sigma_g$. 
\end{theorem}
It is worth remarking that if one removes the assumption of compactness one can still completely classify near-horizon geometries. The extra solution one obtains can be written as a warped product
\bea
g &=& \psi^2( A_0 r^2 \td v^2 +2\td v \td r)+ \frac{\td\psi^2}{P(\psi)}+P(\psi) \td{\phi}^2, \\
{\cal F} &=& e \td r \wedge \td v + b \psi^{-2} \td \psi \wedge \td\phi \,,
\eea 
where $P(\psi)=A_0 -c (2\psi)^{-1}- (e^2+b^2)\psi^{-2}-\Lambda \psi^2/3$, which is an analyticaly continued Reissner--Nordstr\"om-$\Lambda$ solution. 

Non-static near-horizon geometries are not fully classified, except under the additional assumption of axisymmetry.
\begin{theorem}[\cite{Lewandowski:2002ua, Kunduri:2008tk}]
Any axisymmetric, non-static near-horizon geometry in $D=4$ Einstein--Maxwell-$\Lambda$ theory, with a compact horizon cross section, must be given by the near-horizon geometry of an extremal Kerr--Newman-$\Lambda$ black hole.
\end{theorem}
\cite{Lewandowski:2002ua} solved the $\Lambda=0$ in the context of isolated degenerate horizons, where the same equations on $H$ arise. \cite{Kunduri:2008tk} solved the case with $\Lambda \neq 0$. Note that the horizon topology theorem excludes the possibility of toroidal horizon cross sections for $\Lambda \geq 0$. \cite{Li:2013gca} also excluded the possibility of a toroidal horizon cross section if $\Lambda<0$ under the assumptions of the above theorem.

It is worth noting that the results presented in this section, as well as the techniques used to establish them, are entirely analogous to the vacuum case presented in Section~\ref{vac:4d}.

\subsection{Five dimensional Einstein--Maxwell--Chern--Simons theory}
\label{gauge:5d}

The field equations of $D=5$ Einstein--Maxwell theory coupled to a Chern--Simons term are given by Eqs.~(\ref{einstein}) and (\ref{TMax}) with $n=3$ and
\begin{eqnarray}
\td\star{\cal F} +\frac{2 \xi}{\sqrt{3}}{\cal F} \wedge {\cal F} = 0 \label{EMCS2}
\end{eqnarray}
where ${\cal F}$ is the Maxwell two form and $\td {\cal F}=0$. The cases of most interest are $\xi=0$ and $\xi=1$, which correspond to pure Einstein--Maxwell theory and the bosonic sector of minimal supergravity respectively. The near-horizon Maxwell field is given by (\ref{NHmaxwell}). The corresponding near-horizon geometry equations are given by Eqs.~(\ref{horizoneq}) and (\ref{Feq}) and
\begin{eqnarray}
 P_{ab} &=& 2{B}_{ac}{B}_{bd}\gamma^{cd} + \left(\frac{2\Delta^2}{3} - \frac{1}{3}B^2\right)\gamma_{ab}, \label{EMCSabeq} \\
E &=& \frac{4\Delta^2}{3} + \frac{B^2}{3}, \label{EMCSFeq} \\
\label{maxeq}
\td\star_3 B &=& -\star_3 i_h B- \star_3 (\td\Delta-\Delta h)+\frac{4\xi}{\sqrt{3}} \Delta B \,,
\eea
where $\star_3$ is the Hodge dual with respect to the 3D horizon metric $\gamma_{ab}$. In this section we summarise what is known about solutions to these equations, which is mostly restricted to the $\Lambda=0$ case. Hence, unless otherwise stated, we assume $\Lambda=0$ in this section. 

A number of new complications arise that render the classification problem more difficult, most obviously the lack of electro-magnetic duality. Therefore, purely \emph{electric} solutions, which correspond to $\Delta \neq 0$ and $B \equiv 0$, are qualitatively different to purely \emph{magnetic} solutions, which correspond to $\Delta \equiv 0$ and $B \neq 0$.

\subsubsection{Static}
\label{gauge:static}

Perhaps somewhat surprisingly a complete classification of static near-horizon geometries in this theory has not yet been achieved. Nevertheless, a number of results have been proved under various extra assumptions. All the results summarised in this section were proved in~\cite{Kunduri:2009ud}. 

As in other five dimensional near-horizon geometry classifications, the assumption of $U(1)^2$ rotational symmetry proves to be useful. Static near-horizon geometries in this class in general are either warped products of \AdS{2} and $H$, or \AdS{3} and a 2D manifold, see the Corollary~\ref{cor:ads3}. The \AdS{3} near-horizon geometries are necessarily purely magnetic and can be classified for any $\xi$.

\begin{prop} Any static \AdS{3} near-horizon geometry with a $U(1)^2$-rotational symmetry, in Einstein--Maxwell--Chern--Simons theory, with a compact horizon cross section, is the direct product of a quotient of \AdS{3} and a round $S^2$.
\end{prop}
This classifies a subset of purely magnetic geometries. By combining the results of~\cite{Kunduri:2009ud} together with Proposition~\ref{NHminimal5d} of~\cite{Kunduri:2011zr}, it can be deduced that for $\xi=1$ there are no purely magnetic \AdS{2} geometries; therefore, with these symmetries, one has a complete classification of purely magnetic geometries. 
\begin{cor}
Any static, purely magnetic, near-horizon geometry in minimal supergravity, possessing $U(1)^2$-rotational symmetry and compact cross sections $H$, must be locally isometric to $\mathrm{AdS}_3 \times S^2$ with $H =S^1 \times S^2$.
\end{cor}

We now turn to purely electric geometries.

\begin{prop} Consider a static, purely electric, near-horizon geometry in Einstein--Maxwell--Chern--Simons theory, with a $U(1)^2$-rotational symmetry and compact cross section. It must be given by either $\mathrm{AdS}_2 \times S^3$, or a warped product of \AdS{2} and an inhomogeneous $S^3$.
\end{prop}
The latter non-trivial solution is in fact the near-horizon limit of an extremal RN black hole immersed in a background electric field (this can be generated via a Harrison type transformation). 

Finally, we turn to the case where the near-horizon geometry possess both electric and magnetic fields. In fact one can prove a general result in this case, i.e., without the assumption of rotational symmetries.
\begin{prop}\label{prop:staticdyonic}
Any static near-horizon geometry with compact cross sections $H$, in Einstein--Maxwell--Chern--Simons theory with coupling $\xi\neq 0$, with non-trivial electric and magnetic fields, is a direct product of $\mathrm{AdS}_2 \times H$, where the metric on $H$ is not Einstein.
\end{prop}
Explicit examples for $0< \xi^2<1/4$ were also found, which all have $H\cong S^3$ with $U(1)^2$-rotational symmetry. However, we should emphasise that no examples are known for minimal supergravity $(\xi=1)$. Hence there is the possibility that in this case static near-horizon geometries with non-trivial electric and magnetic fields do not exist, although this has not yet been shown. If this is the case, then the above results fully classify static near-horizon geometries with $U(1)^2$-rotational symmetry. For $\xi=0$ the analysis of electro-magnetic geometries is analogous to the purely magnetic case above; in fact there exists a dyonic \AdS{2} geometry that is a direct product $\mathrm{AdS}_2 \times S^2 \times S^1$ and it is conjectured there are no others.

\subsubsection{Homogeneous}
\label{gauge:homo}

The classification of \emph{homogeneous} near-horizon geometries can be completely solved, even including a cosmological constant $\Lambda$. This does not appear to have been presented explicitly before, so for completeness we include it here with a brief derivation. We may define a homogeneous near-horizon geometry as follows. The Riemannian manifold $(H,\gamma_{ab})$ is a homogeneous space, i.e., admits a transitive isometry group $K$, such that the rest of the near-horizon data $(F, h_a, \Delta, B_{ab})$ are invariant under $K$. As discussed in Section~(\ref{vac:5d}), this is equivalent to the near-horizon geometry being a homogeneous spacetime with a Maxwell field invariant under its isometry group.

An immediate consequence of homogeneity is that an invariant function must be a constant and any invariant 1-form must be a Killing field. Hence the 1-forms $h$ and $j \equiv \star_3 B$ are Killing and the functions $F$, $\Delta, h^2, j^2$ must be constants. Thus, the horizon Einstein equations~(\ref{horizoneq}), (\ref{Feq}), (\ref{EMCSabeq}), (\ref{EMCSFeq}) and Maxwell equation~(\ref{maxeq}) simplify. 

Firstly, note that if $h$ and $j$ vanish identically then $H$ is Einstein $R_{ab}= \tfrac{1}{2}(\Delta^2 +2\Lambda ) \gamma_{ab}$, so $H$ is a constant curvature space $S^3, \mathbb{R}^3, \mathbb{H}^3$ (the latter two can only occur if $\Lambda<0$). This family includes the static near-horizon geometry $\mathrm{AdS}_2\times H$.

Now consider the case where at least one of $h$ and $j$ is non-vanishing. By contracting the Maxwell equation~(\ref{maxeq}) with $j^aj^b$ one can show that $(j \cdot h)^2 = j^2 h^2$ and hence by the Cauchy-Schwarz inequality $j$ and $h$ must be parallel as long as they are both non-zero. Thus, if one of $h, j$ is non-vanishing, we can write $h_a = k u_a$ and $j_a = q u_a$ for some constants $k,q$ where $u_a$ is a unit normalised Killing vector field. The near-horizon equations now reduce to
\begin{eqnarray}
R_{ab} &=& \tfrac{1}{2} \left(k^2- 4q^2\right) u_a u_b + \left( \tfrac{4}{3} q^2 +\tfrac{1}{2}\Delta^2 +\Lambda \right) \gamma_{ab}, \\
F &=& \tfrac{1}{2}k^2 -\tfrac{2}{3}q^2 - \Delta^2+ \Lambda, \\
q du &=& \left( \tfrac{\sqrt{3}}{2} k+ 2\xi q\right) \Delta \star u \,. \label{dj}
\end{eqnarray}
This allows one to prove:
\begin{theorem} \label{EMhomo}
Any homogeneous near-horizon geometry in Einstein--Maxwell--CS-$\Lambda$ theory for which one of the constants $k,q$ is non-zero, must be locally isometric to 
\begin{eqnarray}
g &=& \left(-\tfrac{1}{2}k^2 -\tfrac{2}{3}q^2 - \Delta^2+ \Lambda \right) r^2 \td v^2 + 2 \td v \td r + (\hat{\omega}+ k r\td v)^2 + \hat{g}, \\
{\cal F} &=& \Delta \td r \wedge \td v + q \hat{\epsilon} \,,
\end{eqnarray}
where $\hat{\omega}$ is a $U(1)$-connection over a 2D base with metric $\hat{g}$ satisfying $\mathrm{Ric}(\hat{g}) = \hat{\lambda} \hat{g}$, with $\hat{\lambda} = \tfrac{1}{2}k^2+\tfrac{2}{3} q^2+\Delta^2+2\Lambda$, and $\hat{\epsilon}$ is the volume form of the 2D base. The curvature of the connection is $\td \hat{\omega} = [ k^2 -\tfrac{4}{3} q^2+\Delta^2+ 2\Lambda]^{1/2} \,\hat{\epsilon}$ and $k^2 -\tfrac{4}{3} q^2+\Delta^2+ 2\Lambda \geq 0$. If $q\neq 0$ the constants must satisfy
\be
k^2 -\tfrac{4}{3}q^2+\Delta^2+ 2\Lambda=\tfrac{1}{4}q^{-2} \Delta^2 \left( \sqrt{3} k + 4\xi q \right)^2
\ee
whereas if $q=0$ one must have $k\Delta=0$.
\end{theorem}
The proof of this proceeds as in the vacuum case, by reducing the horizon equations to the 2D orbit space defined by the Killing field $u$. The above result contains a number of special cases of interest, which we now elaborate upon. Firstly, note that $q=\Delta=0$ reduces to the vacuum case, see Theorem~\ref{vachomo}. 

Before discussing the general case consider $k=0$, so the near-horizon geometry is static, which connects to Section \ref{gauge:static}. The constraint on the parameters is $(1-4\xi^2) \Delta^2 = \tfrac{4}{3} q^2 -2\Lambda$ and hence, if $\Lambda \leq 0$ one must have $0 \leq \xi^2 <\tfrac{1}{4}$. For $\xi=0$ the connection is trivial and hence the near-horizon geometry is locally isometric to the dyonic $\mathrm{AdS}_2\times S^1 \times S^2$ solution. For $0< \xi^2 <\tfrac{1}{4}$ we get examples of the geometries in Proposition~(\ref{prop:staticdyonic}), where $H\cong S^3$ with its standard homogeneous metric. 

Now we consider $\Lambda=0$ in generality so at least one of $k,q,\Delta$ is non-zero, in which case $\hat{\lambda}>0$ and so $\hat{g}$ is the round $S^2$. The horizon is then either $H\cong S^3$ with its homogeneous metric or $H \cong S^1\times S^2$, depending on whether the connection $\hat{\omega}$ is non-trivial or not, respectively. Notably we have: 

\begin{cor}
Any homogeneous near-horizon geometry of minimal supergravity is locally isometric to $\mathrm{AdS}_3 \times S^2$, or the near-horizon limit of either (i) the BMPV black hole (including $\mathrm{AdS}_2\times S^3$), or (ii) an extremal nonsupersymmetric charged black hole with $SU(2)\times U(1)$ rotational symmetry.
\end{cor}
The proof of this follows immediately from Theorem~\ref{EMhomo} with $\Lambda=0$ and $\xi=1$. In this case the constraint on the parameters factorises to give two branches of possible solutions (a) $k= -2q/\sqrt{3}$ or (b) $\Delta^2 ( k+2 \sqrt{3} q)= 4q^2 ( k-2q/\sqrt{3})/3$. Case (a) gives two solutions. If $\Delta \neq 0$ it must have $H \cong S^3$ and corresponds to the BMPV solution (i) (for $q\to 0$ this reduces to $\mathrm{AdS}_2 \times S^3$), whereas if $\Delta=0$ it is the $\mathrm{AdS}_3\times S^2$ solution with $H \cong S^1\times S^2$. Case (b) also gives two solutions. If $k= 2 q/\sqrt{3}$ then $\Delta=0$, which gives $\mathrm{AdS}_3\times S^2$ with $H\cong S^1 \times S^2$, otherwise we get solution (ii) with $H \cong S^3$. Note that solution (ii) reduces to the vacuum case as $\Delta, q \to 0$. 

The black-hole solution (ii) may be constructed as follows. A charged generalisation of the MP black hole can be generated in minimal supergravity \cite{Cvetic:1996xz}. Generically, the extremal limit depends on 3-parameters with two independent angular momenta $J_1, J_2$ and $\mathbb{R}\times U(1)^2$ symmetry. Setting $|J_1|= |J_2| $ gives two distinct branches of 2-parameter extremal black-hole solutions with enhanced $SU(2)\times U(1)$ rotational symmetry corresponding to the BMPV solution (i) (which reduces to the RN solution if $J=0$) or solution (ii) (which reduces to the vacuum extremal MP black hole in the neutral limit).

It is interesting to note the analogous result for pure Einstein--Maxwell theory:
\begin{cor}
Any homogeneous near-horizon geometry of Einstein--Maxwell theory is locally isometric to either (i)
\begin{eqnarray}
g &=&- \left( \tfrac{1}{2}k^2 +2q^2 \right) r^2 \td v^2 + 2\td v \td r + \left( d\psi + \frac{k \cos \theta \td \phi}{\tfrac{1}{2}k^2+2q^2} + k r \td v \right)^2 + \frac{\td\theta^2 +\sin^2 \theta \td \phi^2}{\tfrac{1}{2}k^2+2q^2}, \nonumber \\
{\cal F} &=& \pm \tfrac{2}{\sqrt{3}} q \td r \wedge \td v+ \frac{q \sin \theta \td\theta \wedge \td \phi}{\tfrac{1}{2}k^2+2q^2},
\end{eqnarray}
or (ii)
\begin{eqnarray}
g &=&- \left( \Delta^2 +\tfrac{4}{3} q^2 \right) r^2 \td v^2 + 2\td v \td r + \left( d\psi + \frac{\Delta \cos \theta \td \phi}{\Delta^2 +\tfrac{4}{3} q^2} \pm \tfrac{2}{\sqrt{3}} q r \td v \right)^2 + \frac{\td\theta^2 +\sin^2 \theta \td \phi^2}{\Delta^2 +\tfrac{4}{3} q^2}, \nonumber \\
{\cal F} &=& \Delta \td r \wedge \td v+ \frac{q \sin \theta \td\theta \wedge \td \phi}{\Delta^2 +\tfrac{4}{3} q^2}.
\end{eqnarray}
\end{cor}
This also follows from application of Theorem~\ref{EMhomo} with $\Lambda=0$ and $\xi=0$. Solution (i) for $q\to 0$ reduces to the vacuum solution of Theorem~\ref{vachomo}, whereas for $k=0$, it gives the static dyonic $\mathrm{AdS}_2 \times S^1 \times S^2$ solution. Solution (ii) for $\Delta=0$ gives $\mathrm{AdS}_3 \times S^2$, whereas for $\Delta \neq 0$ it gives a near-horizon geometry with $H \cong S^3$, which for $q\to 0$ is $\mathrm{AdS}_2 \times S^3$. A charged rotating black-hole solution to Einstein--Maxwell theory generalising MP is not known explicitly. Hence this corollary could be of use for constructing such an extremal charged rotating black hole with $\mathbb{R}\times SU(2)\times U(1)$ symmetry.

For $\Lambda = - 4/\ell^2 <0$ there are even more possibilities, since $\hat{\lambda}$ may be positive, zero, or negative. One then gets near-horizon geometries that generically have $H \cong S^3, \text{Nil}, SL(2,\mathbb{R})$, respectively, equipped with their standard homogeneous metrics. Each of these may have a degenerate limit with $H \cong S^1\times S^2, T^3, S^1\times \mathbb{H}^2$, as occurs in the vacuum and supersymmetric cases. The full space of solutions interpolates between the vacuum case given in Corollary~\ref{cor:vachomo}, and the supersymmetric near-horizon geometries of gauged supergravity of Proposition~\ref{Thm:homogauged}. For example, the supersymmetric horizons~\cite{Gutowski:2004ez} correspond to $k = -2\sqrt{3} q$ and $k^2= 9 /\ell^2$ with $\hat{\lambda} = \Delta^2- 3\ell^{-2}$. We will not investigate the full space of solutions in detail here.

\subsubsection[$U(1)^2$-rotational symmetry]{\boldmath$U(1)^2$-rotational symmetry}

The classification of near-horizon geometries in $D=5$ Einstein--Maxwell--CS theory, under the assumption of $U(1)^2$ symmetry, turns out to be significantly more complicated than the vacuum case. This is unsurprising; solutions may carry two independent angular momenta, electric charge and dipole/magnetic charge (depending on the spacetime asymptotics). As a result, there are several ways for a black hole to achieve extremality. Furthermore, such horizons may be deformed by background electric fields~\cite{Kunduri:2009ud}.

In the special case of minimal supergravity one can show:
\begin{prop}[\cite{Kunduri:2011zr}] \label{NHminimal5d}
Any near-horizon geometry of minimal supergravity with $U(1)^2$-rotational symmetry takes the form of Eqs.~(\ref{enhancedNH}) and (\ref{enhancedF}), where the functional form of $\Gamma(x), B_{ij}(x), b_i(x)$ can be fully determined in terms of rational functions of $x$. In particular, $\Gamma(x)$ is a quadratic function.
\end{prop}
The method of proof is discussed in Section~\ref{gauge:hidden}. Although this solves the problem in principle, it turns out that in practice it is very complicated to disentangle the constraints on the constants that specify the solution. Hence an explicit classification of all possible solutions has not yet been obtained, although in principle it is contained in the above result. 

We now summarise all known examples of five dimensional non-static near-horizon geometries with non-trivial gauge fields, which arise as near-horizon limits of known black hole or black string solutions. All these examples possess $U(1)^2$ rotational symmetry and hence fall into the class of solutions covered by Theorem~\ref{Theorem:genth}, so the near-horizon metric and field strength $(g,{\cal F})$ take the form of Eqs.~(\ref{enhancedNH}) and (\ref{enhancedF}) respectively. We will divide them by horizon topology. 

\subsubsection*{Spherical topology} 

\noindent\emph{Charged Myers--Perry black holes}: This asymptotically-flat solution is known explicitly only for minimal supergravity $\xi=1$ (in particular it is not known in pure Einstein--Maxwell $\xi =0$), since it can be constructed by a solution-generating procedure starting with the vacuum MP solution. It depends on four parameters $M, J_1, J_2$, and $Q$ corresponding to the ADM mass, two independent angular momenta and an electric charge. The extremal limit generically depends on three parameters and gives a near-horizon geometry with $H \cong S^3$. 

\vspace{6pt}
\noindent\emph{Charged Kaluza--Klein black holes}: The most general known solution to date was found in \cite{Tomizawa:2012nk} (see references therein for a list of previously known solutions) and carries a mass, two independent angular momenta, a KK monopole charge, an electric charge and a `magnetic charge'\epubtkFootnote{This is a conserved charge for such asymptotically KK spacetimes.}. The extremal limit will generically depend on five-parameters, however, as for the vacuum case the extremal locus must have more than one connected component. These solutions give a large family of near-horizon geometries with $H \cong S^3$. 

\subsubsection*{\boldmath$S^1 \times S^2$ topology} 

\noindent \textit{Supersymmetric black rings and strings}: The asymptotically-flat supersymmetric black ring~\cite{Elvang:2004rt} and the supersymmetric Taub--NUT black ring~\cite{Elvang:2005sa} both have a near-horizon geometry that is locally $\mathrm{AdS}_3 \times S^2$. There are also supersymmetric black string solutions with such near-horizon geometries~\cite{Gauntlett:2002nw, Bena:2004wv}. 

\vspace{6pt}
\noindent \textit{Dipole black rings}: The singly-spinning dipole black ring~\cite{Emparan:2004wy} is a solution to Einstein--Maxwell--CS for all $\xi$. It is a 3-parameter family with a single angular momentum and dipole charge possessing a 2-parameter extremal limit. The resulting near-horizon geometry with $H \cong S^1 \times S^2$ is parameterised by 4-parameters $(q,\lambda,R_1,R_2)$ with one constraint relating them. Asymptotic flatness of the full black-hole solution imposes one further constraint, although from the viewpoint of the near-horizon geometry, this is strictly an external condition and we will deal with the general case here. The solution is explicitly given by
\begin{eqnarray}\label{dipolering}
g &=& \Gamma(x)\left[ -\frac{r^2 \td v^2}{\ell^2} + 2 \td v \td r\right] + \frac{\ell^2 \Gamma(x) \td x^2}{(1-x^2)} \nonumber \\ &+& \frac{R_1^2 \lambda(1+\lambda) H(x)}{q (1-\lambda) F(x)} \left(\td \phi^1 + \frac{(1-\lambda)}{\lambda R_1 R_2} \sqrt{\frac{1-\lambda}{1+\lambda}} r \td v \right)^2 + \frac{R_2^2 q^2 \omega_0^2(1-x^2)}{H(x)^2} (\td \phi^2)^2, \nonumber \\
 {\cal F}&=& \frac{\sqrt{3}}{2}\td \left[ \sqrt{\frac{1-q}{1+q}}\frac{\omega_0 q R_2 (1+x)}{H(x)} d\phi^2 \right] ,
\end{eqnarray}
where $\Gamma(x) = \sqrt{\frac{q(1-\lambda)}{\lambda(1+\lambda)}}F(x) H(x)$ with $F(x) = 1 + \lambda x$, $H(x) = 1 - qx$ and we have also defined the length scale
$ \ell^2 = R_2^2 \sqrt{\frac{\lambda(1+\lambda) q^3}{1-\lambda}}$. The parameters satisfy $0 < \lambda, q <1$. The local metric induced on spatial cross sections $H$ extends smoothly to a metric on $S^1 \times S^2$ provided $\omega_0 = \sqrt{F(1) H(1)^3} = \sqrt{F(-1)H(-1)^3}$. 

\vspace{6pt}
\noindent \textit{Charged non-supersymmetric black rings}: The dipole ring solution admits a three-parameter charged generalisation with one independent angular momentum and electric and dipole charges \cite{Elvang:2004xi} (the removal of Dirac-Misner string singularities imposes an additional constraint, so this solution has the same number of parameters as that of~\cite{Emparan:2004wy}). The charged black ring has a two-parameter extremal limit with a corresponding two-parameter near-horizon geometry. As in the above case, at the level of near-horizon geometries there is an additional independent parameter corresponding to the arbitrary size of the radius of the $S^1$. 

\vspace{6pt}
\noindent \textit{Electro-magnetic Kerr black strings}: Black string solutions have been constructed carrying five independent charges: mass $M$, linear momentum $P$ along the $S^1$ of the string, angular momentum $J$ along the internal $S^2$, as well as electric $Q_e$ and magnetic charge $Q_m$~\cite{Compere:2010fm}. These solutions admit a four parameter extremal limit, which in turn give rise to a five-parameter family of non-static near horizon geometries (the additional parameter is the radius of the $S^1$ at spatial infinity)~\cite{Kunduri:2011zr}. 

For simplicity we will restrict our attention to the solutions with $P=Q_e=0$. The resulting near-horizon solution is parameterised by $(a,c_\beta,s_\beta)$ with $c_\beta^2 - s_\beta^2 = 1$ and corresponds to an extremal string with non-zero magnetic charge and internal angular momentum:
\begin{eqnarray}
g &=& \Gamma(x)\left[ -\frac{r^2 \td v^2}{\ell^2} + 2 \td v \td r\right] + \frac{\ell^2 \Gamma(x) \td x^2}{(1-x^2)} + \frac{4a^4(c_\beta^4+s_\beta^2)^2 (1-x^2)}{\ell^2 \Gamma(x)} \omega^2 \nonumber \\ &+& a^2\left[ \frac{R \td \phi^1}{a} - \frac{8 a^2 c_\beta^3 s_\beta^3}{\ell^4} r \td v + \frac{2a^2 c_\beta s_\beta(c_\beta^2+s_\beta^2)(1-x^2) }{\ell^2\Gamma(x)} \omega \right]^2, \nonumber \\
{\cal F}&=& \frac{2\sqrt{3}a^3 s_\beta c_\beta (c_\beta^4+s_\beta^4) }{\ell^2} \, \td \left[\frac{x}{\Gamma(x)} \omega \right] ,   \label{kerrstring}
\end{eqnarray}
where we have defined the one-form $\omega = \td \phi^2 - \frac{(c_\beta^2 + s_\beta^2)}{\ell^2(c_\beta^4 + s_\beta^4)} r \td v$, the function $\Gamma(x) = \frac{a^2}{\ell^2} (1+x^2 + 4 c_\beta^2 s_\beta^2 )$ and the length scale $\ell^2 = 2a^2(c_\beta^4 + s_\beta^4)$. The induced metric on cross sections of the horizon $H$ extends smoothly to a cohomogeneity-1 metric on $S^1 \times S^2$.

Although the two solutions (\ref{dipolering}) and (\ref{kerrstring}) share many features, it is important to emphasise that only the former is known to correspond to the near-horizon geometry of an asymptotically-flat black ring. It is conjectured that there exists a general black-ring solution to minimal supergravity that carries mass, two angular momenta, electric and dipole charges all independently. Hence there should exist corresponding 4-parameter families of extremal black rings. \cite{Kunduri:2011zr} discusses the possibility that the tensionless Kerr-string solution is the near-horizon geometry of these yet-to-be-constructed black rings.

\subsection{Theories with hidden symmetry}
\label{gauge:hidden}

Consider $U(1)^{D-3}$-invariant solutions of a general theory of the form (\ref{gentheory}). One may represent such solutions by a three-dimensional metric $h_{\mu\nu}$ and a set of scalar fields $\Phi^M$ (potentials) all defined on a three-dimensional manifold, see, e.g., \cite{Kunduri:2011zr}. Equivalently, such solutions can be derived from the field equations of a three-dimensional theory of gravity coupled to a scalar harmonic map whose target manifold is parameterised by the $\Phi^M$ with metric $G_{MN}(\Phi)$ determined by the specific theory. In certain theories of special interest, the scalar manifold is a symmetric space $G/K$ equipped with the bi-invariant metric 
\be
G_{MN}(\Phi) \td \Phi^M \td \Phi^N = \frac{1}{4m} \text{Tr} ( \Phi^{-1} \td \Phi)^2 \,,
\ee
where $\Phi$ is a coset representative of $G/K$ and $m$ is a normalisation constant dependent on the theory. Then the theory is equivalent to a three-dimensional theory of gravity coupled to a non-linear sigma model with target space $G/K$. 

Consider near-horizon geometries with $U(1)^{D-3}$ isometry in such theories. It can be shown \cite{Kunduri:2007vf, Hollands:2009ng, Kunduri:2011zr} that the classification problem reduces to an ODE on the orbit space $H/U(1)^{D-3}$ for the $\Phi^M$, while $h_{\mu\nu}$ is completely determined. We will assume non-toroidal horizon topology so the orbit space is an interval, which without loss of generality we take to be $[-1,1]$ parameterised by the coordinate $x$. The ODE is the equation of motion for a non-linear sigma model defined on this interval:
\begin{equation}
\frac{\td}{\td x} \left[(1-x^2) \Phi^{-1} \frac{\td \Phi}{\td x} \right] = 0 \,,
\end{equation}
where the a coset representative $\Phi$ depends only on $x$. It is straightforward to integrate this matrix equation and completely solve for the scalar fields $\Phi^M$, which can then be used to reconstruct the full $D$-dimensional solution. Hence in principle one has the full functional form of the solution. However, in practice, reconstructing the near-horizon data is hindered by the non-linearity of the scalar metric.

The most notable example which can be treated in the above formalism is vacuum $D$-dimensional gravity for which $G/K = SL(D-2,\mathbb{R}) / SO(D-2)$. The classification problem for near-horizon geometries has been completely solved using this approach~\cite{Hollands:2009ng}, as discussed earlier in Section~\ref{vac:weyl}. 

Four-dimensional Einstein--Maxwell also possesses such a structure where the coset is now $G/K = SU(2,1)/SU(2)$, although the near-horizon classification discussed earlier in Section~\ref{gauge:4d} does not exploit this fact. 

It turns out that $D=5$ minimal supergravity also has a non-linear sigma model structure with $G/K = G_{2,2}/SO(4)$ ($G_{2,2}$ refers to the split real form of the exceptional Lie group $G_2$). The classification of near-horizon geometries in this case was analysed in \cite{Kunduri:2011zr} using the hidden symmetry and some partial results were obtained, see Proposition~\ref{NHminimal5d}.

It is clear that this method has wider applicability. It would be interesting to use it to classify near-horizon geometries in other theories possessing such hidden symmetry.

We note that near-horizon geometries in this class extremise the energy functional of the harmonic map $\Phi : H/U(1)^{D-3} \to G/K$, with boundary conditions chosen such that the corresponding near-horizon data is smooth. Explicitly, this functional is given by 
\be
E[\Phi] = \int_{-1}^1 \left[ (1-x^2) \frac{1}{4m} \text{Tr} ( \Phi^{-1} \partial_x \Phi)^2 - \frac{2}{1-x^2} \right] \td x \label{eq:functional}
\ee
and from~\cite{Kunduri:2011zr} it can be deduced this vanishes on such extrema. For vacuum gravity, for which $m=1$, it was proved that $E[\Phi]\geq 0$ with equality if and only if $\Phi$ corresponds to a near-horizon geometry, i.e., near-horizon geometries are \emph{global} minimisers of this functional~\cite{Acena:2010ws, Hollands:2011sy}. This result has also been demonstrated for four dimensional Einstein--Maxwell theory~\cite{Clement:2012vb} and $D=4,5$ Einstein--Maxwell-dilaton theory~\cite{Yazadjiev:2012bx, Yazadjiev:2013hk}. It would be interesting if this result could be generalised to other theories with hidden symmetry such as minimal supergravity.

\subsection{Non-Abelian gauge fields}

Much less work has been done on classifying extremal black holes and their near-horizon geometries coupled to \emph{non-Abelian} gauge fields. 

The simplest setup for this is four-dimensional Einstein--Yang--Mills theory. As is well known, the standard four dimensional black-hole uniqueness theorems fail in this case (at least in the non-extremal case), for a review, see~\cite{Volkov:1998cc}. Nevertheless, near-horizon geometry uniqueness theorems analogous to the Einstein--Maxwell case have recently been established for this theory. 

Static near-horizon geometries in this theory have been completely classified.

\begin{theorem}[\cite{Li:2013gca}]
Consider $D=4$ Einstein--Yang--Mills-$\Lambda$ theory with a compact semi-simple gauge group $G$. Any static near-horizon geometry with compact horizon cross section is given by: $\mathrm{AdS}_2\times S^2$ if $\Lambda \geq 0$; $\mathrm{AdS}_2\times H$ where $H$ is one of $S^2, T^2, \Sigma_g$ if $\Lambda<0$.
\end{theorem}

The proof employs the same method as in Einstein--Maxwell theory. However, it should be noted that the Yang--Mills field need not be that of the Abelian embedded solution. The horizon gauge field may be any Yang--Mills connection on $S^2$, or on the higher genus surface as appropriate, with a gauge group $G$ (if there is a non-zero electric field $E$ the gauge group $G$ is broken to the centraliser of $E$). The moduli space of such connections have been previously classified~\cite{Atiyah:1982fa}. Hence, unless the gauge group is $SU(2)$, one may have genuinely non-Abelian solutions. It should be noted that static near-horizon geometries have been previously considered in Einstein--Yang--Mills--Higgs under certain restrictive assumptions~\cite{Bicak:1994tq}.

Non-static near-horizon geometries have also been classified under the assumption of axisymmetry.

\begin{theorem}[\cite{Li:2013gca}]
Any axisymmetric non-static near-horizon geometry with compact horizon cross section, in $D=4$ Einstein--Yang--Mills-$\Lambda$ with a compact semi-simple gauge group, must be given by the near-horizon geometry of an Abelian embedded extreme Kerr--Newman-$\Lambda$ black hole.
\end{theorem}

The proof of this actually requires new ingredients as compared to the Einstein--Maxwell theory. The \AdS{2}-symmetry enhancement theorems discussed in Section~\ref{sec:ads2theorems} do not apply in the presence of a non-Abelian gauge field. Nevertheless, assuming the horizon cross sections are of $S^2$ topology allows one to use a global argument to show the symmetry enhancement phenomenon still occurs. This implies the solution is effectively Abelian and allows one to avoid finding the general solution to the ODEs that result from the reduction of the Einstein--Yang--Mills equations. One can also rule out toroidal horizon cross sections, hence giving a complete classification of horizons with a $U(1)$-symmetry.

Interestingly, Einstein--Yang--Mills theory with a negative cosmological constant is a consistent truncation of the bosonic sector of $D=11$ supergravity on a squashed $S^7$~\cite{Pope:1985bu}. It would be interesting if near-horizon classification results could be obtained in more general theories with non-Abelian Yang--Mills fields, such as the full ${\cal N}=8$, $D=4$, $SO(8)$-gauged supergravity that arises as a truncation of $D=11$ supergravity on $S^7$.



\newpage
\section{Applications and Related Topics}
\label{section:applications}

\subsection{Black-hole uniqueness theorems}

One of the main motivations for classifying near-horizon geometries is to prove uniqueness theorems for the corresponding extremal black-hole solutions. This turns out to be a much harder problem and has only been achieved when extra structure is present that constrains certain global aspects of the spacetime. The role of the near-horizon geometry is to provide the correct boundary conditions near the horizon for the global--black-hole solution. 

\subsubsection{Supersymmetric black holes}

Uniqueness theorems for supersymmetric black holes have been proved in the simplest four and five-dimensional supergravity theories, by employing the associated near-horizon classifications described in Section~\ref{sec:susy}.

In four dimensions, the simplest supergravity theory that admits supersymmetric black holes is ${\cal N}=2$ minimal supergravity; its bosonic sector is simply Einstein--Maxwell theory.

\begin{theorem}[\cite{Chrusciel:2005ve}]
Consider an asymptotically-flat, supersymmetric black-hole solution to ${\cal N}=2,D=4$ minimal supergravity. Assume that the supersymmetric Killing vector field is timelike everywhere outside the horizon. Then it must belong to the Majumdar--Papapetrou family of black holes. If the horizon is connected it must be the extremal RN black hole.
\end{theorem} 

It would be interesting to remove the assumption on the supersymmetric Killing vector field to provide a complete classification of supersymmetric black holes in this case.

In five dimensions, the simplest supergravity theory that admits supersymmetric black holes is ${\cal N}=1$ minimal supergravity. Using general properties of supersymmetric solutions in this theory, as well as the near-horizon classification discussed in Section~\ref{sec:susy}, the following uniqueness theorem has been obtained.

\begin{theorem}[\cite{Reall:2002bh}]
Consider an asymptotically-flat, supersymmetric black-hole solution to ${\cal N}=1, D=5$ minimal supergravity, with horizon cross section $H\cong S^3$. Assume that the supersymmetric Killing vector field is timelike everywhere outside the horizon. Then it must belong to the BMPV family of black holes \cite{Breckenridge:1996is}. 
\end{theorem}

\noindent \textit{Remarks}:
\begin{itemize}
\item The BMPV solution is a stationary, non-static, non-rotating black hole with angular momentum $J$ and electric charge $Q$. For $J=0$ it reduces to the extremal RN black hole. 
\item It would be interesting to investigate the classification of supersymmetric black holes without the assumption on the supersymmetric Killing vector field.
\item An analogous uniqueness theorem for supersymmetric \emph{black rings}~\cite{Elvang:2004ds}, i.e., for $H \cong S^1 \times S^2$, remains an open problem.
\item An analogous result has been obtained for minimal supergravity theory coupled to an arbitrary number of vector multiplets \cite{Gutowski:2004bj}.
\end{itemize}
Analogous results for asymptotically AdS black holes in \emph{gauged} supergravity have yet-to-be-obtained and this remains a very interesting open problem. This is particularly significant due to the lack of black-hole uniqueness theorems even for pure gravity in AdS. However, it is worth mentioning that the analysis of~\cite{Gutowski:2004ez} used the homogeneous near-horizon geometry of Theorem~\ref{Thm:homogauged} with $H\cong S^3$, together with supersymmetry, to explicitly integrate for the full cohomogeneity-1 \AdS{5} black hole solution. It would be interesting to prove a uniqueness theorem for supersymmetric \AdS{5} black holes assuming $\mathbb{R}\times U(1)^2$ symmetry.

\subsubsection{Extremal vacuum black holes}

The classic black-hole uniqueness theorem of general relativity roughly states that any stationary, asymptotically-flat black-hole solution to the vacuum Einstein equations must be given by the Kerr solution. Traditionally this theorem assumed that the event horizon is non-degenerate, at a number of key steps. Most notably, the rigidity theorem, which states that a stationary rotating black hole must be axisymmetric, is proved by first showing that the event horizon is a Killing horizon. Although the original arguments~\cite{Hawking:1971vc} assumed non-degeneracy of the horizon, in four dimensions this assumption can be removed~\cite{IM, Moncrief:2008mr, Hollands:2008wn}.

This allows one to reduce the problem to a boundary value problem on a two-dimensional domain (the orbit space), just as in the case of a non-degenerate horizon. However, the boundary conditions near the boundary corresponding to the horizon depend on whether the surface gravity vanishes or not. Unsurprisingly, the boundary conditions near a degenerate horizon can be deduced from the near-horizon geometry. Curiously, this has only been realised rather recently. This has allowed one to extend the uniqueness theorem for Kerr to the degenerate case.

\begin{theorem}[\cite{Meineletal, Amsel:2009et, Figueras:2009ci, Chrusciel:2010gq}] The only four-dimensional, asymptotically-flat, stationary and axisymmetric, rotating, black-hole solution of the Einstein vacuum equations, with a connected degenerate horizon with non-toroidal horizon sections, is the extremal Kerr solution.
\end{theorem}

We note that~\cite{Meineletal} employs methods from integrability and the inverse scattering method. The remaining proofs employ the near-horizon geometry classification theorem discussed in Section~\ref{vac:4d} together with the standard method used to prove uniqueness of non-extremal Kerr. The above uniqueness theorem has also been established for the extremal Kerr--Newman black hole in Einstein--Maxwell theory~\cite{Amsel:2009et, Chrusciel:2010gq, Meinel:2011wu}.

The assumption of a non-toroidal horizon is justified by the black-hole--horizon topology theorems. Similarly, as discussed above, axisymmetry is justified by the rigidity theorem under the assumption of analyticity. Together with these results, the above theorem provides a complete classification of rotating vacuum black holes with a single degenerate horizon, under the stated assumptions. The proof that a non-rotating black hole must be static has only been established for non-degenerate horizons, so the classification of non-rotating degenerate black holes remains an open problem.

Of course, in higher dimensions, there is no such simple general uniqueness theorem. For spacetimes with $\mathbb{R}\times U(1)^{D-3}$ symmetry though, one has a mathematical structure analogous to $D=4$ stationary and axisymmetric spacetimes. Namely, one can reduce the problem to an integrable boundary-value problem on a 2D orbit space. However in this case the boundary data is more complicated. It was first shown that non-degenerate black-hole solutions in this class are uniquely specified by certain topological data, which specifies the $U(1)^{D-3}$-action on the manifold, referred to as interval data (i.e., rod structure)~\cite{Hollands:2007aj,Hollands:2008fm}. The proof is entirely analogous to that for uniqueness of Kerr amongst stationary and axisymmetric black holes. This has been extended to cover the degenerate case, again by employing the near-horizon geometry to determine the correct boundary conditions.

\begin{prop}[\cite{Figueras:2009ci}]
Consider a five-dimensional, asymptotically-flat, stationary black-hole solution of the vacuum Einstein equations, with $\mathbb{R}\times U(1)^2$ isometry group and a connected degenerate horizon (with non-toroidal sections). There exists at most one such solution with given angular momenta and a given interval structure.
\end{prop}

It is worth emphasising that although there is no near-horizon uniqueness theorem in this case, see Section~\ref{vac:5d}, this result actually only requires the general $SO(2,1)\times U(1)^2$ form for the near-horizon geometry and not its detailed functional form.

It seems likely that the results of this section could be extended to $\mathbb{R} \times U(1)^{D-3}$ invariant extremal black holes in Einstein--Maxwell type theories in higher dimensions. In $D=5$ it is known that coupling a CS term in such a way to give the bosonic sector of minimal supergravity, implies such solutions are determined by a non-linear sigma model analogous to the pure vacuum case. Hence it should be straightforward to generalise the vacuum uniqueness theorems to this theory.

\subsection{Stability of near-horizon geometries and extremal black holes}

All known near-horizon geometries are fibrations of the horizon cross section $H$ over an \AdS{2} base (see Section~\ref{sec:ads2theorems}). By writing the \AdS{2} in global coordinates one obtains examples of smooth complete spacetimes that solve the Einstein equations. Explicitly, by converting the near-horizon geometry (\ref{nhads2gen}) to \AdS{2} global coordinates, such spacetimes take the general form
\begin{eqnarray}
\td s^2 &=&\ell^2 \Gamma(y) \left[ - \cosh^2\rho \,\td t^2+ \td \rho^2 \right]+ \gamma_{mn}(y) \td y^m \td y^n \nonumber \\ &&\quad + \gamma_{IJ}(y)(\td\phi^I+\ell^2 k^I \sinh\rho \, \td t)(\td\phi^J+\ell^2 k^J \sinh\rho \, \td t) \,, \label{globalads2}
\end{eqnarray}
where we have written the constant $A_0=-\ell^{-2}$ in terms of the radius of \AdS{2}. These spacetimes possess two timelike boundaries, and of course do not contain a horizon. It is of interest to consider the stability of such ``global'' near-horizon geometries, as spacetimes in their own right. It turns out that this problem is rather subtle.

In fact, general arguments suggest that any near-horizon geometry must be unstable when backreaction is taken into account and the nearby solution must be singular~\cite{Bardeen:1999px}. This follows from the fact that $H$ is marginally trapped, so there exist perturbations that create a trapped surface and by the singularity theorems the resulting spacetime must be geodesically incomplete. If the perturbed solution is a black hole sitting inside the near-horizon geometry, then this need not be an issue.\epubtkFootnote{Similarly, higher-dimensional pure-AdS spacetime is unstable to the formation of small black holes~\cite{Bizon:2011gg}.} For $\mathrm{AdS}_2\times S^2$, heuristic arguments also indicating its instability have been obtained by dimensional reduction to an \AdS{2} theory of gravity~\cite{Maldacena:1998uz}. In particular, this suggests that the backreaction of matter in $\mathrm{AdS}_2 \times S^2$, is not consistent with a fall-off preserving both of the \AdS{2} boundaries.

So far we have been talking about the \emph{non-linear} stability of near-horizon geometries. Of course, linearised perturbations in these backgrounds can be analysed in more detail. A massless scalar field in the near-horizon geometry of an extremal Kerr black hole (NHEK) reduces to a massive charged scalar field in \AdS{2} with a homogeneous electric field~\cite{Bardeen:1999px}. This turns out to capture the main features of the Teukolsky equation for NHEK, which describes linearised \emph{gravitational} perturbations of NHEK~\cite{Amsel:2009ev,Dias:2009ex}. In contrast to the above instability, these works revealed the stability of NHEK against linearised gravitational perturbations. In fact, one can prove a non-linear uniqueness theorem in this context: any stationary and axisymmetric solution that is asymptotic to NHEK, possibly containing a smooth horizon, must in fact be NHEK~\cite{Amsel:2009ev}.

So far we have discussed the stability of near-horizon geometries as spacetimes in their own right. A natural question is what information about the stability of an extremal black hole can be deduced from the stability properties of its near horizon geometry. Clearly, stability of the near-horizon geometry is insufficient to establish stability of the black hole, but it has been argued that certain instabilities of the near-horizon geometry imply instability of the black hole~\cite{Durkee:2010ea}. For higher dimensional vacuum near-horizon geometries, one can construct gauge-invariant quantities (Weyl scalars), whose perturbation equations decouple, generalising the Teukolsky equation \cite{Durkee:2010ea}.\epubtkFootnote{This stems from the fact that such spacetimes admit null geodesic congruences with vanishing expansion, rotation, and shear (i.e., they are Kundt spacetimes and hence algebraically special).} One can then perform a KK reduction on $H$ to find that linearised gravitational (and electromagnetic) perturbations reduce to an equation for a massive charged scalar field in \AdS{2} with a homogeneous electric field (as for the NHEK case above). The authors of \cite{Durkee:2010ea} \emph{define} the near-horizon geometry to be unstable if the effective Breitenlohner--Freedman bound for this charged scalar field is violated. They propose the following conjecture: an instability of the near-horizon geometry (in the above sense), implies an instability of the associated extreme black hole, provided the unstable mode satisfies a certain symmetry requirement. This conjecture is supported by the linear stability of NHEK and was verified for the known instabilities of odd dimensional cohomogeneity-1 MP black holes~\cite{Dias:2010eu}. It is also supported by the known stability results for the five-dimensional MP black hole~\cite{Murata:2011my} and the Kerr-\AdS{4} black hole~\cite{Dias:2012pp}. This conjecture suggests that even-dimensional near-extremal MP black holes, which are more difficult to analyse directly, are also unstable~\cite{Tanahashi:2012si}.

Recently, it has been shown that extremal black holes exhibit linearised instabilities at the horizon. This was first observed for a massless scalar field in an extremal RN and extremal Kerr--black-hole background~\cite{Aretakis:2011gz, Aretakis:2011ha, Aretakis:2011hc, Aretakis:2012ei}. The instability is somewhat subtle. While the scalar decays everywhere on and outside the horizon, the first transverse derivative of the scalar does not generically decay on the horizon, and furthermore the $k$th-transverse derivative blows up as $v^{k-1}$ along the horizon. These results follow from the existence of a non-vanishing conserved quantity on the horizon linear in the scalar field. If the conserved quantity vanishes it has been shown that a similar, albeit milder, instability still occurs on the horizon~\cite{Dain:2012qw, Bizon:2012we, Aretakis:2012bm, Lucietti:2012xr}. It should be noted that this instability is not in contradiction with the above linear stability of the near-horizon geometry. From the point of view of the near-horizon geometry, it is merely a coordinate artefact corresponding to the fact that the Poincar\'e horizon of \AdS{2} is not invariantly defined~\cite{Lucietti:2012xr}.

The horizon instability has been generalised to a massless scalar in an arbitrary extremal black hole in any dimension, provided the near-horizon geometry satisfies a certain assumption~\cite{Lucietti:2012sf}. This assumption in fact follows from the \AdS{2}-symmetry theorems and hence is satisfied by all known extremal black holes. Furthermore, by considering the Teukolsky equation, it was shown that a similar horizon instability occurs for linearised gravitational perturbations of extremal Kerr~\cite{Lucietti:2012sf}. This was generalised to certain higher-dimensional vacuum extremal black holes~\cite{Murata:2012ct}. Similarly, using Moncrief's perturbation formalism for RN, it was shown that coupled gravitational and electromagnetic perturbations of extremal RN within Einstein--Maxwell theory also exhibit such a horizon instability~\cite{Lucietti:2012sf}. An interesting open question is what is the fate of the \emph{non-linear} evolution of such horizon instabilities. To this end, an analogous instability has been established for certain non-linear wave equations~\cite{Aretakis:2013dpa}.

\subsection{Geometric inequalities} 

Interestingly, near-horizon geometries saturate certain geometric bounds relating the area and conserved angular momentum and charge of dynamical axisymmetric horizons, see~\cite{Dain:2011mv} for a review. 

In particular, for four-dimensional dynamical axially-symmetric spacetimes, the area of black-hole horizons with a given angular momentum is minimised by the extremal Kerr black hole.
 The precise statement is:
\begin{theorem}[\cite{Dain:2011pi, Jaramillo:2011pg}] Consider a spacetime satisfying the Einstein equations with a non-negative cosmological constant and matter obeying the dominant energy condition. The area of any axisymmetric closed (stably outermost) marginally--outer-trapped surface $S$ satisfies
\begin{equation} \label{Dainbound}
A \geq 8\pi |J| \,,
\end{equation} 
where $J$ is the angular momentum of $S$. Furthermore, this bound is saturated if and only if the metric induced on $S$ is that of the (near-)horizon geometry of an extreme Kerr black hole.
\end{theorem}
Furthermore, it has been shown that if $S$ is a section of an isolated horizon the above equality is saturated if and only if the surface gravity vanishes~\cite{Jaramillo:2012zi} (see also~\cite{Mars:2012sb}). An analogous area-angular momentum-charge inequality has been derived in Einstein--Maxwell theory, which is saturated by the extreme Kerr--Newman black hole~\cite{Clement:2012vb}.

The above result can be generalised to higher dimensions, albeit under stronger symmetry assumptions.

\begin{theorem}[\cite{Hollands:2011sy}]
Consider a $D$-dimensional spacetime satisfying the vacuum Einstein equations with non-negative cosmological constant $\Lambda$ that admits a $U(1)^{D-3}$-rotational isometry. Then the area of any (stably outer) marginally--outer-trapped surface satisfies $A \geq 8\pi |J_+ J_-|^{1/2}$ where $J_\pm$ are distinguished components of the angular momenta associated to the rotational Killing fields, which have fixed points on the horizon. Further, if $\Lambda=0$ then equality is achieved if the spacetime is a near-horizon geometry and conversely, if the bound is saturated, the induced geometry on spatial cross sections of the horizon is that of a near-horizon geometry. 
\end{theorem} 

In particular, for $D=4$, the horizon is topologically $S^2$ and $J_+=J_-$ and one recovers \eqref{Dainbound}.

Other generalisations of such inequalities have been obtained in $D=4,5$ Einstein--Maxwell-dilaton theories~\cite{Yazadjiev:2012bx, Yazadjiev:2013hk}.

The proof of the above results involve demonstrating that axisymmetric near-horizon geometries are global minimisers of a functional of the form (\ref{eq:functional}) that is essentially the energy of a harmonic map, as discussed in Section~\ref{gauge:hidden}.

\subsection{Analytic continuation}

In this section we discuss analytic continuation of near-horizon geometries to obtain other Lorentzian or Riemannian metrics. As we will see, this uncovers a number of surprising connections between seemingly different spacetimes and geometries.

As discussed in Section~\ref{sec:ads2theorems}, typically near-horizon geometries have an $SO(2,1)$ isometry. Generically, the orbits of this isometry are three-dimensional line or circle bundles over \AdS{2}. One can often analytically continue these geometries so $\mathrm{AdS}_2 \to S^2$ and $SO(2,1)\to SU(2)$ (or $SO(3)$) with orbits that are circle bundles over $S^2$. It is most natural to work with the near-horizon geometry written in global \AdS{2} coordinates (\ref{globalads2}). Such analytic continuations are then obtained by setting $\rho \to i (\theta -\tfrac{\pi}{2} )$ and $k^I \to i k^I$.

First we discuss four dimensions. One can perform an analytic continuation of the near-horizon geometry of extremal Kerr to obtain the zero mass Lorentzian Taub--NUT solution~\cite{Kunduri:2007vf}. More generally, there is an analytic continuation of the near-horizon geometry of extremal Kerr--Newman-$\Lambda$ to the zero mass Lorentzian RN--Taub--NUT-$\Lambda$ solution~\cite{Kunduri:2008tk}. In fact, Page constructed a smooth compact Riemannian metric on the non-trivial $S^2$-bundle over $S^2$ with $SU(2)\times U(1)$ isometry, by taking a certain limit of the Euclidean Kerr-dS metric~\cite{Page}. He showed that locally his metric is the Euclidean Taub--NUT-$\Lambda$ with zero mass. Hence, it follows that there exists an analytic continuation of the near-horizon geometry of extremal Kerr-$\Lambda$ to Page's Einstein manifold. (Also we deduce that Page's limit is a Riemannian version of a combined extremal and near-horizon limit). 

Explicitly, the analytic continuation of the near-horizon extremal Kerr-$\Lambda$ metric~(\ref{NHKerrAdS}) to the Page metric, can be obtained as follows. First write the near-horizon geometry in global coordinates (\ref{globalads2}), then analytically continue $\rho\to i (\theta -\tfrac{\pi}{2} )$ and $\ell^2 k= 2i \alpha^2 , \ell^2 \beta = 4\alpha^2$ and redefine the coordinates $(t,\phi) \to (\phi, \psi)$ appropriately, to find 
\begin{equation} 
\label{Pagemet}
\td s^2 =\frac{\alpha^2(1 - x^2) \td x^2}{P(x)} +\frac{4\alpha^2P(x)}{(1- x^2)}\left(\td \psi + \cos\theta \td \phi \right)^2 + \alpha^2(1-x^2) \left(\td \theta^2 + \sin^2\theta \td\phi^2\right) 
\end{equation} with 
\begin{equation}
P(x) = 1+x^2 - (1+2x^2 - \tfrac{1}{3} x^4)\alpha^2\Lambda \,.
\end{equation} 
By an appropriate choice of parameters, this metric extends to a smooth global, inhomogeneous Einstein metric on the non-trivial $S^2$ bundle over $S^2$, as follows. Compactness and positive definiteness requires one to take $\Lambda>0$ and $-x_1 < x<x_1$ where $\pm x_1$ are two adjacent roots of $P(x)$ such that $|x_1|<1$. The $(x,\psi)$ part of the metric has conical singularities at $x=\pm x_1$, which are removed by imposing
\begin{equation}
\Delta \psi = \frac{2\pi x_1}{1-\Lambda \alpha^2(1-x_1^2)}
\end{equation} 
resulting in an $S^2$ fibre. This fibration is globally defined if $m \Delta \psi = 4\pi$, where $m \in \mathbb{N}$, so combining these results implies
\begin{equation}\label{Pagecond}
m = \frac{4x_1(3+x_1^2)}{3 + 6x_1^2-x_1^4} \,.
\end{equation} 
As Page showed, the only solution is $m=1$, which implies the $S^2$-bundle is non-trivial. This manifold is diffeomorphic to the first del~Pezzo surface $\mathbb{CP}^2 \# \, \overline{\mathbb{CP}^2}$. 

Similarly, there is an analytic continuation of the near-horizon geometry of the extremal Kerr--Newman-$\Lambda$ that gives a family of smooth Riemannian metrics on $S^2$-bundles over $S^2$ that satisfy the Riemannian Einstein--Maxwell equations (and hence have constant scalar curvature). Interestingly, this leads to an infinite class of metrics (i.e., there are an infinite number of possibilities for the integer $m$). The local solution can be derived directly by classifying solutions of the Riemannian Einstein--Maxwell equations with $SU(2) \times U(1)$ isometry group acting on three-dimensional orbits (see, e.g., \cite{Martelli:2012sz}). One finds the geometry is given by \eqref{Pagemet} but with $P(x)$ now given by 
\begin{equation}
P(x) = 1+x^2 + c - (1+2x^2 - \tfrac{1}{3} x^4)\alpha^2 \Lambda \,,
\end{equation}
where $c$ is a constant related to the electric and magnetic charges of the analytically continued extremal Kerr--Newmann black hole. The regularity condition now becomes
\begin{equation}
m = \frac{4x_1(3+x_1^2)}{3 + 6x_1^2 - x_1^4} + \frac{8c(3-x_1^2)x_1}{(1-x_1^2)(3+6x_1^2 - x_1^4)}\,,
\end{equation}
which for $c=0$ reduces to Eq.~\eqref{Pagecond}. One can check the 2nd term above is monotonically increasing in the range $0< x_1 < 1$ and unbounded as $x_1 \to 1$. If $c>0$ then there is a unique solution for every integer $m\geq 1$. For $c<0$ the only allowed solution is $m=1$, and in fact there exist values of $c<0$ such that there are \emph{two} solutions with $m=1$. For $m$ even, the metric extends to a global metric on the trivial $S^2$ bundle over $S^2$, whereas for $m$ odd, globally the space is $\mathbb{CP}^2 \# \, \overline{\mathbb{CP}^2}$. Hence one obtains a generalisation of the Page metric. 

Curiously, in five dimensions there exist analytic continuations of near-horizon geometries to stationary black-hole solutions~\cite{Kunduri:2007vf}. For example, one can perform an analytic continuation of the near-horizon geometry of an extremal MP black hole with $J_1 \neq J_2$ to obtain the full cohomogeneity-1 MP black hole with $J_1=J_2$ (which need not be extremal). In this case the $S^1$ bundle over $S^2$ that results after analytic continuation is the homogenous $S^3$ of the resulting black hole. This generalises straightforwardly with the addition of a cosmological constant and/or charge. Interestingly, this also works with the near-horizon geometries of extremal black rings. For example, there is an analytic continuation of the near-horizon geometry of the extremal dipole black ring that gives a static charged squashed KK black hole with $S^3$ horizon topology. In these five-dimensional cases, the isometry of the near-horizon geometries is $SO(2,1)\times U(1)^2$, which has 4D orbits; hence one can arrange the new time coordinate to lie in these orbits in such a way it is not acted upon by the $SU(2)$. This avoids NUT charge, which is inevitable in the four-dimensional case discussed above. As in the four-dimensional case, there are analytic continuations which result in Einstein metrics on compact manifolds. For example, there is an analytic continuation of the near-horizon geometry of extremal MP-$\Lambda$ that gives an infinite class of Einstein metrics on $S^3$-bundles over $S^2$, which were first found by taking a Page limit of the MP-dS black hole~\cite{Hashimoto:2004kc}.

In higher than five dimensions one can similarly perform analytic continuations of Einstein near-horizon geometries to obtain examples of compact Einstein manifolds. The near-horizon geometries of MP-$\Lambda$ give the Einstein manifolds that have been constructed by a Page limit of the MP-dS metrics~\cite{Gibbons:2004js}. On the other hand, the new families of near-horizon geometries~\cite{Kunduri:2010vg, Kunduri:2012hi}, discussed in Section~\ref{exotic}, analytically continue to new examples of Einstein metrics on compact manifolds that have yet to be explored.
		
So far we have discussed analytic continuations in which the \AdS{2} is replaced by $S^2$. Another possibility is to replace the \AdS{2} with hyperbolic space $\mathbb{H}^2$. For simplicity let us focus on static near-horizon geometries. Such an analytic continuation is then easily achieved by replacing the global \AdS{2} time with imaginary time, i.e., $ t \to i t$ in Eq.~(\ref{globalads2}). In this case, instead of a horizon, one gets a new asymptotic region corresponding to $\rho \to \infty$. General static Riemannian manifolds possessing an end that is asymptotically extremal in this sense were introduced in~\cite{Figueras:2011va}. Essentially, they are defined as static manifolds possessing an end in which the metric can be written as an Euclideanised static spacetime containing a smooth degenerate horizon. It was shown that Ricci flow preserves this class of manifolds, and furthermore asymptotically-extremal Ricci solitons must be Einstein spaces~\cite{Figueras:2011va}. These results were used to numerically simulate Ricci flow to find a new Einstein metric that has an interesting interpretation in the AdS/CFT correspondence~\cite{Figueras:2011va}, which we briefly discuss in Section \ref{app:branes}. It would be interesting to investigate non-static near-horizon geometries in this context.

\subsection{Extremal branes}
\label{app:branes}

Due to the applications to black-hole solutions, we have mostly focused on the near-horizon geometries of degenerate horizons with cross sections $H$ that are \emph{compact}. However, as we discussed in Section~\ref{section:nearhorizongeometry}, the concept of a near-horizon geometry exists for any spacetime containing a degenerate horizon, independent of the topology of $H$. In particular, extremal black branes possess horizons with non-compact cross sections $H$. Hence the general techniques discussed in this review may be used to investigate the classification of the near-horizon geometries of extremal black branes. In general, this is a more difficult problem, since as we have seen, compactness of $H$ can often be used to avoid solving for the general local metric by employing global arguments. Since it is outside the scope of this review, we will not give a comprehensive overview of this topic; instead we shall select a few noteworthy examples.

First consider \AdS{D} space written in Poincar\'e coordinates
\be
\td s^2 = y^2( -\td t^2 + \td x^i \td x^i) + \frac{\td y^2}{y^2} \,,
\ee
where $i=1, \dots, D-2$. The surface $y=0$, often called the Poincar\'e horizon, is a degenerate Killing horizon of the Killing field $K=\partial/ \partial t$. However, these coordinates are not valid at $y=0$ (the induced metric is singular) and hence are unsuitable for extracting the geometry of the Poincar\'e horizon. One may introduce coordinates adapted to the Poincar\'e horizons by constructing Gaussian null coordinates as described in Section~\ref{section:nearhorizongeometry}. We need to find null geodesics $\gamma(\lambda)$ that in particular satisfy $K \cdot \dot{\gamma} =1$. Explicitly, this condition is simply $\dot{t}= - y^{-2}$. Now, since $\partial/\partial x^i$ are Killing fields, along any geodesic the quantities $(\partial/ \partial x^i) \cdot \dot{\gamma}$ must be constant; this gives $\dot{x}^i = - k^i y^{-2}$, where $k^i$ are constant along the geodesics. The null constraint now simplifies to $\dot{y}^2 = 1- k^i k^i$ and so $k^ik^i<1$. This latter equation is easily integrated to give $y(\lambda)$ and using the above we may now integrate for $t(\lambda)$ and $x^i(\lambda)$. The result is
\be
y = \sqrt{(1- k^ik^i)} \,\lambda \,, \qquad 
t= v+ \frac{1}{(1-k^ik^i) \lambda} \,, \qquad 
x^i = \frac{k^i}{(1- k^i k^i) \lambda} \,,
\ee
where $v$ is an integration constant and we have set the other integration constants to zero to ensure the horizon is at $\lambda=0$. This gives a family of null geodesics parameterised by $(v,k^i)$, which shoot out from every point on the horizon; hence we may take the $(v,k^i)$ as coordinates on the horizon. We can then change from Poincar\'e coordinates $(t, x^i, y)$ to the desired Gaussian null coordinates $(v, \lambda, k^i)$. Indeed, one can check that in the coordinates $(v,\lambda, k^i)$, the Killing field $K= \partial /\partial v$ and the metric takes the form (\ref{degenerate}) (with $r=\lambda$). In fact, as is often the case, it is convenient to use a different affine parameter $r= \lambda (1-k^ik^i)$. Also, since $k^ik^i<1$ we may write $k^i = \tanh \eta \, \mu^i$, where $\mu^i \mu^i=1$ parameterise a unit $(D-3)$-sphere. Now the coordinate transformation becomes
\be
y= r \cosh \eta \,, \qquad 
t = v+\frac{1}{r} \,, \qquad 
x^i = \frac{\tanh \eta \, \mu^i}{r} \,,
\ee
and the \AdS{D} metric in these coordinates is
\be
\label{poincare}
\td s^2 = \cosh^2\eta \left(- r^2 \td v^2+ 2 \td v \td r \right) + \td \eta^2 +\sinh^2 \eta \, \td \Omega^2_{D-3} \,.
\ee
It is now manifest that the surface $r=0$ is a smooth degenerate Killing horizon of $\partial /\partial v$, corresponding to the Poincar\'e horizon, which we may now extend onto and through by taking $r\leq 0$. Cross sections of the Poincar\'e horizon are non-compact and of topology $\mathbb{R}^{D-2}$ with a (non-flat) induced metric given by the standard Einstein metric on hyperbolic space $\mathbb{H}^{D-2}$. Observe that the above expresses \AdS{D} as a warped product of \AdS{2} and hyperbolic space $\mathbb{H}^{D-2}$, i.e., as a static near-horizon geometry (with no need to take a near-horizon limit!), as observed in~\cite{Figueras:2011va}. 

The BPS extremal $D3, M2, M5$ black branes of 10,11-dimensional supergravity are well known to have a near-horizon geometry given by $\mathrm{AdS}_5 \times S^5$, $\mathrm{AdS}_4 \times S^7$ and $\mathrm{AdS}_7 \times S^4$ respectively with their horizons corresponding to a Poincar\'e horizon in the AdS factor. However, as discussed above, the standard Poincar\'e coordinates are not valid on the horizon and hence unsuitable if one wants to extend the brane geometry onto and beyond the horizon. To this end, it is straightforward to construct Gaussian null coordinates adapted to the horizon of these black branes and check their near-horizon limit is indeed given by Eq.~(\ref{poincare}) plus the appropriate sphere.\epubtkFootnote{We would like to thank Carmen Li for verifying this.}

Of course, extremal branes occur in other contexts. For example, the Randall--Sundrum model posits that we live on a 3\,+\,1 dimensional brane in a 4\,+\,1-dimensional bulk spacetime with a negative cosmological constant. A longstanding open problem has been to construct solutions to the five-dimensional Einstein equations with a black hole localised on such a brane, the \emph{brane-world} black hole.\epubtkFootnote{There is a vast literature on this problem, which we will not attempt to review here.} In the five-dimensional spacetime, the horizons of such black holes extend out into the bulk. \cite{Kaus:2009cg} constructed the near-horizon geometry of an extremal charged black hole on a brane. This involved constructing (numerically) the most general five dimensional static near-horizon geometry with $SO(3)$ rotational symmetry, which turns out to be a 1-parameter family generalisation of Eq.~(\ref{poincare}) (this is the 5D analogue of the 4D general static near-horizon geometry with a non-compact horizon, see Section~\ref{vac:static}). 

Notably, \cite{Figueras:2011gd} numerically constructed the first example of a brane-world black-hole solution. This corresponds to a Schwarzschild-like black hole on a brane suspended above the Poincar\'e horizon in \AdS{5}. An important step towards this solution was the construction of a novel asymptotically AdS Einstein metric with a Schwarzschild conformal boundary metric and an extremal Poincar\'e horizon in the bulk (sometimes called a black droplet)~\cite{Figueras:2011va}. This was found by numerically simulating Ricci flow on a suitable class of stationary and axisymmetric metrics. This solution is particularly interesting since by the AdS/CFT duality it is the gravity dual to a strongly coupled CFT in the Schwarzschild black hole, thus allowing one to investigate strongly coupled QFT in black-hole backgrounds. Recently, generalisations in which the boundary black hole is rotating have been constructed, in which case there is also the possibility of making the black hole on the boundary extremal~\cite{Figueras:2013jja, Fischetti:2013hja}.


\section*{Acknowledgements}

HKK is supported by an NSERC Discovery Grant. JL is supported by an ESPRC Career Acceleration Fellowship. We would especially like to thank Harvey Reall for several fruitful collaborations on this topic and numerous discussions over the years. We would also like to thank Pau Figueras and Mukund Rangamani for a collaboration on this topic. JL would also like to thank Pau Figueras, Keiju Murata, Norihiro Tanahashi and Toby Wiseman for collaborations on related topics.


\newpage


\bibliography{refs}

\end{document}